\documentclass[apj]{emulateapj}

\usepackage{amsmath,amssymb}
\usepackage{mathptmx}
\usepackage{graphicx}
\usepackage{datetime}

\usepackage[backref,breaklinks,colorlinks,citecolor=blue,linkcolor=blue]{hyperref}

\usepackage[all]{hypcap} 
\def\sectionautorefname~#1\null{Section #1\null}
\def\subsectionautorefname~#1\null{Section #1\null}
\def\subsubsectionautorefname~#1\null{Section #1\null}

\newcommand{\ltsim}{\raisebox{-.5ex}{$\;\stackrel{<}{\sim}\;$}}
\newcommand{\gtsim}{\raisebox{-.5ex}{$\;\stackrel{>}{\sim}\;$}}

\newcommand{\todo}{\ifmmode {\Huge \bullet} \else {\Huge$\bullet$}\fi}
\newcommand{\til}{\ifmmode \sim \else $\sim$\fi}
\newcommand{\tido}{\ifmmode {{\color{red}\bullet}} \else {\color{red}$\bullet$}\fi}

\newcommand{\mic}{\ifmmode {\rm \mu m} \else $\mu$m\fi}
\newcommand{\kms}{\ifmmode {\rm km\,s}^{-1} \else km\,s$^{-1}$\fi}
\newcommand{\ergs}{\ifmmode {\rm erg\,s}^{-1} \else erg s$^{-1}$\fi}
\newcommand{\kpc}{\ifmmode {\rm kpc} \else kpc\fi}
\newcommand{\Msun}{\ifmmode M_{\odot} \else $M_{\odot}$\fi}
\newcommand{\Lsun}{\ifmmode L_{\odot} \else $L_{\odot}$\fi}
\newcommand{\mpyr}{\ifmmode \Msun\,{\rm yr}^{-1} \else $\Msun\,{\rm yr}^{-1}$\fi}
\newcommand{\Msol}{\Msun}
\newcommand{\Lsol}{\Lsun}
\newcommand{\NDunit}{\ifmmode {\rm Mpc}^{-3} \else Mpc$^{-3}$\fi}

\newcommand{  \Lya      }{\ifmmode {\rm Ly}\alpha \else Ly$\alpha$\fi}
\newcommand{  \civ      }{\ifmmode {\rm C}\,\textsc{iv}  \else C\,\textsc{iv}\fi}
\newcommand{  \CIV      }{\ifmmode {\rm C}\,\textsc{iv}\,\lambda1549 \else C\,\textsc{iv}\,$\lambda1549$\fi}
\newcommand{  \mgii     }{\ifmmode {\rm Mg}\,\textsc{ii} \else Mg\,\textsc{ii}\fi}
\newcommand{  \MgII     }{\ifmmode {\rm Mg}\,\textsc{ii}\,\lambda2798 \else Mg\,\textsc{ii}\,$\lambda2798$\fi}

\newcommand{  \CII }{\ifmmode \left[{\rm C}\,\textsc{ii}\right]\,\lambda157.74\,\mu{\rm m} \else [C\,{\sc ii}]\ $\lambda157.74\,\mu{\rm m}$\fi}
\newcommand{  \cii }{\ifmmode \left[{\rm C}\,\textsc{ii}\right] \else [C\,{\sc ii}]\fi}
\newcommand{ \Lcii }{\ifmmode L_{\cii} \else $L_{\cii}$\fi}
\newcommand{ \fwcii }{\ifmmode {\rm FWHM}\cii \else FWHM\cii\fi}

\newcommand{ \fwhm  }{\ifmmode {\rm FWHM} \else FWHM\fi} 
\newcommand{ \fwhb  }{\ifmmode {\rm FWHM}\left(\hb\right) \else FWHM(\hb)\fi}
\newcommand{ \fwmg  }{\ifmmode {\rm FWHM}\left(\mgii\right) \else FWHM(\mgii)\fi}
\newcommand{  \lamLlam  }{\ifmmode \lambda L_{\lambda} \else $\lambda L_{\lambda}$\fi}
\newcommand{  \Lop      }{\ifmmode L_{5100} \else $L_{5100}$\fi}
\newcommand{\fbol}{\ifmmode f_{\rm bol} \else $f_{\rm bol}$\fi}
\newcommand{\fbolwv}{\ifmmode f_{\rm bol}\left(\lambda\right) \else $f_{\rm bol}\left(\lambda\right)$\fi}
\newcommand{\fbolopt}{\ifmmode f_{\rm bol}\left(5100{\rm \AA}\right) \else $f_{\rm bol}\left(5100{\rm \AA}\right)$\fi}
\newcommand{\fbolthree}{\ifmmode f_{\rm bol}\left(3000{\rm \AA}\right) \else $f_{\rm bol}\left(3000{\rm \AA}\right)$\fi}
\newcommand{ \vmax  }{\ifmmode v_{\rm max} \else $v_{\rm max}$\fi} 

\newcommand{  \mbh      }{\ifmmode M_{\rm BH} \else $M_{\rm BH}$\fi}
\newcommand{  \lledd    }{\ifmmode L/L_{\rm Edd} \else $L/L_{\rm Edd}$\fi}
\newcommand{  \mmedd    }{\ifmmode \dot{m}/\dot{m}_{\rm \,Edd} \else $\dot{m}/\dot{m}_{\rm \,Edd}$\fi}
\newcommand{  \Lbol     }{\ifmmode L_{\rm bol} \else $L_{\rm bol}$\fi}
\newcommand{  \Lthree   }{\ifmmode L_{3000} \else $L_{3000}$\fi}
\newcommand{  \Lagn     }{\ifmmode L_{\rm AGN} \else $L_{\rm AGN}$\fi}

\newcommand{  \tacc    }{\ifmmode t_{\rm acc} \else $t_{\rm acc}$\fi}
\newcommand{  \tgrow     }{\ifmmode t_{\rm growth} \else $t_{\rm growth}$\fi}
\newcommand{  \tUni      }{\ifmmode t_{\rm U} \else $t_{\rm U}$\fi}
\newcommand{  \Mdotbh	}{\ifmmode \dot{M}_{\rm BH} \else $\dot{M}_{\rm BH}$\fi}
\newcommand{  \Mdotdisk	}{\ifmmode \dot{M}_{\rm disk} \else $\dot{M}_{\rm disk}$\fi}
\newcommand{  \Mdotacc	}{\ifmmode \dot{M}_{\rm acc} \else $\dot{M}_{\rm acc}$\fi}
\newcommand{  \Mdotthin	}{\ifmmode \dot{M}_{\rm thin} \else $\dot{M}_{\rm thin}$\fi}

\newcommand{  \as	}{\ifmmode a_{\rm *} 		\else $a_{\rm *}$\fi}
\newcommand{  \avec	}{\ifmmode \vec{a}_{\rm *} 	\else $\vec{a}_{\rm *}$\fi}
\newcommand{  \re	}{\ifmmode \eta      	\else $\eta$\fi}
\newcommand{  \mseed    }{\ifmmode M_{\rm seed} \else $M_{\rm seed}$\fi}

\newcommand{  \mbul     }{\ifmmode M_{\rm bulge} \else $M_{\rm bulge}$\fi} 
\newcommand{  \mstar    }{\ifmmode M_{*} \else $M_{*}$\fi} 
\newcommand{  \mgal     }{\ifmmode M_{*} \else $M_{*}$\fi} 
\newcommand{  \mhost    }{\ifmmode M_{\rm host} \else $M_{\rm host}$\fi}
\newcommand{  \mmsmall  }{\ifmmode M_{\rm BH}/M_{*} \else $M_{\rm BH}/M_{*}$\fi}
\newcommand{  \mmlarge  }{\ifmmode M_{*}/M_{\rm BH} \else $M_{*}/M_{\rm BH}$\fi}

\newcommand{  \mmdotlarge}{\ifmmode \dot{M}_*/\Mdotbh \else $\dot{M}_*/\Mdotbh$\fi}
\newcommand{  \mmdotsmall}{\ifmmode \Mdotbh/\dot{M}_* \else $\Mdotbh/\dot{M}_*$\fi}

\newcommand{  \mmwp     }{\ifmmode \left(M_{*}/M_{\rm BH}\right) \else $\left(M_{*}/M_{\rm BH}\right)$\fi}
\newcommand{  \ml       }{\ifmmode M_{*}/L_{*} \else $M_{*}/L_{*}$\fi}
\newcommand{  \mlwp     }{\ifmmode \left(M_{*}/L\right) \else $\left(M_{*}/L\right)$\fi}
\newcommand{  \mlk      }{\ifmmode \left(M_{*}/L_{K}\right) \else $\left(M_{*}/L_{K}\right)$\fi}
\newcommand{  \sigs     }{\ifmmode \sigma_{*} \else $\sigma_{*}$\fi}
\newcommand{  \LSF      }{\ifmmode L_{\rm SF}  \else $L_{\rm SF}$\fi}
\newcommand{  \LFIR     }{\ifmmode L_{\rm FIR} \else $L_{\rm FIR}$\fi}
\newcommand{  \Lfir     }{\ifmmode L_{\rm FIR} \else $L_{\rm FIR}$\fi}
\newcommand{  \LTIR     }{\ifmmode L_{\rm TIR} \else $L_{\rm TIR}$\fi}
\newcommand{  \Ltir     }{\ifmmode L_{\rm TIR} \else $L_{\rm TIR}$\fi}

\newcommand{  \mdyn     }{\ifmmode M_{\rm dyn} \else $M_{\rm dyn}$\fi} 
\newcommand{  \mgas     }{\ifmmode M_{\rm gas} \else $M_{\rm gas}$\fi} 
\newcommand{  \sfr      }{\ifmmode {\rm SFR} \else SFR\fi}

\newcommand{  \jwst    }  {{\it JWST}}

\newcommand{  \Spitzer }  {{\it Spitzer}}

\newcommand{  \herschel} {{\it Herschel}}

\newcommand{\Ntot  }{six}
\newcommand{\Nbright  }{three}
\newcommand{\Nfaint   }{three}
\newcommand{\Nsmg     }{three}
\newcommand{\LLfir}{\ifmmode \Lcii/\Lfir \else $\Lcii/\Lfir$\fi}
\newcommand{\zrange}{\ifmmode z\simeq4.6-4.9 \else $z\simeq4.6-4.9$\fi}
\newcommand{\zpaper}{\ifmmode z\simeq4.8 \else $z\simeq4.8$\fi}
\newcommand{\zfpe}{$z \simeq 4.8$}
\newcommand{\zsix}{$z \simeq 6.2$}

\newcommand{\obsband}{band-7}
\newcommand{\obscycle}{2}

\begin{document}

\title{
ALMA Observations Show Major Mergers Among the Host Galaxies of Fast-growing, High-redshift,​ Supermassive​ Black Holes
}

\shorttitle{Merger-Driven Growth of High-$z$ SMBHs with ALMA}
\shortauthors{Trakhtenbrot et al.}

\author{
Benny Trakhtenbrot\altaffilmark{1,8},
Paulina Lira\altaffilmark{2},
Hagai Netzer\altaffilmark{3},
Claudia Cicone\altaffilmark{1,4},
Roberto Maiolino\altaffilmark{5,6},
Ohad Shemmer\altaffilmark{7}
}

\altaffiltext{1}{Institute for Astronomy, Department of Physics, ETH Zurich,\\ Wolfgang-Pauli-Strasse 27, CH-8093 Zurich, Switzerland}
\altaffiltext{2}{Departamento de Astronomia, Universidad de Chile,\\ Camino del Observatorio 1515, Santiago, Chile}
\altaffiltext{3}{School of Physics and Astronomy and the Wise Observatory,\\ The Raymond and Beverly Sackler Faculty of Exact Sciences,\\ Tel-Aviv University, Tel-Aviv 69978, Israel}
\altaffiltext{4}{INAF-Osservatorio Astronomico di Brera,\\ via Brera 28, I-20121, Milano, Italy}
\altaffiltext{5}{Cavendish Laboratory, University of Cambridge,\\ 19 J.\ J.\ Thomson Avenue, Cambridge CB3 0HE, UK}
\altaffiltext{6}{Kavli Institute for Cosmology, University of Cambridge,\\ Madingley Road, Cambridge CB3 0HA, UK}
\altaffiltext{7}{Department of Physics, University of North Texas,\\ Denton, TX 76203, USA}
\altaffiltext{8}{Zwicky Postdoctoral Fellow}

\slugcomment{Published as \textbf{ ApJ, 836, 8} }

\email{benny.trakhtenbrot@phys.ethz.ch}

\begin{abstract}
We present new ALMA band-7 data for a sample of six luminous quasars at \zfpe, powered by fast-growing supermassive black holes (SMBHs) with rather uniform properties: the typical accretion rates and black hole masses are $\lledd\simeq0.7$ and $\mbh\simeq10^{9}\,\Msun$. 
Our sample consists of three ``FIR-bright'' sources which were individually detected in previous \herschel/SPIRE observations, with star formation rates of $\sfr>1000\,\mpyr$, and three ``FIR-faint'' sources for which \herschel\ stacking analysis implies a typical SFR of $\sim$400 \mpyr. 
The dusty interstellar medium in the hosts of all six quasars is clearly detected in the ALMA data and resolved on scales of $\sim$2 kpc, in both continuum ($\lambda_{\rm rest}\sim150\,\mic$) and \CII\ line emission. 
The continuum emission is in good agreement with the expectations from the \herschel\ data, confirming the intense SF activity in the quasar hosts. 
Importantly, we detect companion sub-millimeter galaxies (SMGs) for three sources -- one FIR-bright and two FIR-faint, separated by $\sim14-45\,\kpc$ and $<450\,\kms$ from the quasar hosts. 
The \cii-based dynamical mass estimates for the interacting SMGs are within a factor of $\sim$3 of the quasar hosts' masses, while the continuum emission implies ${\rm SFR}_{\rm quasar} \sim \left(2-11\right)\times {\rm SFR}_{\rm SMG}$. 
Our ALMA data therefore clearly support the idea that major mergers are important drivers for rapid early SMBH growth. 
However, the fact that not all high-SFR quasar hosts are accompanied by interacting SMGs and the gas kinematics as observed by ALMA suggest that other processes may be fueling these systems. 
Our analysis thus demonstrates the diversity of host galaxy properties and gas accretion mechanisms associated with early and rapid SMBH growth.
\end{abstract}

\section{Introduction}
\label{sec:intro}

\setcounter{footnote}{0}

Several lines of evidence suggest that the growth histories of Supermassive Black Holes (SMBHs) are closely linked to that of their host galaxies. 
These include 
the well-known scaling relations between the SMBH mass (\mbh) and several properties of the (bulge component of the) hosts, observed in local relic systems (see \citealt{KormendyHo2013_MM_Rev} for a recent review). 
and the coincidence of intense star formation (SF) and SMBH growth, at higher redshifts.
Indeed, for systems dominated by accretion onto the SMBH -- identified as active galactic nuclei (AGN) -- the luminosities associated with SMBH accretion (\Lagn) and with star formation (\LSF) are correlated over several orders of magnitude (e.g., \citealt{Netzer2009_SF_AGN,Lutz2010,Shao2010,RosarioTrakht2013}, but see also \citealt{Page2012}). 
This suggests that the phase of fast SMBH growth occurs in tandem with intense SF activity, reaching star formation rates (SFRs) of $\sim$1000 \mpyr\ for SMBHs with $\Mdotbh \sim15\,\mpyr$ (i.e. $\Lagn\sim10^{47}\,\ergs$). 
All this supports a general idea that both processes (SF and AGN) are fed by a common reservoir of cold gas that collapses, forms stars, and (eventually) reaches the central region of the host galaxy to be accreted by the SMBH.

A particularly popular framework for the co-evolution of SMBHs and their hosts focuses on major mergers between massive, gas-rich galaxies.
Theoretical studies highlight the ability of such mergers to account for both the observed properties of AGN and SF galaxies, and for the SMBH-host relations in relic systems (see, e.g., \citealt{Sanders1988,DiMatteo2005,Hopkins2006} and the review by \citealt{Alexander_Hickox_Durham_2012}).
In particular, many simulations of such mergers show a relatively short episode (of order 100 Myr) of parallel intense SF and AGN activity, with SFRs reaching several hundred \mpyr\ \cite[or exceeding $\sim1000\,\mpyr$ in some simulations; see, e.g.,][]{Blecha2011,DeBuhr2011,Sijacki2011,Capelo2015_mergers,Volonteri2015_SF_BH_mergers,Gabor2016_GASOLINE_RAMSES}.
Observationally, however, the relevance of mergers to fast SMBH growth, and indeed to the co-evolutionary framework, is not yet well established.
While some studies of low-$z$ luminous AGN have reported a high occurrence of mergers \cite[e.g.,][]{Bahcall1997,Veilleux2009_QUEST_HST,Koss2011_BAT_hosts}, several other studies have demonstrated that interacting AGN hosts are relatively rare at $1\ltsim z\ltsim 2$ and their occurrence rate does not exceed what was found for non-active galaxies \cite[e.g.,][]{Gabor2009_AGN_morph_COS,Bennert2011,Cisternas2011_nomerger,Mainieri2011,Schawinski2011_disks,Schawinski2012_disks,Kocevski2012_CANDELS_mergers}.
A possible explanation for these apparently contradictory results, as put forward by \cite{Treister2012_mergers}, is that the merger-driven scenario may only be relevant for the epochs of fastest growth of the most massive BH (that is, highest \Lagn), and at $z>2$ -- when the overall rate of major mergers is expected to be higher \cite[e.g.,][]{Genel2009,Hopkins2010_merger_rate} and the amount of gas in the relevant halos is considerably larger \cite[but see also][]{Dubois2015_SN_BH_fb}.

As the relevance of mergers to SMBH growth is still debated, several recent studies have highlighted the importance of alternatives to the merger scenario.
Direct flows of cold gas from the intergalactic medium \cite[IGM; e.g.,][]{Dekel2009_cold_streams,DiMatteo2012,Dubois2012_hiz_inflows}, which may also trigger ``secular'' instabilities of the gas or the stars in the close environment of the SMBH \cite[e.g.,][]{Springel2005_sims,Bournaud2011_VDI_sims,Bournaud2012_VDI_obs}, are claimed to be most relevant for the early and fast growth of high-redshift SMBHs.
Such models, however, usually produce SFRs of only a few hundred \mpyr.
All this suggests that the best way to test and understand the relevance of the merger-driven scenario for SMBH growth is to focus on well-defined samples of fast-growing SMBHs, preferably at early cosmic epochs ($z>4$), when the most massive BHs were growing at maximal rates \cite[][]{Trakhtenbrot2011,TrakhtNetzer2012_Mg2,DeRosa2014}.
Another consequence of the aforementioned scenarios is that such fast-growing SMBHs would be predominantly found in significantly over-dense large-scale environments, where the rate of mergers is yet higher and where cold gas streams are expected to converge and provide ample gas supply to the SMBHs \cite[e.g.,][]{DiMatteo2008,Sijacki2009,Dubois2012_hiz_inflows,Costa2014_z6_env_sims}.

Testing these ideas observationally is, however, extremely challenging. 
The AGN-related emission dominates over most of the optical-NIR spectral regime, significantly limiting the prospects of determining the host properties.
The cosmic environments of the SMBHs are often characterized by searching for nearby (rest-frame) UV-bright galaxies, without precise redshift determinations, and possibly missing dusty obscured SF galaxies. 
Indeed, several studies provided ambiguous evidence for over-densities around some, but definitely not all $z\gtrsim5$ quasars \cite[e.g.,][and references therein]{Willott2005_z6_comp,Overzier2006_z52,Kim2009_idrops_z6,Utsumi2010_env_J2329,Banados2013_z57_env,Husband2013,Simpson2014_ULASJ1120_env}.

The advent of large and sensitive sub-millimeter (sub-mm) interferometric arrays, such as the IRAM Plateau de Bure interferometer (now NOEMA) and the Atacama Large Millimeter/sub-millimeter Array (ALMA), has enabled the direct observation of the hosts of high-redshift quasars in a spectral regime that is mostly uncontaminated by the AGN emission.
The early study of \citet[][using the SMA]{Iono2006_z47} demonstrated the ability of such data to reveal major mergers among quasar hosts, presenting a close merger between two $z=4.7$ SF galaxies powering a \emph{pair} of AGN, one of which is a luminous quasar. 
The ALMA study of the same system \cite[][]{Wagg2012_z47_QSO} showcased the revolutionary increase of spatial resolution and sensitivity provided by ALMA.
In recent years, a growing number of $z\gtrsim5$ quasars were studied with various sub-mm facilities  \cite[e.g.,][]{Maiolino2005_CII_J1148,Walter2009_CII_J1148,Gallerani2012,Venemans2012_z71_CII,Venemans2016_z6_cii,Wang2013_z6_ALMA,Willott2013_z6_ALMA,Willott2015_CFHQS_ALMA,Cicone2015_J1148}, 
focusing on the continuum emission, which originates from SF-heated dust in the interstellar medium (ISM) of the hosts, and the \CII\ emission line, which is among the most efficient ISM coolants \cite[e.g.,][and references therein]{Stacey1991,Stacey2010,Carilli_ARAA_2013}.
These studies have provided additional support for the coexistence of fast-growing SMBHs and intense SF activity in their hosts, but mostly could not address the questions related to the merger-driven scenario, due to limited sensitivity, resolution, and/or field of view. 
Indeed, no major mergers were identified in the aforementioned studies of high-redshift AGN.

In this study, we focus on a sample of quasar hosts at \zpaper, for which we have accumulated a wealth of multi-wavelength data.
In our previous studies, we have shown that the SMBHs powering these quasars trace an epoch of fast, Eddington-limited growth from massive BH seeds, which is expected to form the most massive BHs known by $z\sim4$ \cite[][T11 hereafter]{Trakhtenbrot2011}.
Our \herschel/SPIRE campaign \cite[][M12 and N14 hereafter]{Mor2012_z48,Netzer2014_z48_SFR} showed that these fast-growing SMBHs are hosted by SF galaxies, with $\sim1/4$ of the systems exceeding $\sfr\sim1000\,\mpyr$, while a stacking analysis of the other $\sim3/4$ of the systems suggested a typical SFR of $\sim$400\,\mpyr \cite[see also][]{Netzer2016_herschel_hiz}.
The extremely high SFRs found for the FIR-bright sources were interpreted as tracing major merger activity, while the lower SFRs found for most systems were thought to be tracing the early stages of SF suppression by AGN-driven ``feedback''. 
The poor spatial resolution of the \herschel\ data ($\sim18\arcsec$, or $\gtrsim100\,\kpc$) was, however, insufficient to test these ideas, which can be now addressed directly with the ALMA FIR continuum and \CII\ emission line observations presented in this study.
In \autoref{sec:obs_data_analysis} we describe the sample, the ALMA observations, data reduction, and analysis.
In \autoref{sec:res_and_disc} we compare the ISM properties and the occurrence of close (interacting) companions to those found in other samples of high-redshift AGN.
We summarize our main findings in \autoref{sec:summary}.
Throughout this work, we assume a cosmological model with $\Omega_{\Lambda}=0.7$, $\Omega_{\rm M}=0.3$, and
$H_{0}=70\,\kms\,{\rm Mpc}^{-1}$, which provides an angular scale of about $6.47\,\kpc/\arcsec$ at $z=4.7$ - the typical redshift of our sources. 
We further assume the stellar initial mass function (IMF) of \cite{Chabrier2003_IMF_rev}.

%
\section{Sample, ALMA Observations, and Data Analysis}
\label{sec:obs_data_analysis}

\capstartfalse
\begin{deluxetable*}{lccccccl}
\tablecolumns{8}
\tablewidth{0pt}
\tablecaption{Observations Log \label{tab:obs_log}}
\tablehead{
\colhead{sub-sample} &
\colhead{Target}  &
\colhead{$N_{\rm Ant}$ \tablenotemark{a}} &
\colhead{$T_{\rm exp}$} &
\colhead{$F_\nu$ rms \tablenotemark{b}} &
\colhead{Beam Size} &
\colhead{Pixel Size} & 
\colhead{ALMA Companions} \\
       &   ID                      &       & sec  & mJy/beam           & \sq\arcsec       & \arcsec & 
}
\startdata
Bright & SDSS J033119.67-074143.1  & 29    &  792 & $9.2\times10^{-2}$ & $0.41\times0.31$ & 0.06 & ...  \\
~~~~~~ & SDSS J134134.20+014157.7  & 35    &  697 & $5.6\times10^{-2}$ & $0.38\times0.30$ & 0.06 & ...  \\
~~~~~~ & SDSS J151155.98+040803.0  & 30    &  729 & $8.7\times10^{-2}$ & $0.53\times0.31$ & 0.06 & 1 SMG (w/ \cii) and 1 ``blob'' (w/o \cii)\\ 
\hline \\ [-1.75ex]
Faint  & SDSS J092303.53+024739.5  & 38    & 2978 & $4.3\times10^{-2}$ & $0.51\times0.29$ & 0.06 & 1 SMG (w/ \cii) \\
~~~~~~ & SDSS J132853.66-022441.6  & 36    & 2852 & $4.2\times10^{-2}$ & $0.48\times0.31$ & 0.06 & 1 SMG (w/ \cii) \\
~~~~~~ & SDSS J093508.49+080114.5  & 35-33 & 3230 & $5.1\times10^{-2}$ & $0.54\times0.29$ & 0.06 & ...
\enddata
\tablenotetext{a}{Number of antennae used, averaging after antennae flagging.}
\tablenotetext{b}{Flux densities, averaging over the three continuum (line-free) spectral windows. \texttt{CLEAN} performed with weighting=``natural''.}
\end{deluxetable*}
\capstarttrue

\subsection{Sample Selection and Properties}
\label{subsec:sample}

Our targets are drawn from a flux-limited sample of 40 luminous, unobscured quasars at \zfpe, which we have studied in detail in a series of previous publications.
Here, we only briefly mention the sample selection and the ancillary data available for our targets and refer the reader to our previous papers (T11, M12, N14) for additional details.
This \zfpe\ quasar sample was originally selected from the sixth data release of the Sloan Digital Sky Survey \cite[SDSS/DR6;][]{York2000,AdelmanMcCarthy2008_SDSS_DR6}.
The sample spans a narrow redshift range of $z\sim4.65-4.92$, to enable follow-up near-IR spectroscopy of the broad \MgII\ emission line. 
Such near-IR spectroscopy was indeed performed using the VLT/SINFONI and Gemini-North/NIRI instruments and has provided reliable estimates of the masses (\mbh) and normalized accretion rates (\lledd) of the quasars (T11).
These clearly showed that the \zfpe\ quasars are powered by fast-growing SMBHs with typical masses of $\mbh \simeq 8\times10^{8}\,\Msol$ and accretion rates of $\lledd \simeq 0.6$ (median values), representing the epoch of fastest growth for the most massive BHs.

A follow-up \herschel/SPIRE campaign targeted these quasars, probing the peak of the SF-heated dust continuum emission (M12, N14).
These data provided robust continuum detections for $\sim$1/4 of the quasars, with FIR luminosities of $\Lfir\sim2.4\times10^{13}\,\Lsol$, suggesting SFRs in the range $\sim1000-4000\,\mpyr$. 
A stacking analysis of the remaining $\sim$3/4 of the quasars revealed a median SFR of $\sim400\,\mpyr$ \cite[see the refined stacking analysis in][]{Netzer2016_herschel_hiz}.
Importantly, these ``FIR-bright'' and ``FIR-faint'' sources are highly uniform in terms of the SMBH-related properties that drive the AGN emission (i.e., \Lagn, \mbh, and hence \lledd), with a tendency of the FIR-bright sources to have somewhat higher \mbh\ and \Lagn\ (by $\lesssim$0.4 dex). 
Thus, the \herschel\ data presented in M12 and N14 suggests a wide variety of SF activity among an almost uniform sample of fast-growing SMBHs in the early universe, with SFRs that range over an order of magnitude, compared with a significantly narrower range of basic SMBH properties.
The N14 study also presented the results of a dedicated \Spitzer/IRAC campaign targeting almost all of the T11 quasars, at observed-frame 3.6 and 4.5 \mic. 
These data were used to derive positional priors that allowed us to de-blend the low-resolution \herschel/SPIRE data.

A comparison of the FIR luminosities of the quasar hosts (implied from the \herschel\ data) to the AGN-related luminosities (i.e., the bolometric luminosities from T11), implies that the 
FIR-bright sources are so FIR-luminous, that they reach $\LSF\simeq\Lagn$.
The FIR-faint systems, on the other hand, have $\Lagn\gtrsim5\times\LSF$, making them ``AGN dominated''.

The current study focuses on \Ntot\ objects selected from our parent T11 quasar sample, split equally between \Nbright\ FIR-bright and \Nfaint\ FIR-faint objects.
We chose to focus on the lower-redshift sources among our parent sample ($z<4.8$) to avoid the \cii\ emission line to be redshifted into the low-atmospheric transmission region near 325 GHz.
The redshifts of the selected targets are in the range $z=4.658-4.729$, with a median of $z_{\rm median}=4.6703$ (as determined from the \mgii\ lines; see \autoref{subsec:lines} for more details). 
We will nonetheless refer to them here as ``\zfpe\ quasars'', following our previous studies.
The targets are all equatorial, so to allow efficient ALMA observations.
In practice, this implies that the NIR data for all the \Ntot\ quasars studied here were obtained with the VLT/SINFONI.
These redshift and declination restrictions leave only 4 of the 10 \herschel-detected, FIR-bright quasars reported in N14, of which we finally chose 3.
As for the 3 FIR-faint targets, we focused on those with higher quality \mgii\ fits, as assessed in T11.\footnote{In terms of both flux calibration and spectral fit quality; corresponding to ``$L$-quality'' and ``\fwmg-quality'' flags of 1 or 2 in Table 2 of T11.}
We stress that all these selection criteria do not introduce any biases, in terms of the a-priori known SMBH- and host-related properties of the systems under study.
In particular, the \mbh\ values of the two sub-groups are consistent, within the errors (except for the relatively high-mass object J1341; see \autoref{tab:gal_props}).

\subsection{Observations, Data Reduction and Analysis}
\label{subsec:obs_red}

\subsubsection{ALMA Observations}
\label{subsubsec:obs}

The \Ntot\ targets were observed in ALMA \obsband, as part of cycle-\obscycle\ (project code 2013.1.01153.S), during the period 2014 July 18 to 2015 June 13. 
As the capabilities of ALMA were expanded during this period, the data presented here were collected with a varying number of 12m antennas, between 29-38.
The observations were set up to use the extended C34-4 configuration, which provides a resolution of about 0\farcs3 at 330 GHz (corresponding to about 2 kpc at \zfpe).
We aimed at spectrally resolving the \cii\ emission line, which is expected to have a width of several hundred \kms\ and also to cover a wide spectral range that includes line-free continuum regions.

We chose to use the TDM correlator mode, providing four spectral windows, each covering an effective bandwidth of 1875 MHz, corresponding to $\sim1650\,\kms$ at the observed frequencies.
This spectral range is sampled by 128 channels, providing a spectral resolution of $\sim30\,\kms$.
One such spectral window was centered on the frequency corresponding to the expected peak of the \cii\ line, given the \mgii-based redshifts of our targets (as determined in T11).
The other three bands extended to higher frequencies, with the first being adjacent to the \cii-centered band and the other two separated from this first pair by about 12 GHz.
Each of these two pairs of bands included a small overlapping spectral region, of roughly 50 MHz.
In addition, the rejection of a few of the channels at the edge of the spectral bands, due to divergent flux values (a common flagging procedure in ALMA data reduction), implies that in some cases a small gap is seen between windows.
Given this spectral setup of four bands, the ALMA observations could in principle probe \cii\ line emission over a spectral region with a width corresponding to roughly $\Delta z \simeq 0.06$.
Additional details regarding the ALMA observations, including the full object names of our sources, are given in \autoref{tab:obs_log}.
For clarity, we use abbreviated object names (i.e., ``JHHDD'') throughout the rest of the paper.

\begin{figure*}[!ht]
\begin{center}
\includegraphics[width=0.33\textwidth]{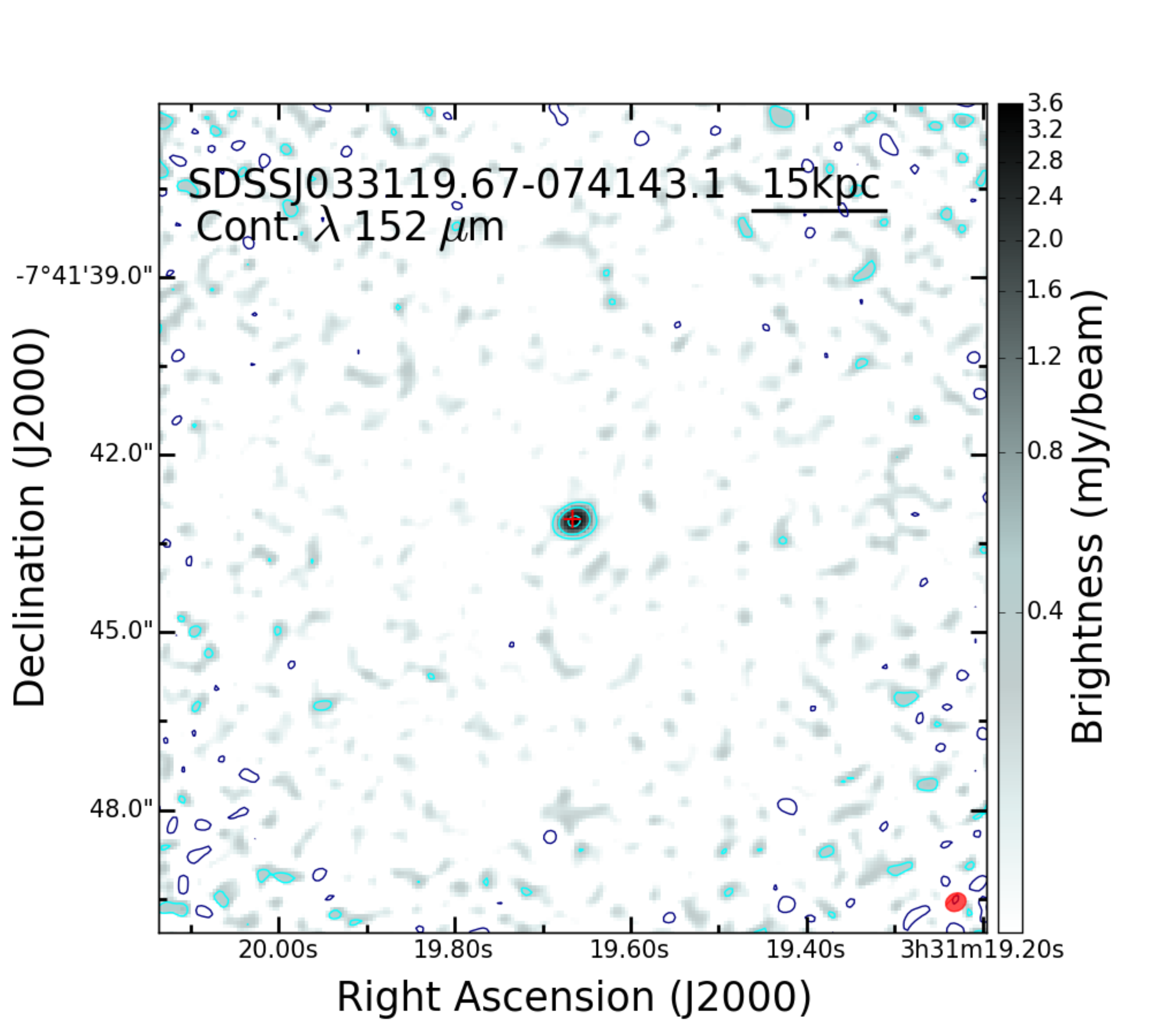} 
\includegraphics[width=0.33\textwidth]{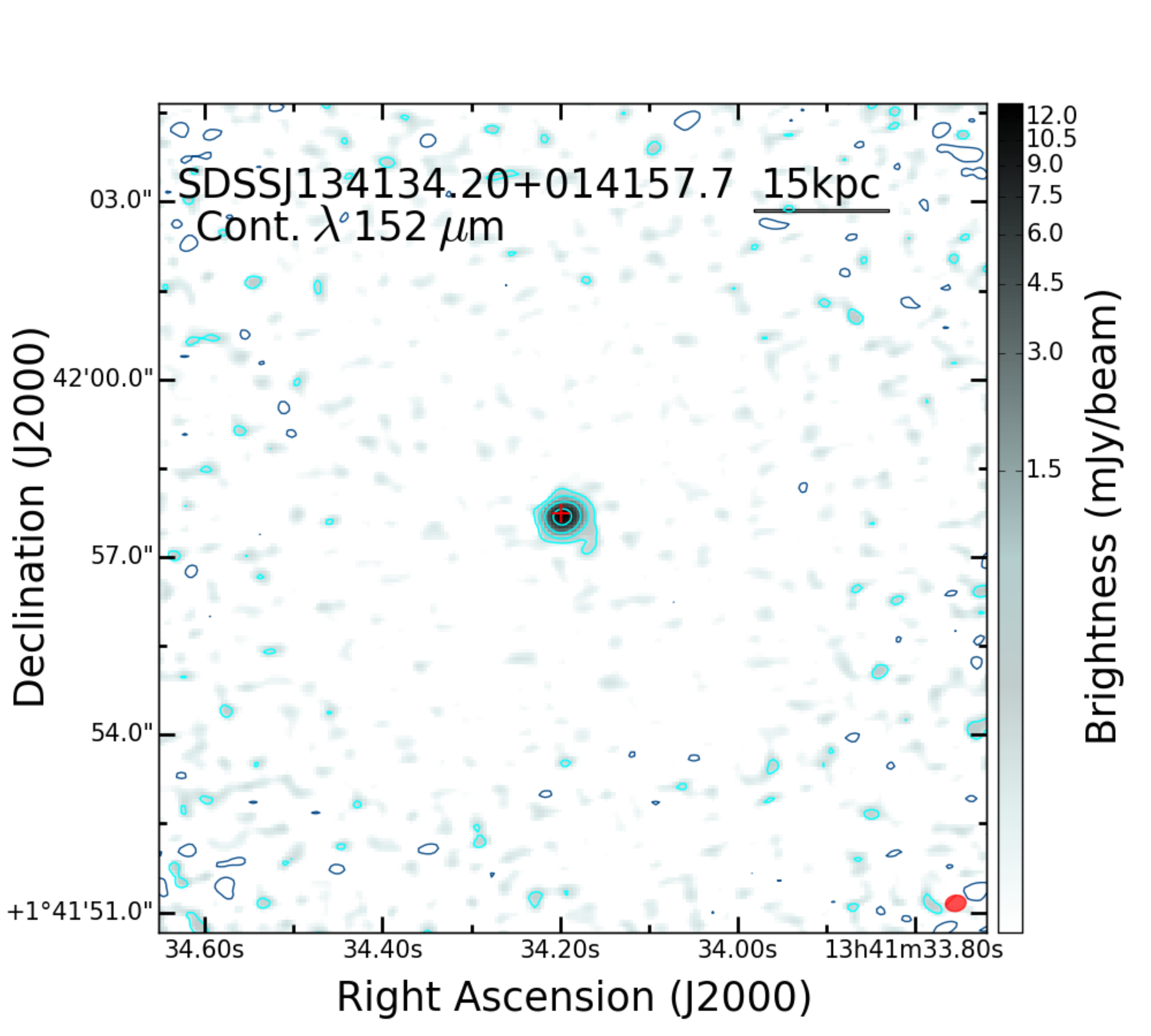}
\includegraphics[width=0.33\textwidth]{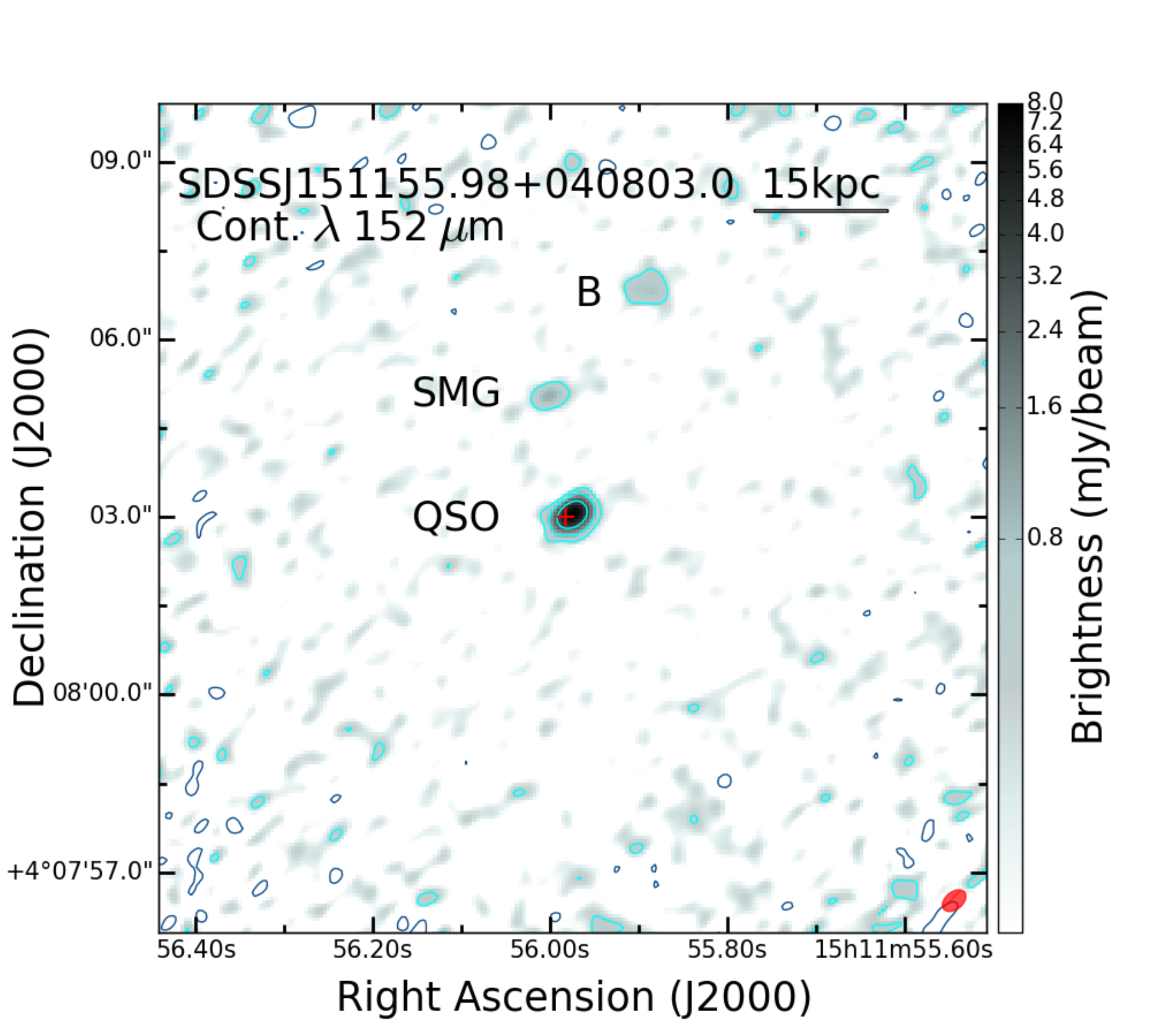} \\
%
%
\includegraphics[width=0.33\textwidth]{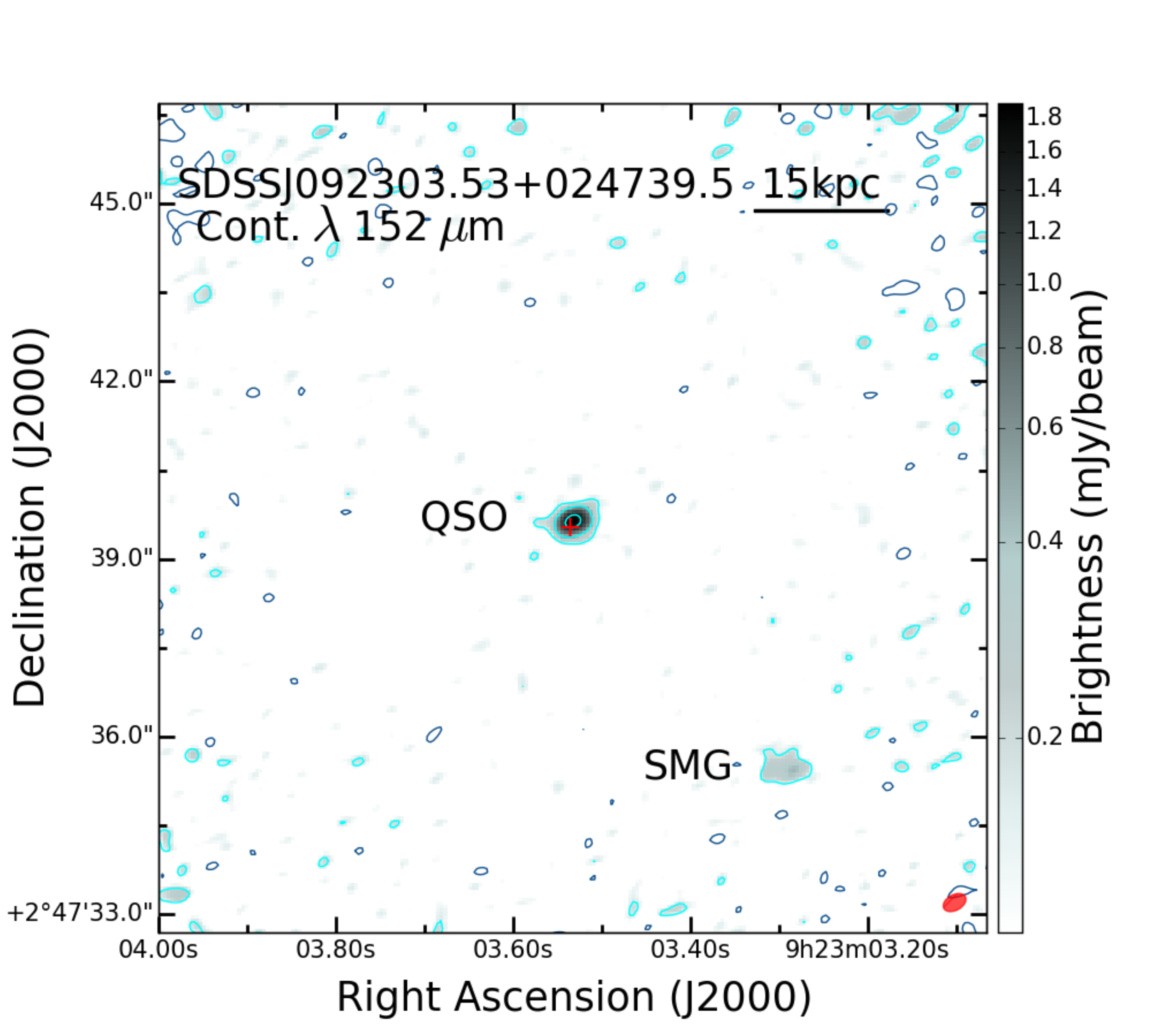}
\includegraphics[width=0.33\textwidth]{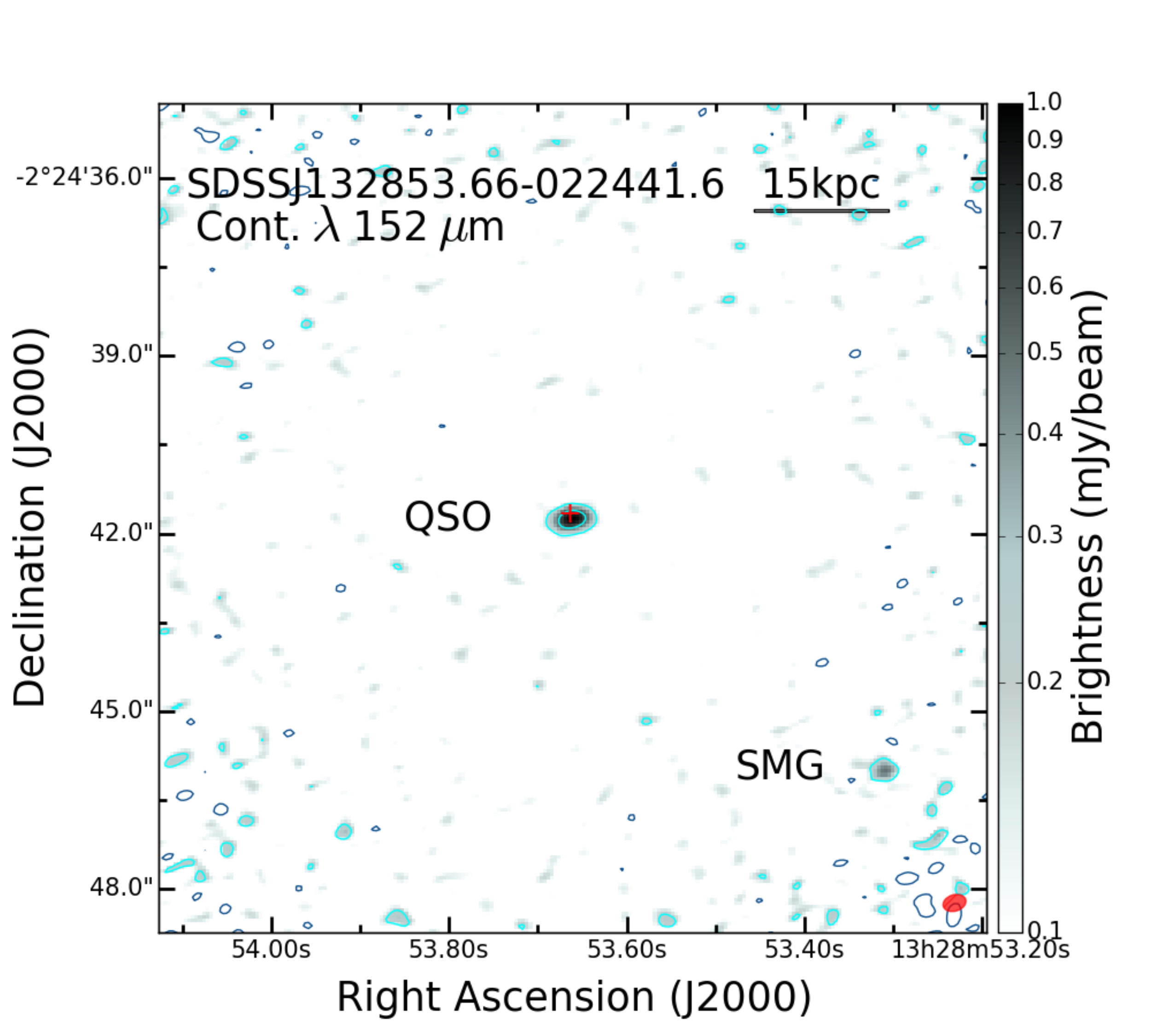}
\includegraphics[width=0.33\textwidth]{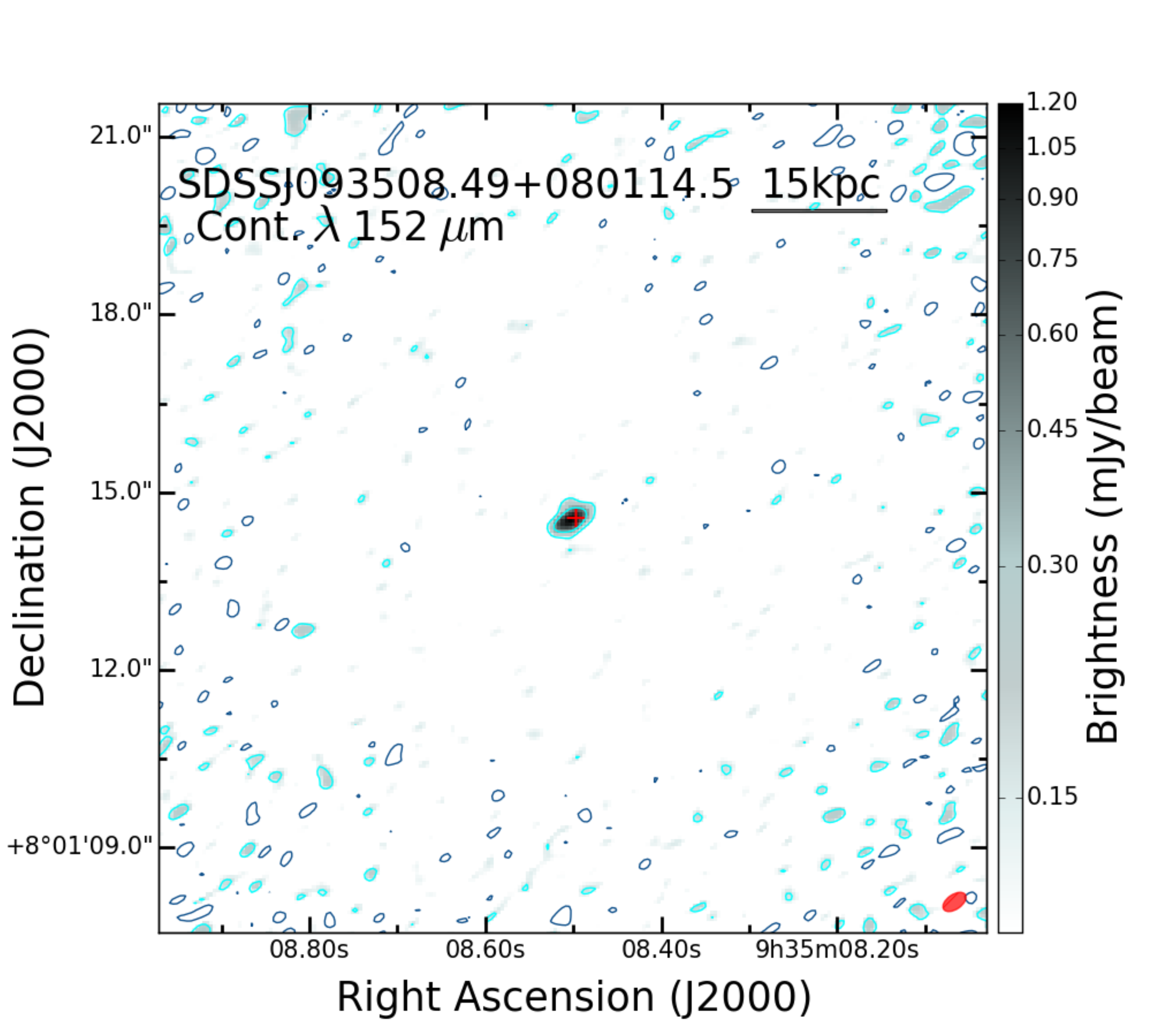}\\
\caption{
Large-scale \emph{continuum} emission maps derived from the new ALMA data, for the \Ntot\ quasar systems in our sample: 
the \Nbright\ FIR-bright sources (top row), and the \Nfaint\ FIR-faint sources (bottom row).
In each panel, the quasar and any accompanying sub-millimeter galaxies are marked as ``QSO'' and ``SMG'', respectively.
The gray-scale map shows the continuum emission, determined from the line-free ALMA spectral windows.
Cyan and blue contours trace emission levels at different positive and negative significance levels, respectively, with the first contour tracing the region where the continuum emission exceeds 2.5$\sigma$, and consecutive contours plotted in steps of $2.5\sigma$.
The ALMA beams are shown as red ellipses near the bottom-right of each panel.
Red crosses mark the locations of the quasars' optical emission (taken from the SDSS).
Interacting companions, i.e. sources that have clear detections of \cii\ with redshifts consistent with those of the quasars, are marked as ``SMG''.
The continuum source accompanying J1511 that lacks significant \cii\ emission is marked as ``B''.
}
\label{fig:cont_maps_lg}
\end{center}
\end{figure*}

The ALMA \obsband\ flux densities reach a depth of $F_{\nu}\simeq\left(4.2-9.2\right)\times10^{-2}$ mJy/beam (rms).
This flux limit can be translated to a limit on our ability to detect dusty, SF galaxies, at the redshift range of interest.
If we assume the same gray-body FIR spectral energy distribution (SED) as we do in our analysis of the quasars hosts' -- a dust temperature of $T_{\rm d}=47\,{\rm K}$ and a power-law exponent of $\beta=1.6$, and the scaling between FIR luminosity and SFR appropriate for our chosen IMF (see \autoref{subsec:hosts}), this corresponds to $3\,\sigma$ upper limits on the SFR in the range $\sim20-50\,\mpyr$, per beam.
Given the beam sizes of our ALMA observations (\autoref{tab:obs_log}), these translate to $\sim4-11\,\mpyr/\kpc^2$.

\subsubsection{Data Reduction and Spectral Measurements}
\label{subsubsec:reduction}

Data reduction was performed using CASA package, version 4.5.0 \cite[][]{Mcmullin2007_CASA}.
The scripts provided by the observatory were used to generate the visibilities. 
We then applied the \texttt{CLEAN} algorithm, using a ``natural'' weighting to determine the noise level for each observation by averaging over the three line-free spectral windows. 
The resulting flux density sensitivities and synthesized beam sizes are presented in \autoref{tab:obs_log}.
We note that self calibration for the brightest sources did not improve the signal-to-noise ratios and therefore was not implemented.

Both continuum and emission line images were created by applying the \texttt{CLEAN} algorithm using a ``briggs'' weighting (with robustness parameter set to 0.5), to obtain images with the best possible spatial resolution.
Continuum emission images were constructed using the two highest frequency, line-free spectral windows. 
\cii\ emission line images were constructed by subtracting the continuum emission from the other two, lower frequency spectral windows, in the UV space (using standard CASA procedures).
We verified that the resulting \cii\ emission line images have no residual continuum signal in them.
The sizes of the continuum- and line-emitting regions were determined from the respective images by fitting two-dimensional Gaussians.
These sizes are given in \autoref{tab:lum_cii}.
Velocity and velocity dispersion maps (i.e., second- and third-moment maps) were obtained from the line images using the standard CASA procedure.

\begin{figure*}[ht!]
\begin{center}
\includegraphics[width=0.33\textwidth]{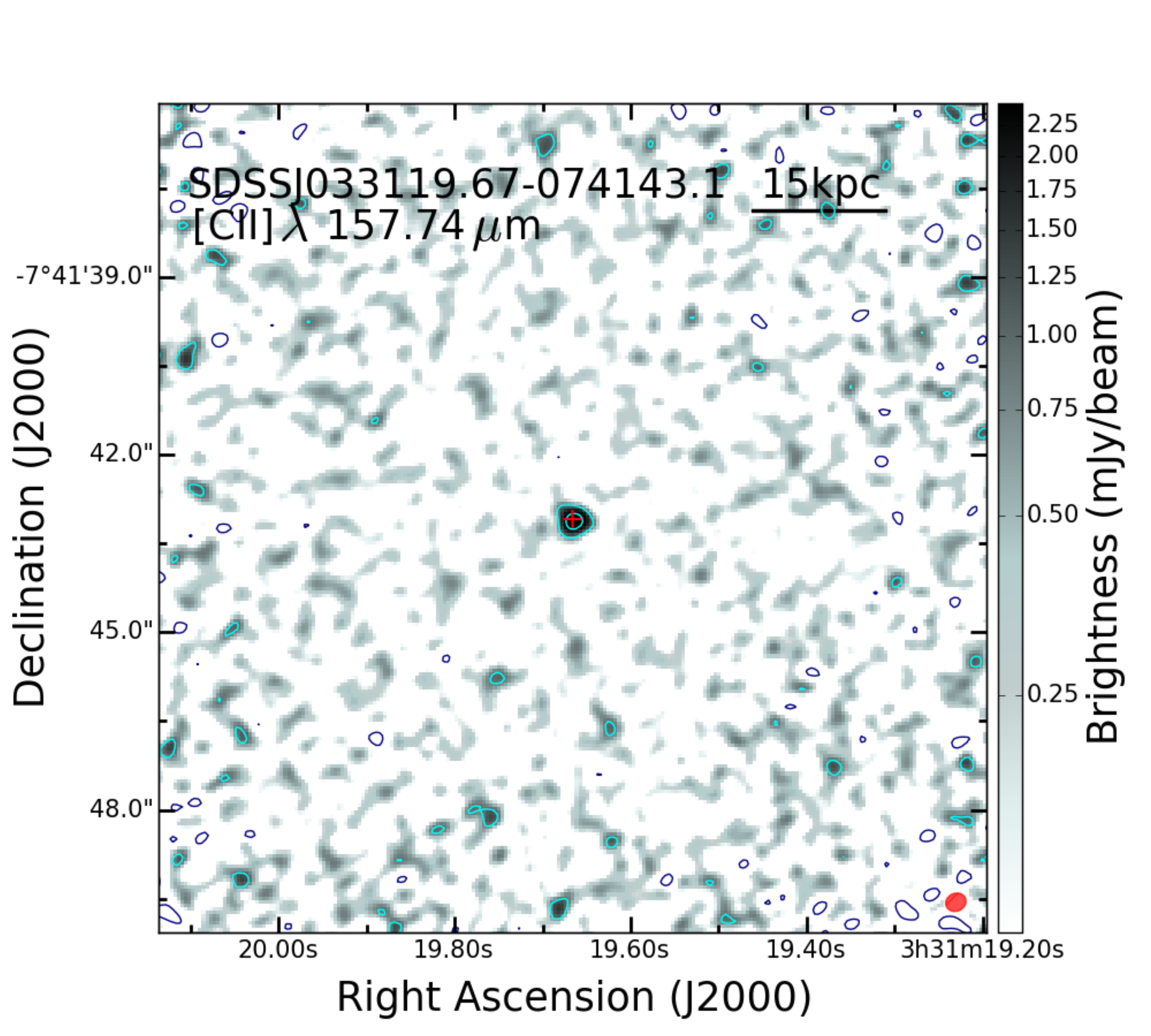} 
\includegraphics[width=0.33\textwidth]{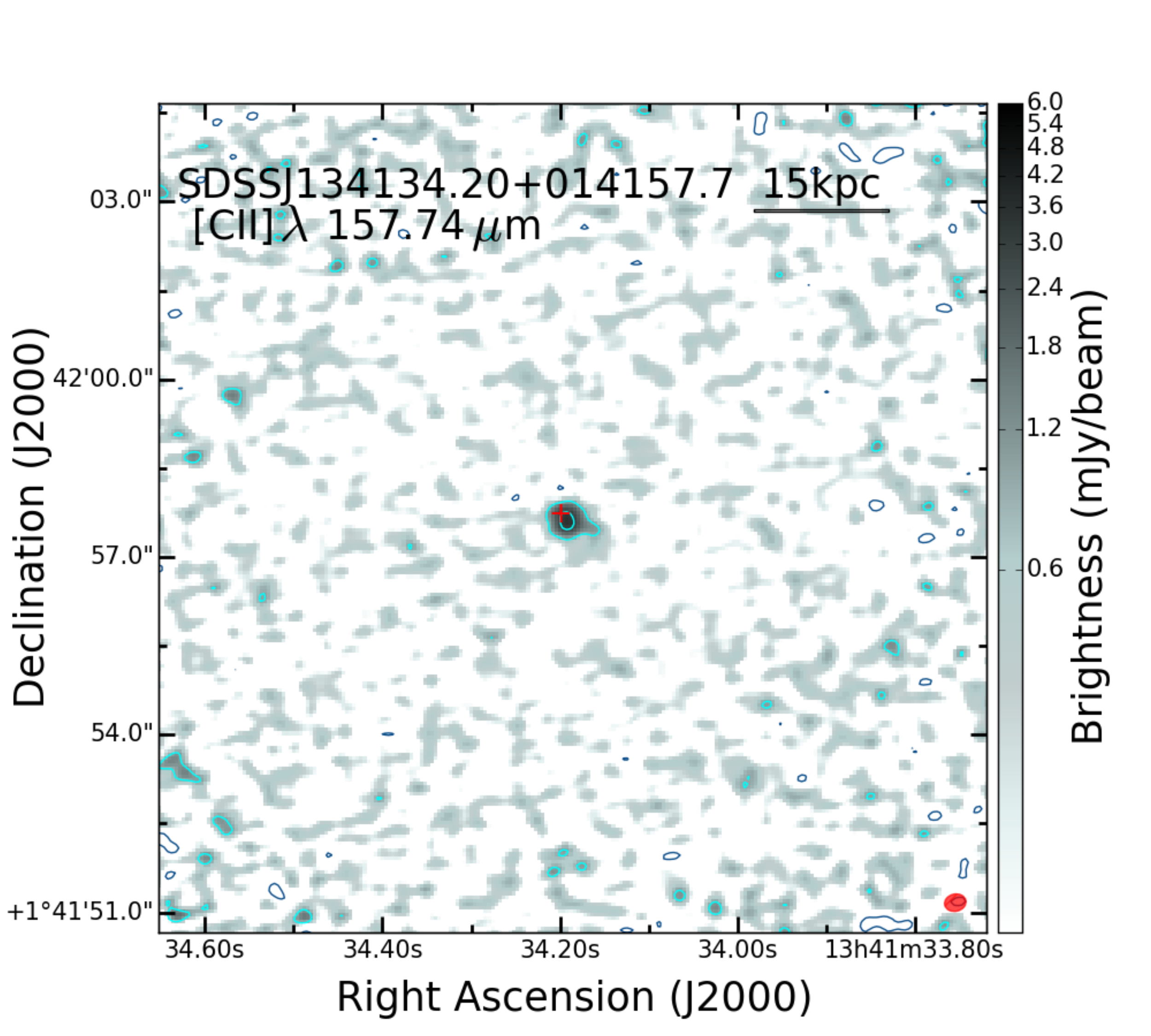}
\includegraphics[width=0.33\textwidth]{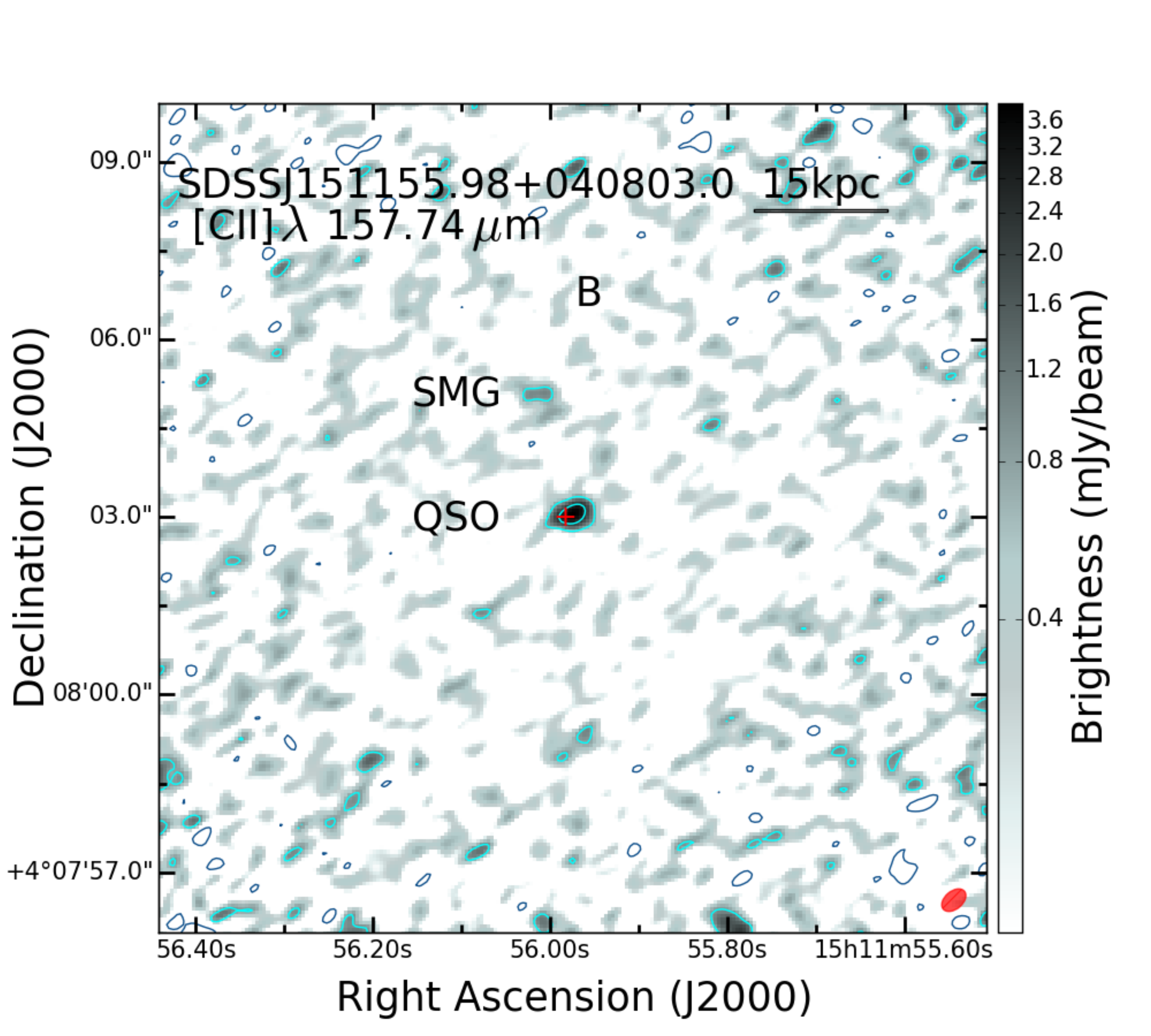} \\
%
%
\includegraphics[width=0.33\textwidth]{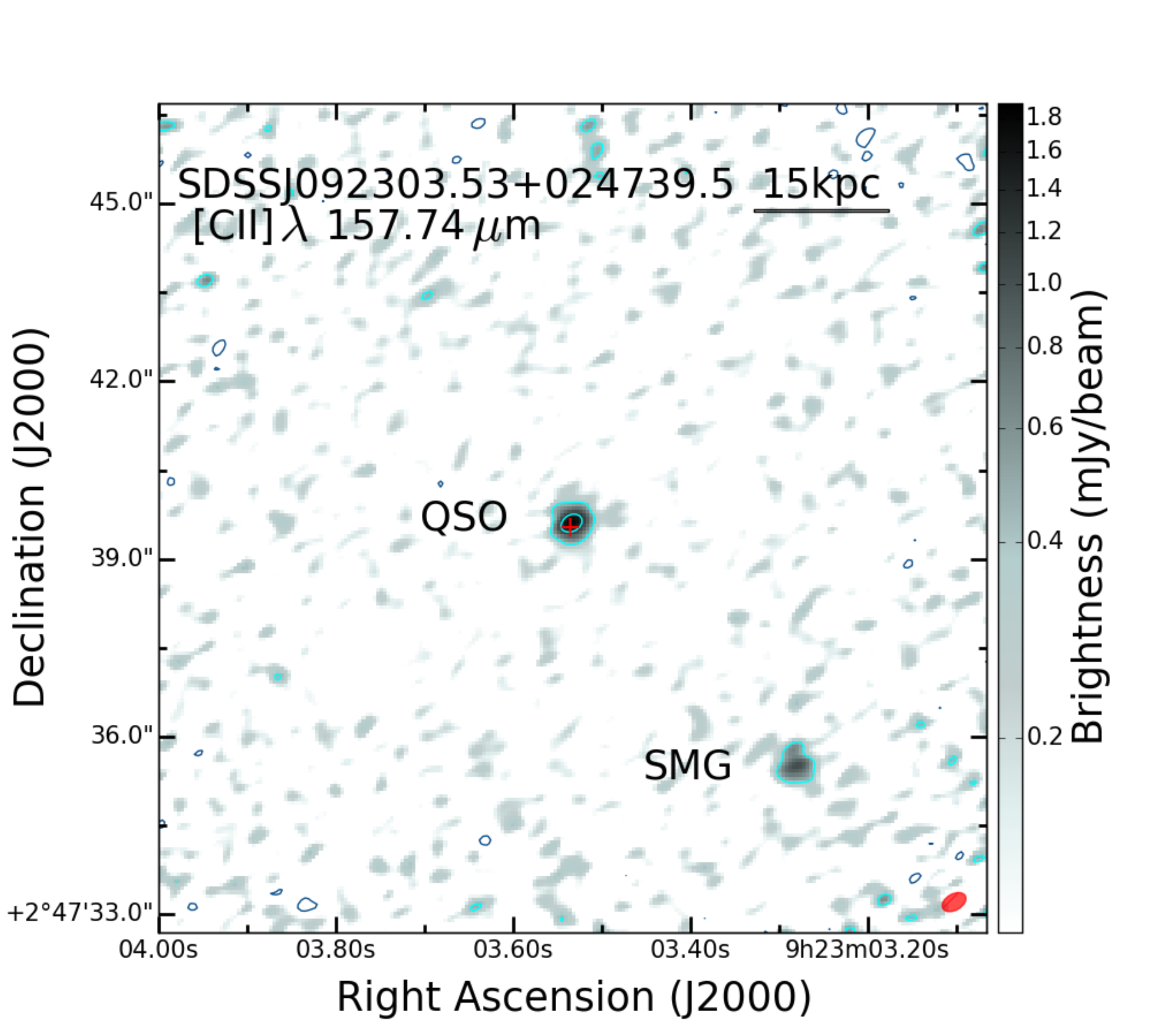}
\includegraphics[width=0.33\textwidth]{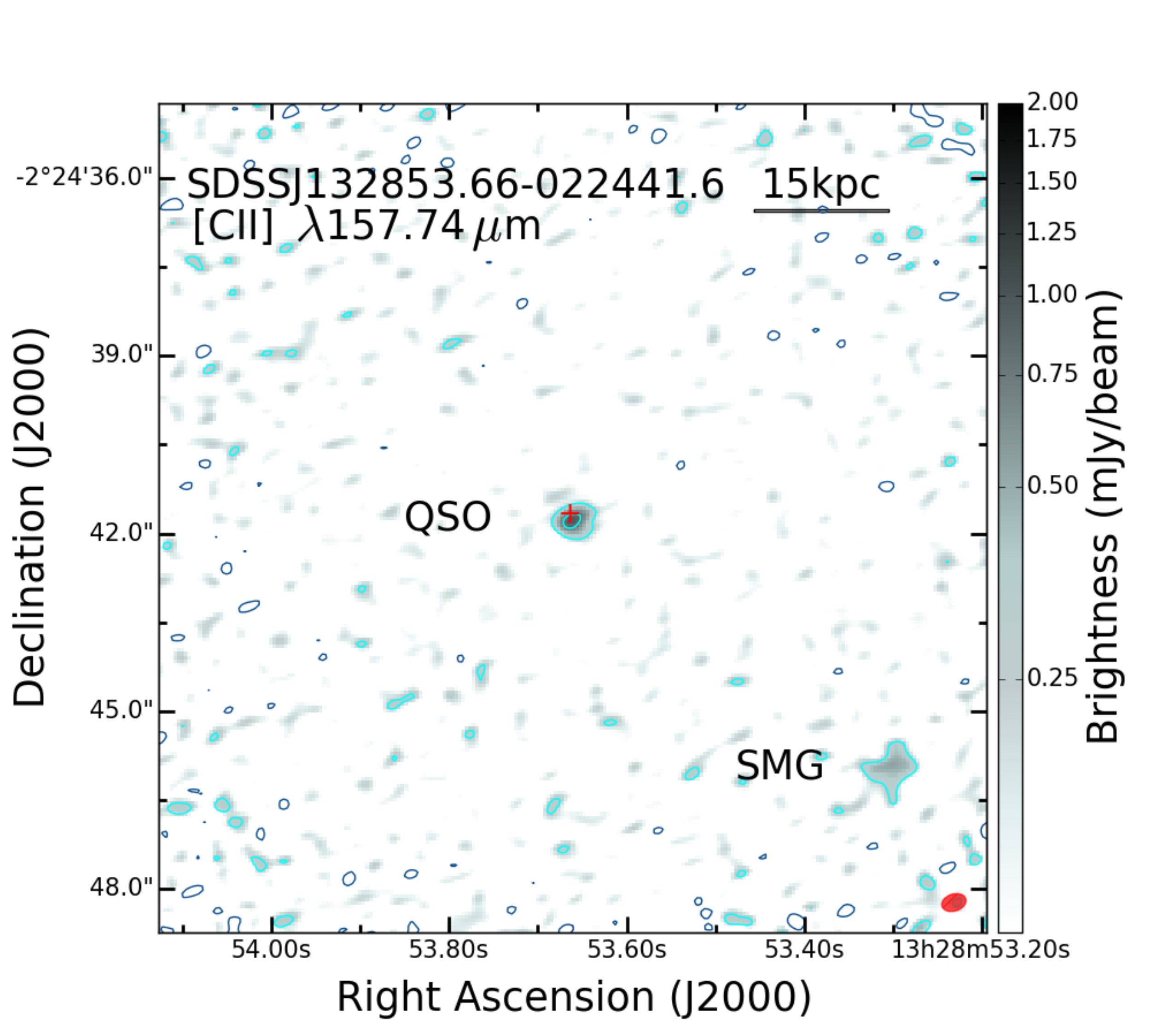}
\includegraphics[width=0.33\textwidth]{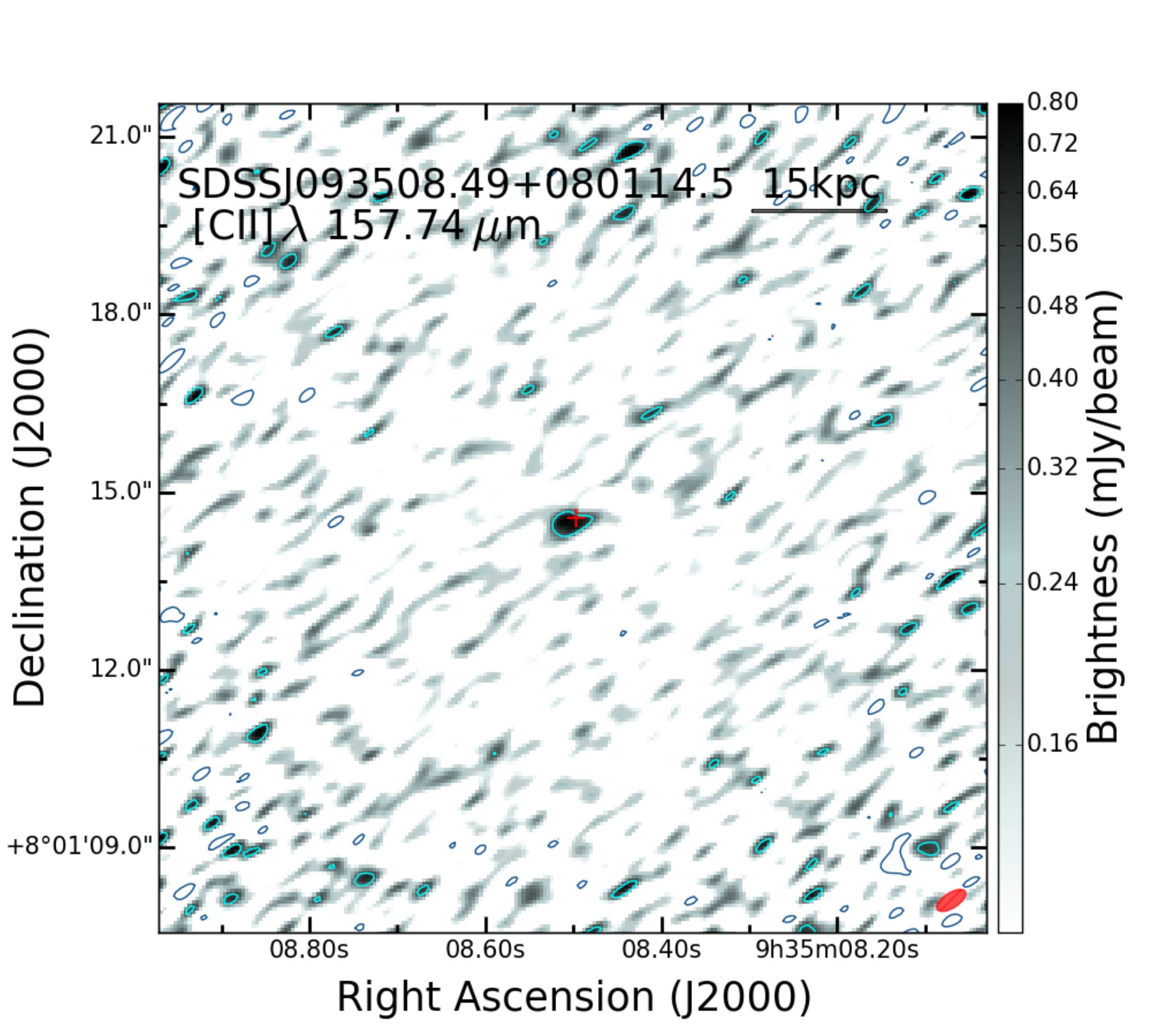}\\
\caption{
Large-scale \cii\ \emph{line} emission maps derived from the new ALMA data for the \Ntot\ quasar systems in our sample:
the \Nbright\ FIR-bright sources (top row), and the \Nfaint\ FIR-faint sources (bottom row).
In each panel, the quasar and any accompanying sub-millimeter galaxies are marked as ``QSO'' and ``SMG'', respectively.
The gray-scale map shows the continuum emission, determined from the line-free ALMA spectral windows.
Cyan and blue contours trace emission levels at different positive and negative significance levels, respectively, with the first contour tracing the region where the line emission exceeds 2.5$\sigma$, and consecutive contours plotted in steps of $2.5\sigma$.
For each source, the line fluxes used for the contours were extracted from a spectral band spanning $\pm500\,\kms$ around the \cii\ line peak, for the three quasars lacking a robust companion (i.e., J0331, J1341, and J0935). 
For systems with a companion SMGs, the spectral band spans $\pm\left(500+\left|\Delta v\right|\right)\,\kms$ around the mean \cii\ peak frequency of the quasar and the SMG, where $\Delta v$ is the velocity separation between the line peaks.
The ALMA beams are shown as red ellipses near the bottom-right of each panel.
Red crosses mark the locations of the quasars' optical emission (taken from the SDSS).
Interacting companions, i.e. sources that have clear detections of \cii\ with redshifts consistent with those of the quasars, are marked as ``SMG''.
The continuum source accompanying J1511 that lacks significant \cii\ emission is marked as ``B''.
}
\label{fig:cii_maps_lg}
\end{center}
\end{figure*}

The properties of the \cii\ emission lines were measured from the two lower frequency spectral windows of the continuum-subtracted ``briggs'' weighted cubes.
The integrated line fluxes were measured through two different procedures.
In the first (``spatial'') approach, we created zero-moment images (i.e., integrated over the spectral axes) for all sources and fitted the spatial distribution of line emission with 2D Gaussian profiles, which are characterized by a peak line flux, semi-major and semi-minor axes, and a position angle. 
The line fluxes were measured by integrating over these spatial 2D Gaussians.
In the second (``spectral'') approach, we extracted 1D spectra of the \cii\ line from the continuum-subtracted cubes.
We used apertures that are based on the aforementioned spatial 2D Gaussian profiles, with varying radii ranging $1.5-2 \times \sigma$, where $\sigma$ is the width (i.e., equivalent to standard deviation) of the spatial 2D fitted Gaussian profiles.
We then fitted the emission line profiles with a simple model of a single Gaussian, from which we obtained the integrated line flux.  
The \cii\ line fluxes obtained using the two methods are in good agreement, with differences of no more than 0.1 dex for the \Ntot\ quasar hosts (median difference of 0.05 dex).
We eventually chose to adopt the line fluxes measured through the former, ``spatial'' approach, as it is less sensitive to the low-$S/N$ regions in the outer extended regions of the sources and/or the wings of the line profiles, and since it provides more realistic uncertainties. 
These line fluxes are reported in \autoref{tab:lum_cii}.

The best-fit emission line profiles obtained through the latter (``spectral'') approach allow us to obtain the \cii\ line centers, and therefore \cii-based redshifts, as well as the line widths (see \autoref{tab:lum_cii} and \autoref{tab:line_shifts}).
We stress that the centers of the \cii\ line profiles were treated as free parameters and \emph{not} constrained to reflect the previously known redshifts of the quasars. 
We also note that for four of the quasar hosts, the \cii\ line is observed near the edge of the spectral band, due to the differences between the redshifts based on AGN-line- and host-ISM-related emission regions (see \autoref{subsec:lines}).
The line widths we measure for the quasar hosts are in the range $\fwcii\sim240-510\,\kms$, with a tendency for broader lines among the FIR-bright sources (see discussion in \autoref{subsec:hosts}).
The formal uncertainties on line width measurements are of order of 10\% (i.e., $\sim10-40\,\kms$).

\section{Results and Discussion}
\label{sec:res_and_disc}

\subsection{Source detection and identification}
\label{subsec:sources}

\begin{figure*}[ht!]
\includegraphics[trim={0cm 0cm 1.9cm 0.2cm},clip,width=0.33\textwidth]{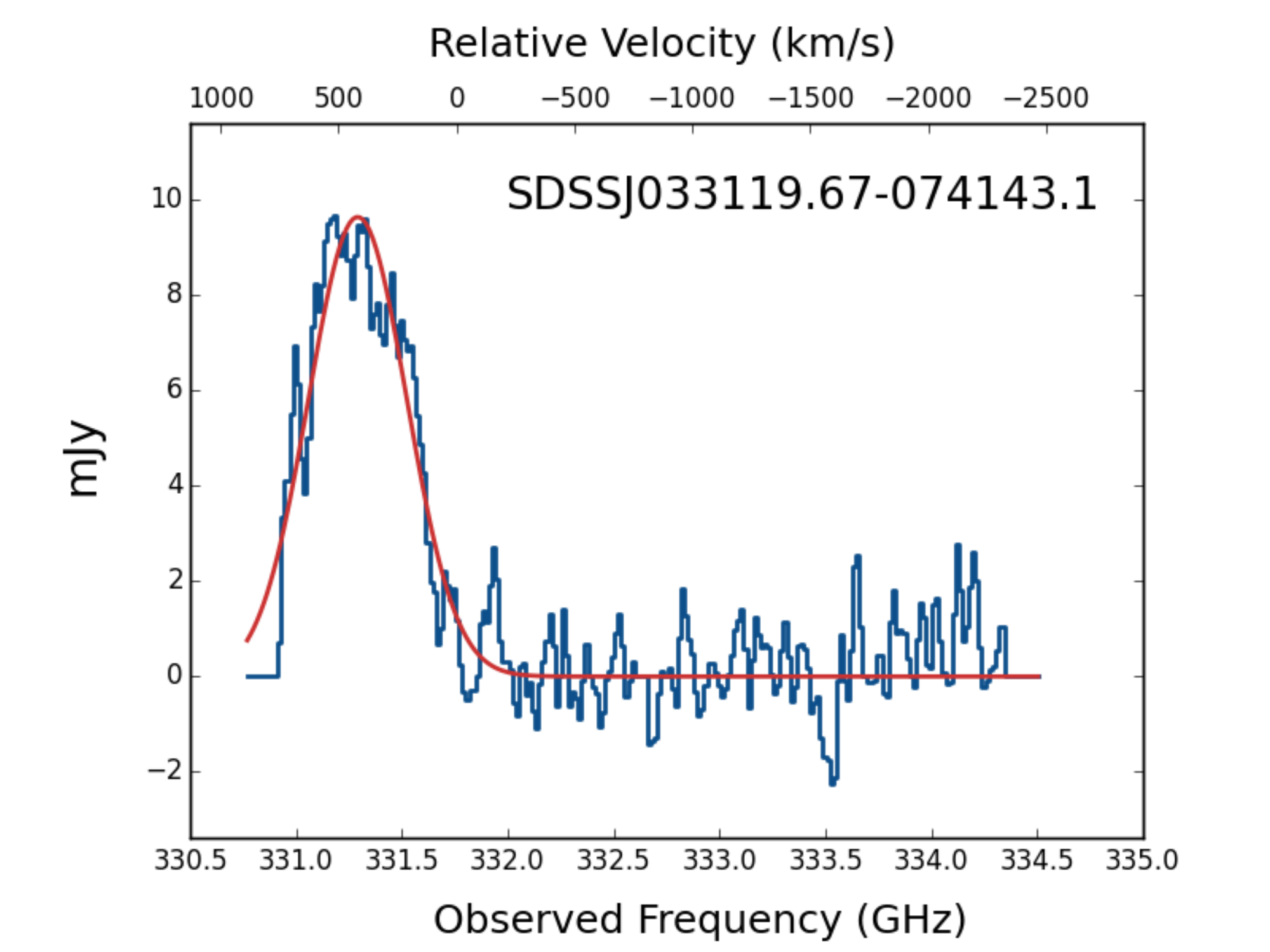} 
\includegraphics[trim={0cm 0cm 1.9cm 0.2cm},clip,width=0.33\textwidth]{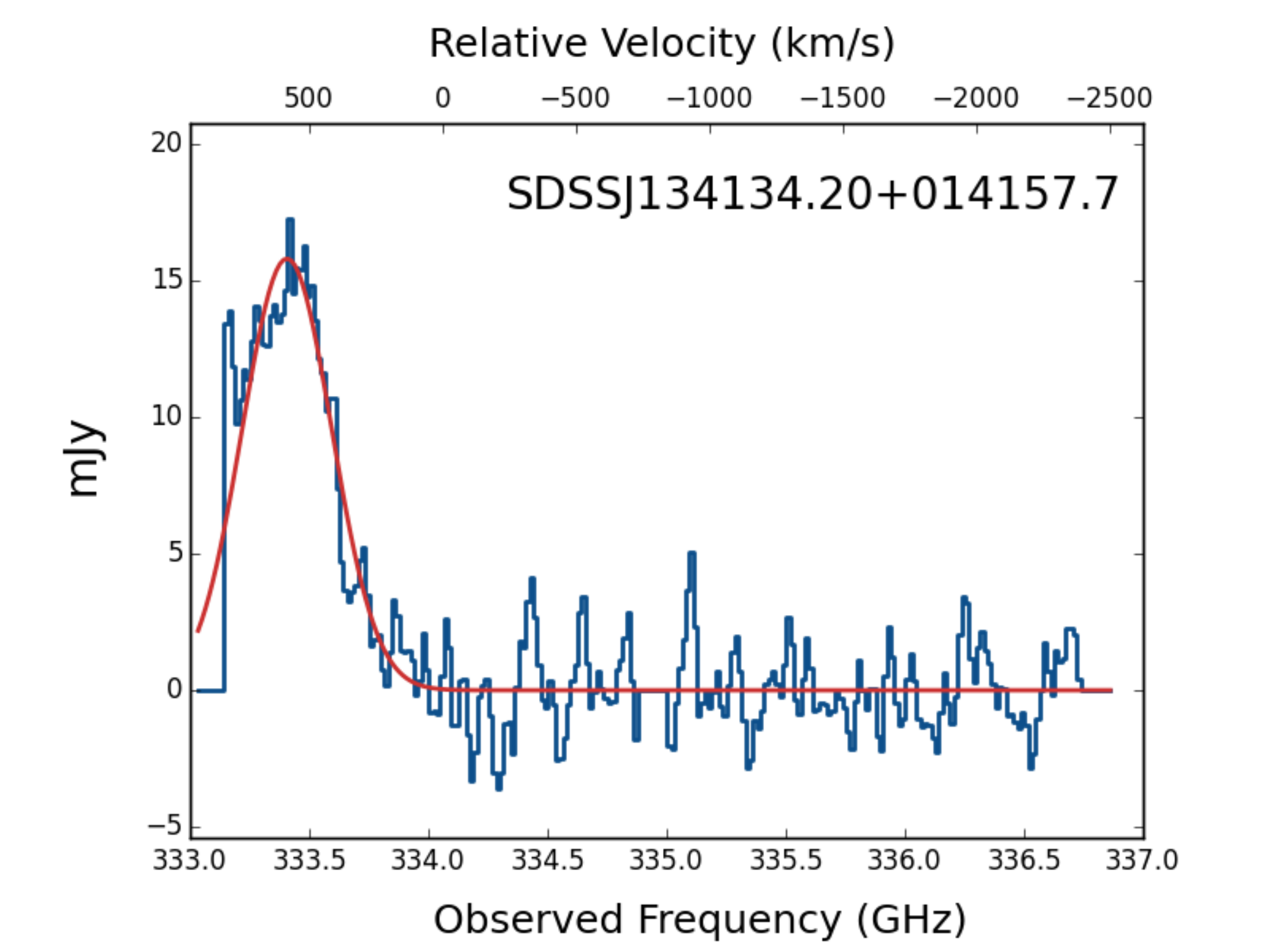}
\includegraphics[trim={0cm 0cm 1.9cm 0.2cm},clip,width=0.33\textwidth]{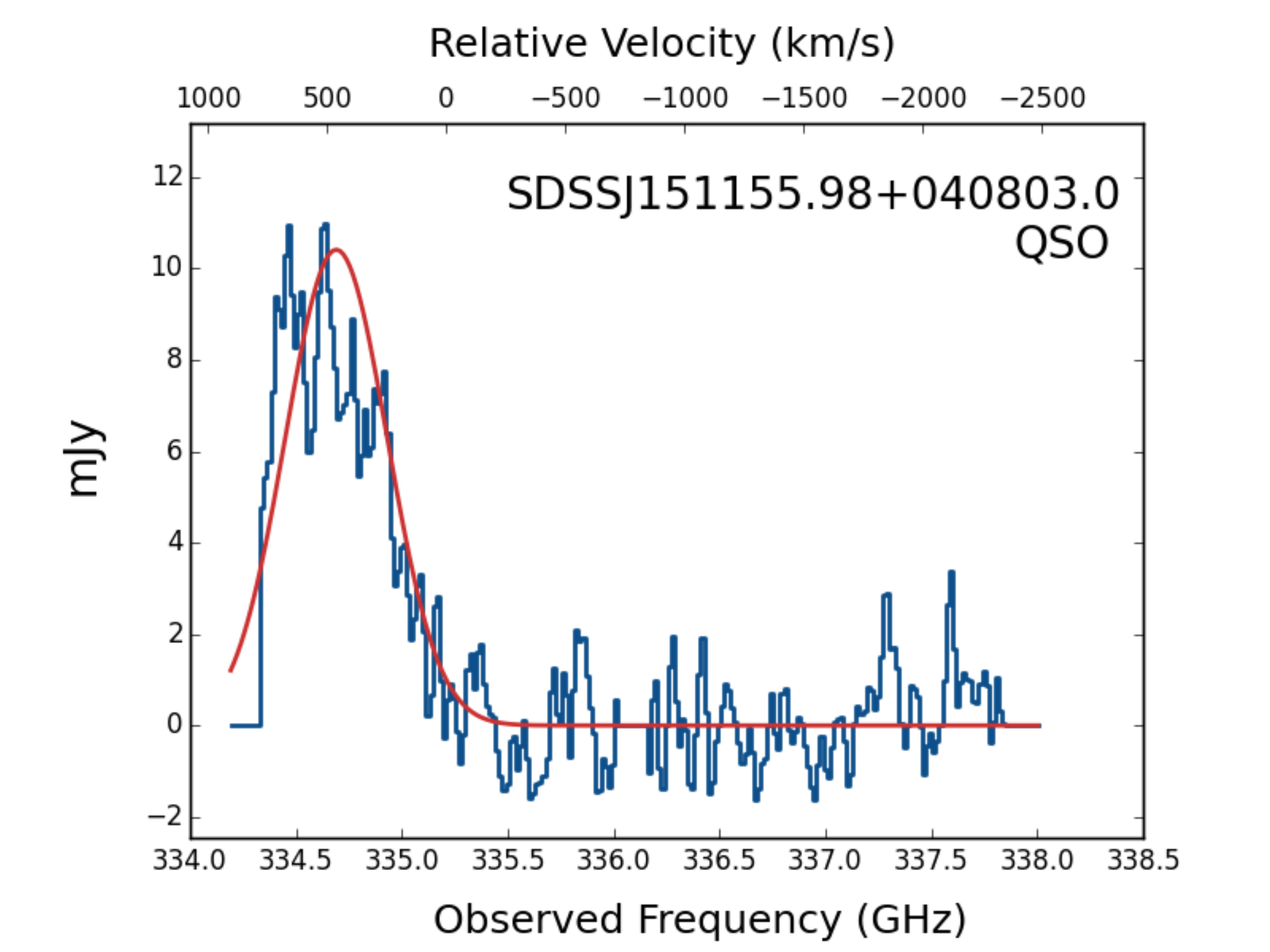} \\
%
%
\includegraphics[trim={0cm 0cm 1.9cm 0.2cm},clip,width=0.33\textwidth]{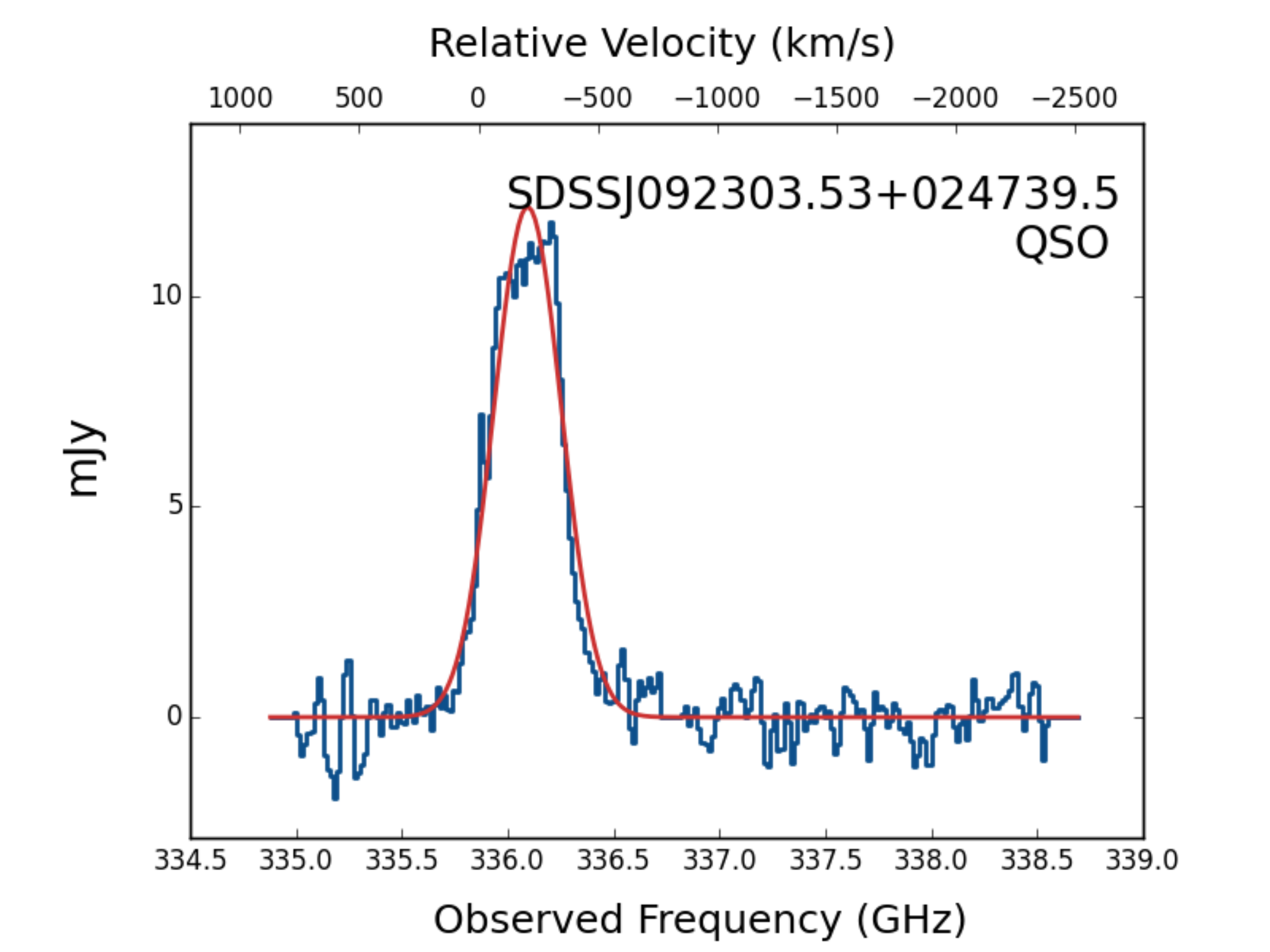}
\includegraphics[trim={0cm 0cm 1.9cm 0.2cm},clip,width=0.33\textwidth]{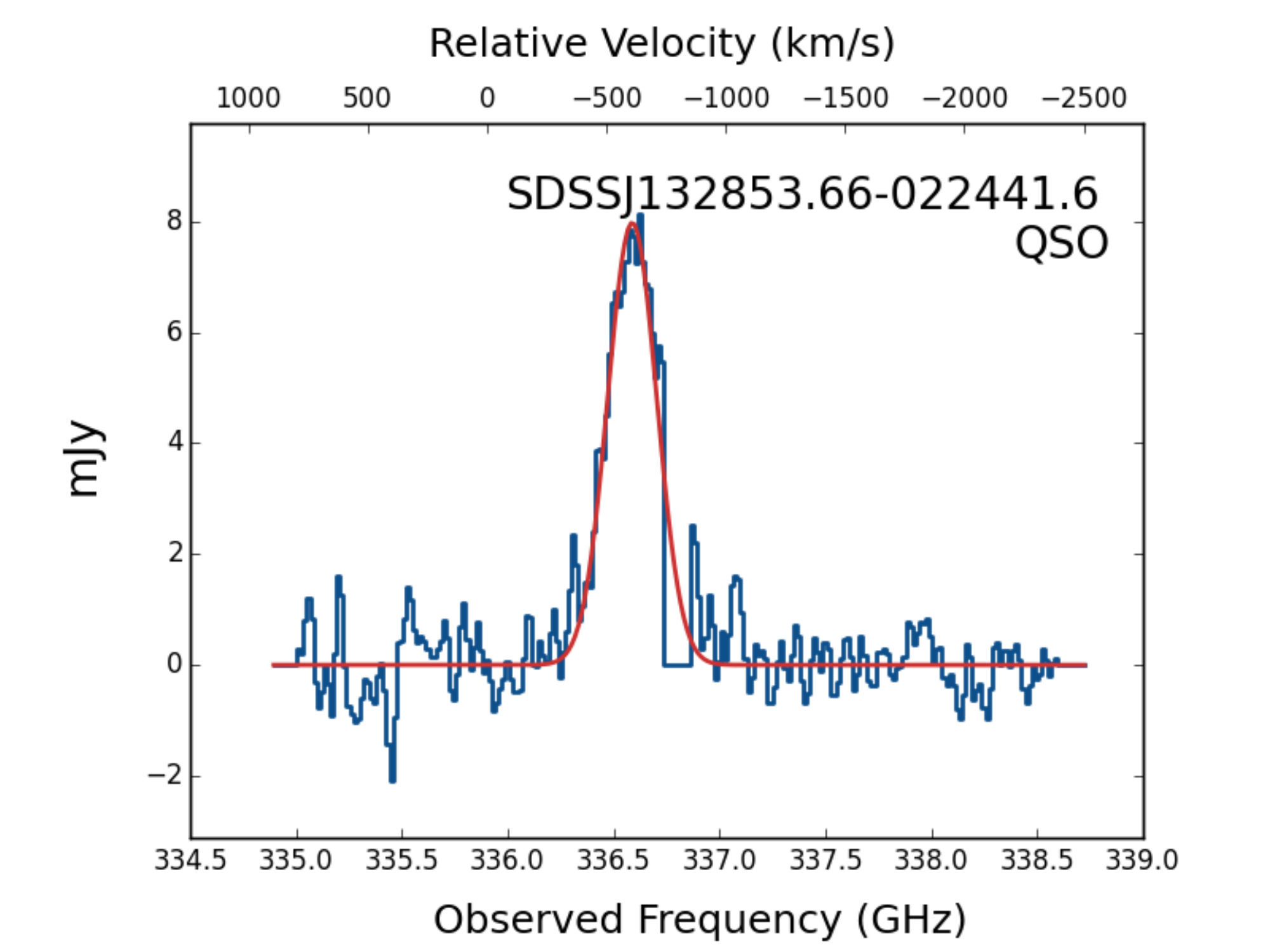}
\includegraphics[trim={0cm 0cm 1.9cm 0.2cm},clip,width=0.33\textwidth]{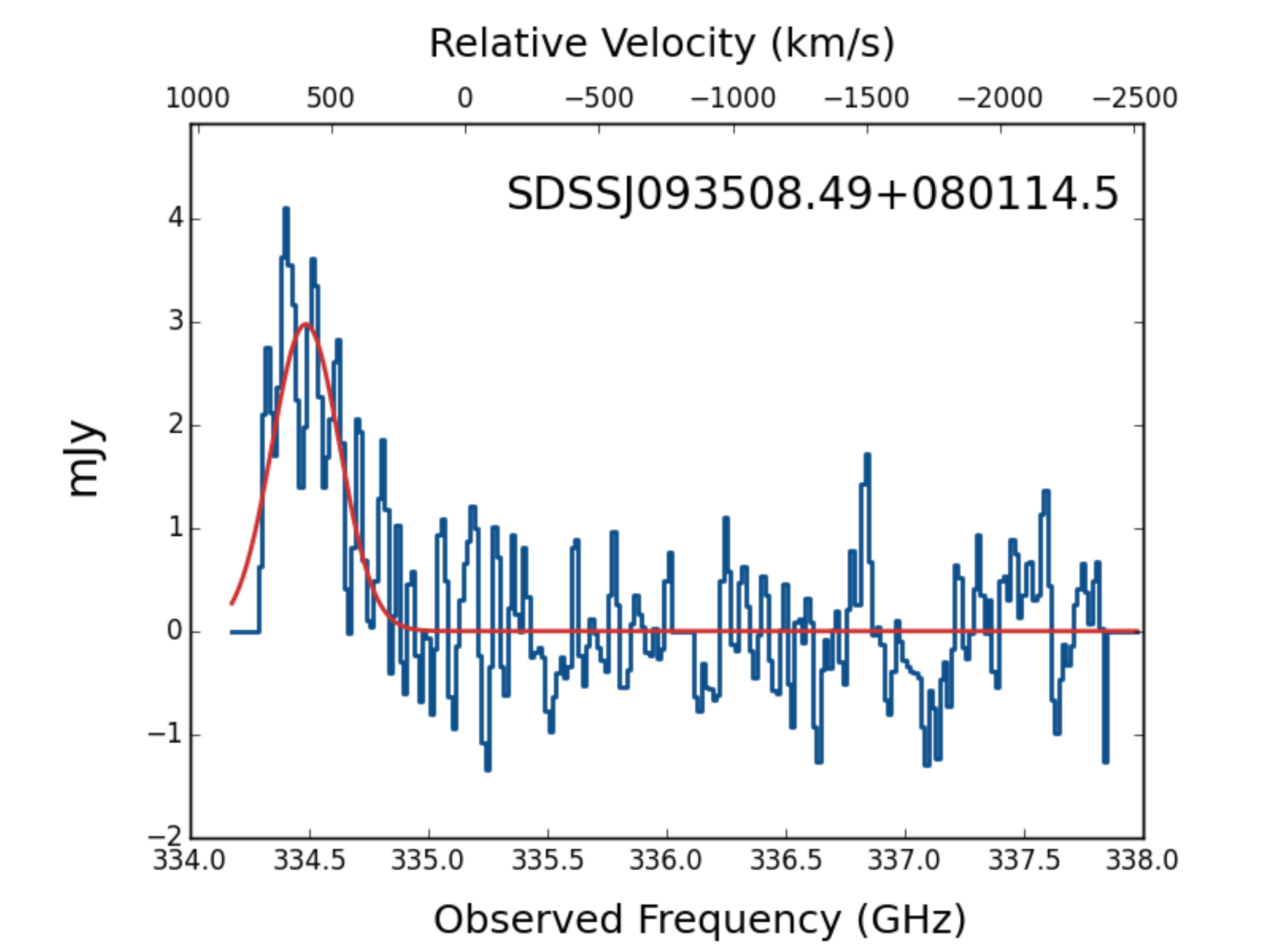}\\
\vspace*{-0.5cm}
\begin{center}
\hspace{1cm}\rule[1ex]{1\textwidth}{0.5pt}\\
\end{center}
\includegraphics[trim={0cm 0cm 1.9cm 0.2cm},clip,width=0.33\textwidth]{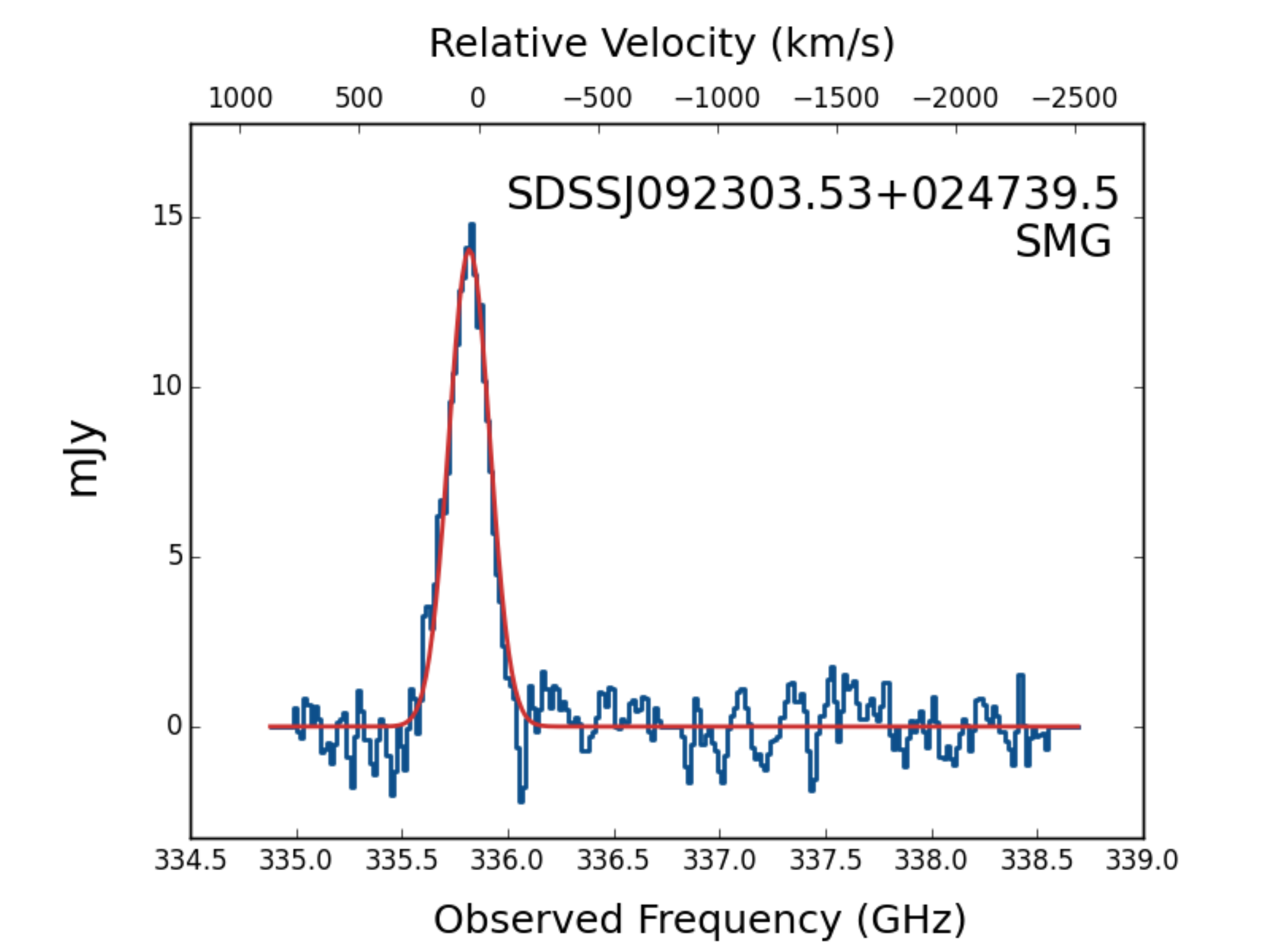}
\includegraphics[trim={0cm 0cm 1.9cm 0.2cm},clip,width=0.33\textwidth]{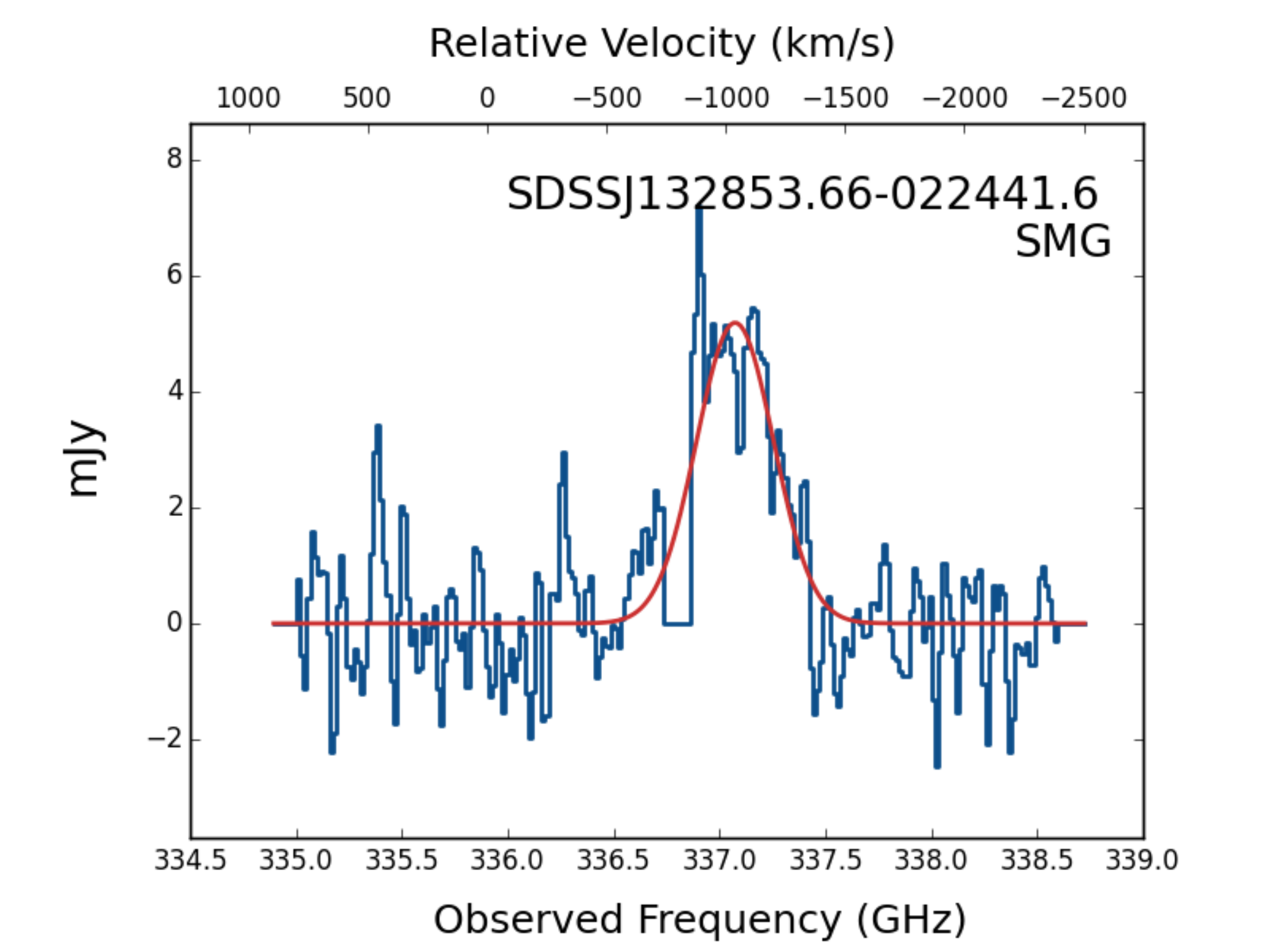}
\includegraphics[trim={0cm 0cm 1.9cm 0.2cm},clip,width=0.33\textwidth]{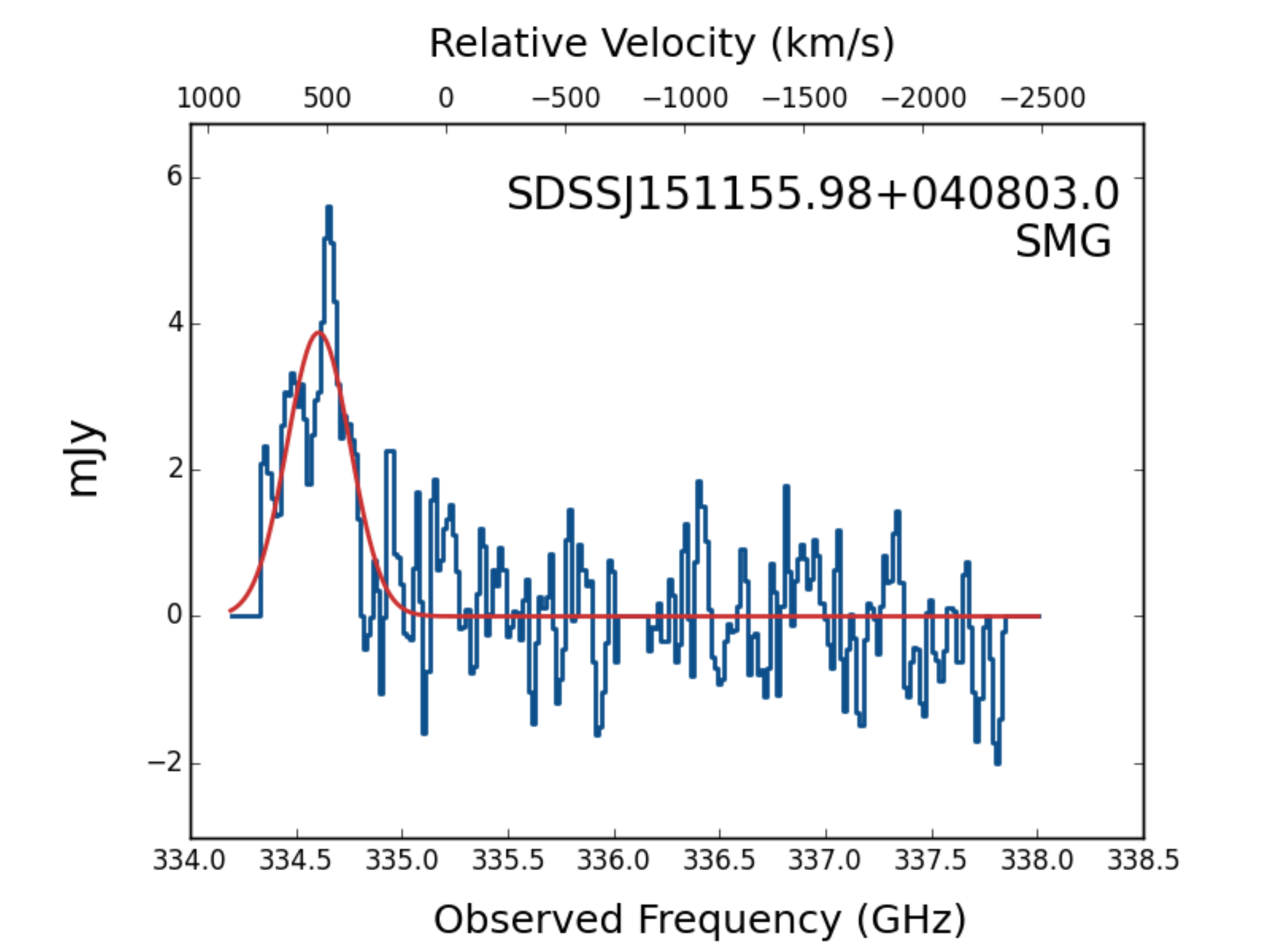}\\
\caption{
Spectra of the \CII\ emission line for all the sources with clear line detections: 
the \Nbright\ ``FIR-bright'' sources in our sample (top row), the \Nfaint\ ``FIR-faint'' sources (middle row), and the accompanying SMGs (bottom row).
The upper $x$-axes denote the velocity offsets with respect to the quasars systemic redshifts, derived from the \mgii\ broad emission lines (T11).
Red lines show the Gaussian fits to the line profiles. 
}
\label{fig:spec_cii}
\end{figure*}

\autoref{fig:cont_maps_lg} presents the full-scale continuum emission maps of the sources, extending to about 6\farcs8 ($\sim$45 \kpc) from the quasars' locations,
and \autoref{fig:cii_maps_lg} shows the equivalent \cii\ line emission maps.
%
All \Ntot\ quasar hosts are clearly detected in our new ALMA data, both in continuum and in \cii\ line emission, with the least significant \cii\ detection having S/N$\sim$4.5 (for J0935; see \autoref{tab:lum_cii}).
The \cii\ spectra of the quasar hosts are shown in \autoref{fig:spec_cii} (together with the companions we discuss below).
We note that the central frequencies of the \cii\ lines are shifted, by several hundreds of \kms, from the redshifts determined using the quasars' broad UV emission lines \cite[i.e.,][]{Hewett2010_SDSS_z}. 
We discuss this in more detail in \autoref{subsec:lines} below.
%

\capstartfalse
\begin{deluxetable*}{lllcccccccccc}
\tablecolumns{13}
\tablewidth{0pt}
\tablecaption{Spectral Measurements \label{tab:lum_cii}}
\tablehead{
\colhead{sub-sample} &
\multicolumn{2}{c}{Target} &
\colhead{Cont. Flux} &
\colhead{$\nu$} &
\colhead{Cont. size \tablenotemark{a}} &
\colhead{$F_{\cii}$} &
\colhead{FWHM$_{\cii}$} &
\colhead{$\nu_{0,\cii}$} &
\colhead{\cii\ size \tablenotemark{a}} &
\colhead{$L_{\cii}$} &
\colhead{$\Delta d$ \tablenotemark{b}} &
\colhead{$\Delta v$ \tablenotemark{c}} \\
       & ID     & comp. & mJy          & GHz    & \sq\arcsec       & Jy \kms     & \kms       & GHz             & \sq\arcsec        & $10^9\,\Lsun$  & kpc & \kms  
}
\startdata
Bright & J0331  & QSO   & $ 4.3\pm0.2$ & 344.47 & $0.18\times0.11$ & $5.2\pm0.5$ & $495\pm31$ & $331.29\pm0.006$ & $ 0.25\times0.15$  & 3.58 &  ...    & ...    \\
~~~~~~ & J1341  & QSO   & $18.5\pm0.5$ & 346.84 & $0.23\times0.21$ & $6.1\pm0.8$ & $384\pm16$ & $333.41\pm0.006$ & $ 0.47\times0.28$  & 4.51 &  ...    & ...    \\
~~~~~~ & J1511  & QSO   & $10.4\pm0.3$ & 347.99 & $0.22\times0.19$ & $5.3\pm0.4$ & $507\pm36$ & $334.69\pm0.011$ & $<0.39\times0.12$  & 3.42 &  ...    & ...    \\
~~~~~~ &        & SMG   & $ 1.6\pm0.2$ & 347.99 & $0.43\times0.11$ & $2.4\pm0.7$ & $313\pm38$ & $334.61\pm0.021$ & $<0.94\times0.16$  & 1.53 & 13.9    & $ +75$ \\
~~~~~~ &        & B     & $ 2.0\pm0.3$ & 347.99 & $0.52\times0.28$ & ...         & ...        & ...              & ...                & ...  & 25.2    & ...    \\
\hline \\ [-1.75ex]
Faint  & J0923  & QSO   & $3.0\pm0.1$  & 348.65 & $0.28\times0.24$ & $4.1\pm0.3$ & $363\pm11$ & $336.09\pm0.005$ & $ 0.45\times0.34$  & 3.21 &  ...    & ...    \\
~~~~~~ &        & SMG   & $1.2\pm0.3$  & 348.65 & $0.57\times0.36$ & $4.1\pm0.7$ & $214\pm10$ & $335.82\pm0.004$ & $ 0.60\times0.31$  & 2.16 & 36.5    & $+246$ \\
~~~~~~ & J1328  & QSO   & $1.7\pm0.1$  & 348.74 & $0.40\times0.21$ & $3.1\pm0.3$ & $221\pm14$ & $336.59\pm0.008$ & $ 0.35\times0.31$  & 1.63 &  ...    & ...    \\
~~~~~~ &        & SMG   & $0.7\pm0.2$  & 348.74 & $<0.56\times0.26$& $2.0\pm0.3$ & $423\pm43$ & $337.07\pm0.020$ & $ 0.71\times0.44$  & 2.01 & 44.5    & $-432$ \\
~~~~~~ & J0935  & QSO   & $1.6\pm0.1$  & 347.93 & $0.26\times0.16$ & $0.9\pm0.2$ & $338\pm42$ & $334.49\pm0.016$ & $ 0.44\times0.25$  & 0.80 &  ...    & ...     %
\enddata
\tablenotetext{a}{Note that the \emph{deconvolved} sizes of the \cii\ emitting regions are smaller than those reported in \autoref{tab:obs_log}. \texttt{CLEAN} performed with weighting=``briggs''.}
\tablenotetext{b}{Distances between the companions and the quasar hosts, calculated based on the centroids of the ALMA continuum emission, and assuming the redshifts of the quasar hosts' \cii\ emission lines.}
\tablenotetext{c}{Velocity offsets of the accompanying SMGs with respect to the quasar hosts, calculated from the central frequencies of the \cii\ emission lines.}
\end{deluxetable*}
\capstarttrue

\begin{figure*}[t!]
\includegraphics[width=0.34\textwidth]{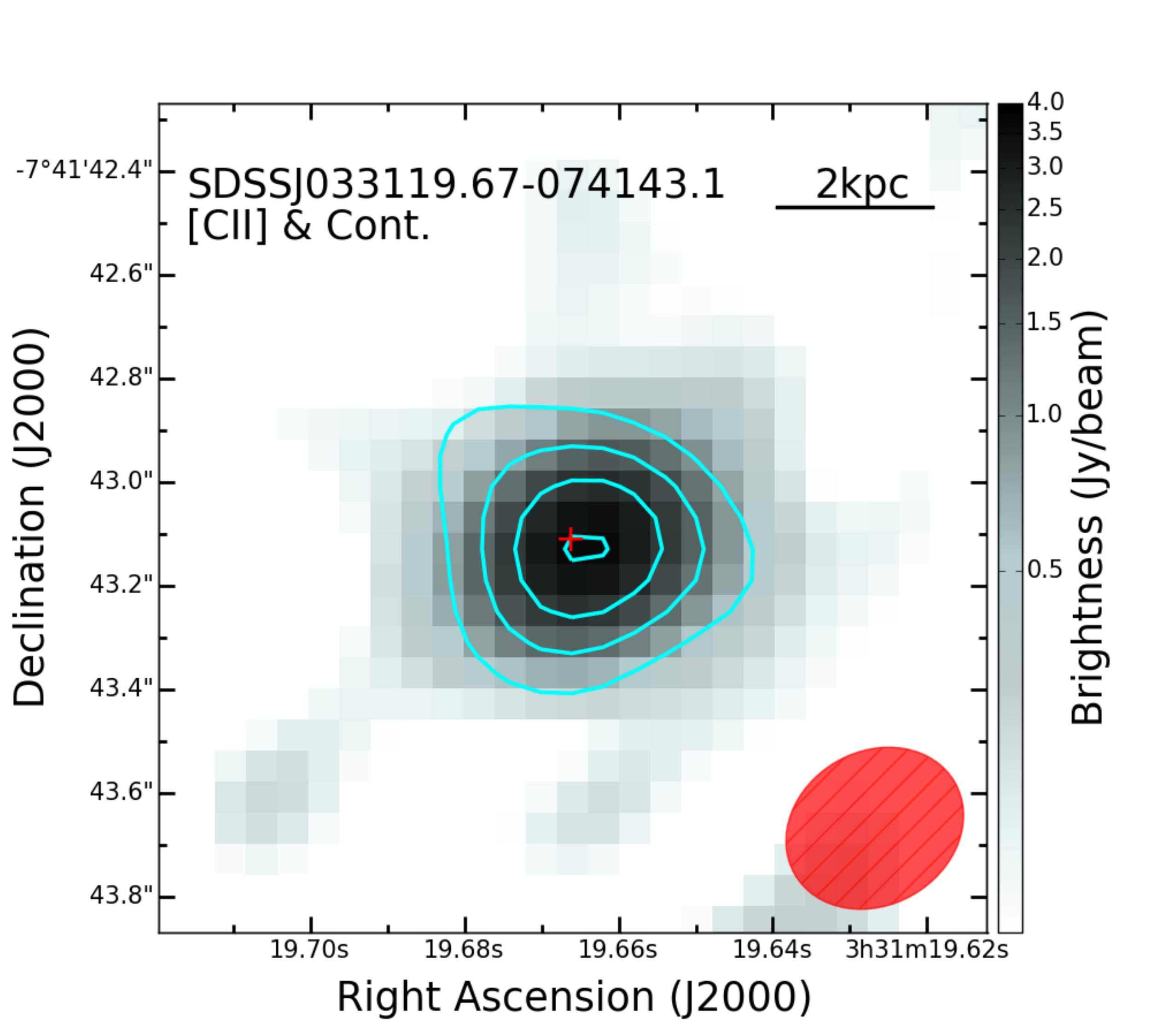} 
\includegraphics[width=0.34\textwidth]{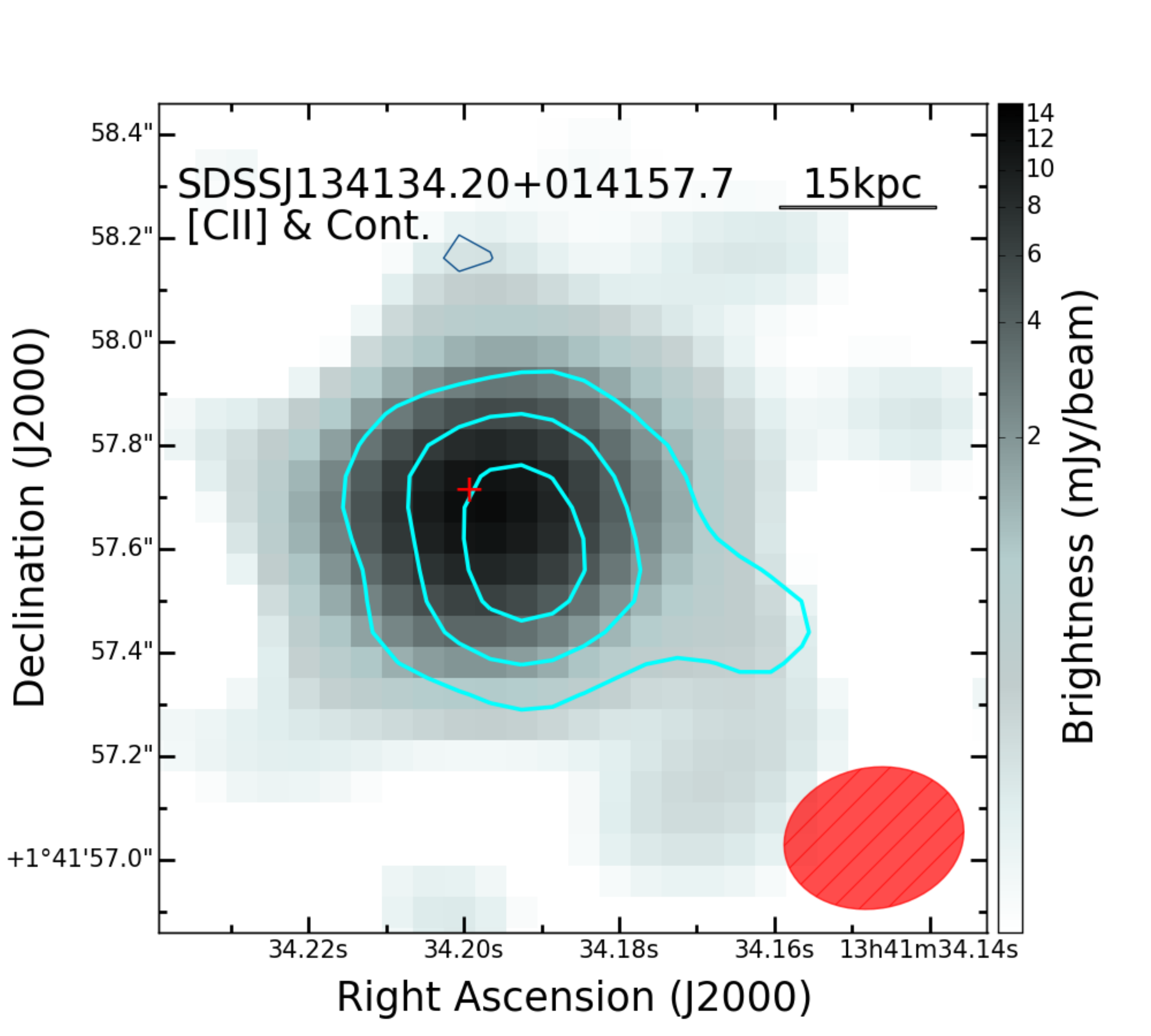}
\includegraphics[width=0.34\textwidth]{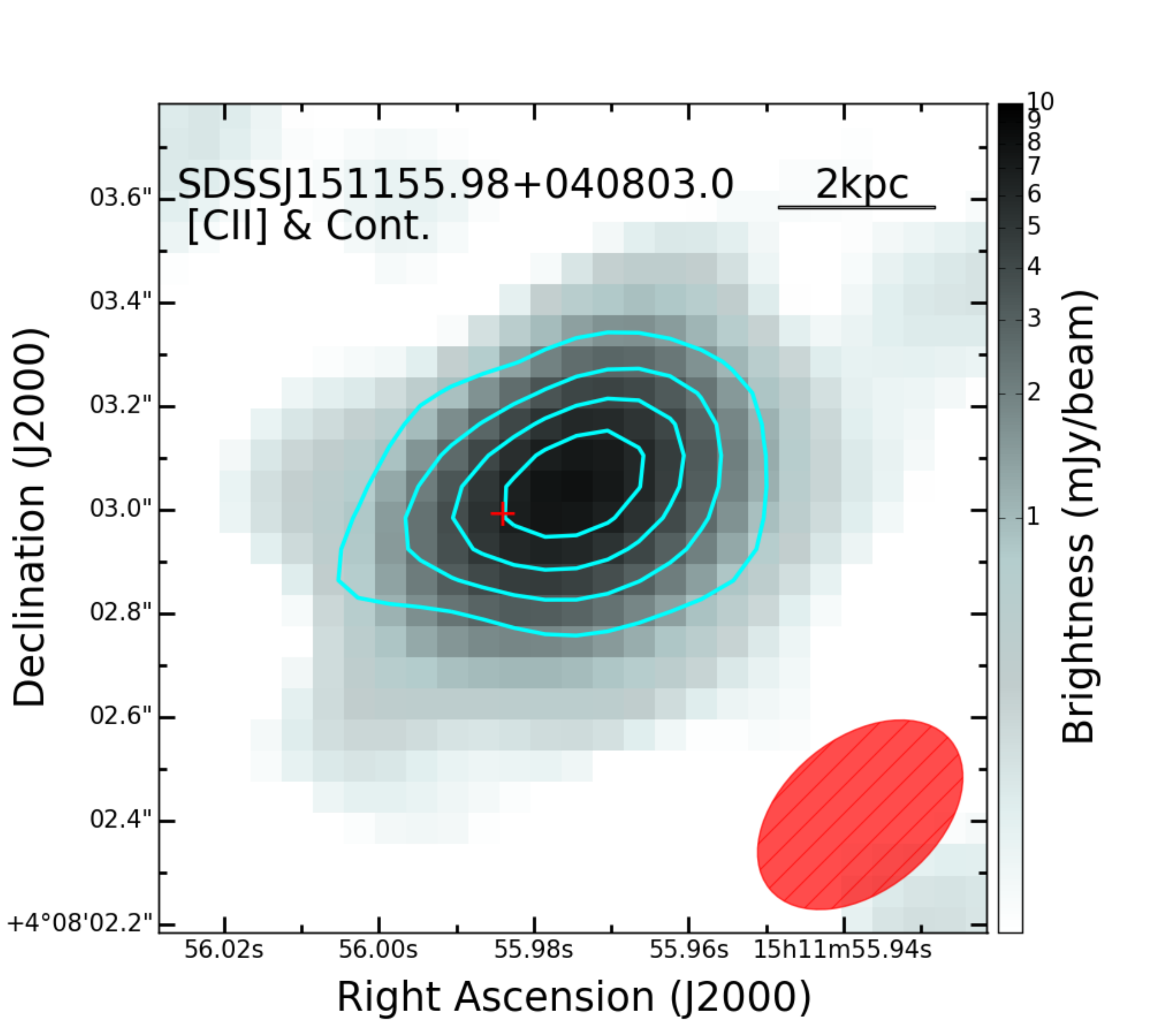} \\
%
%
\includegraphics[width=0.34\textwidth]{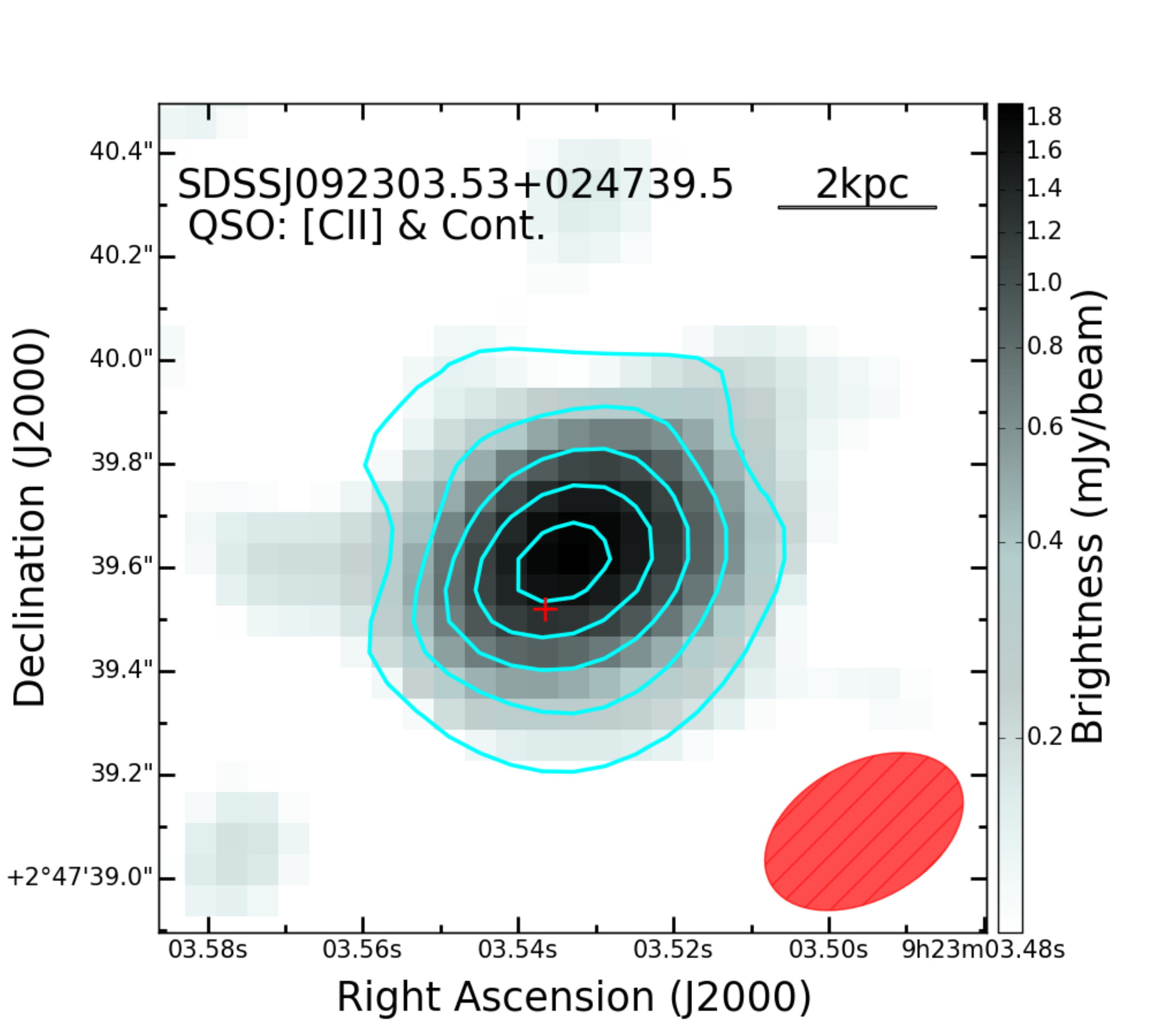}
\includegraphics[width=0.34\textwidth]{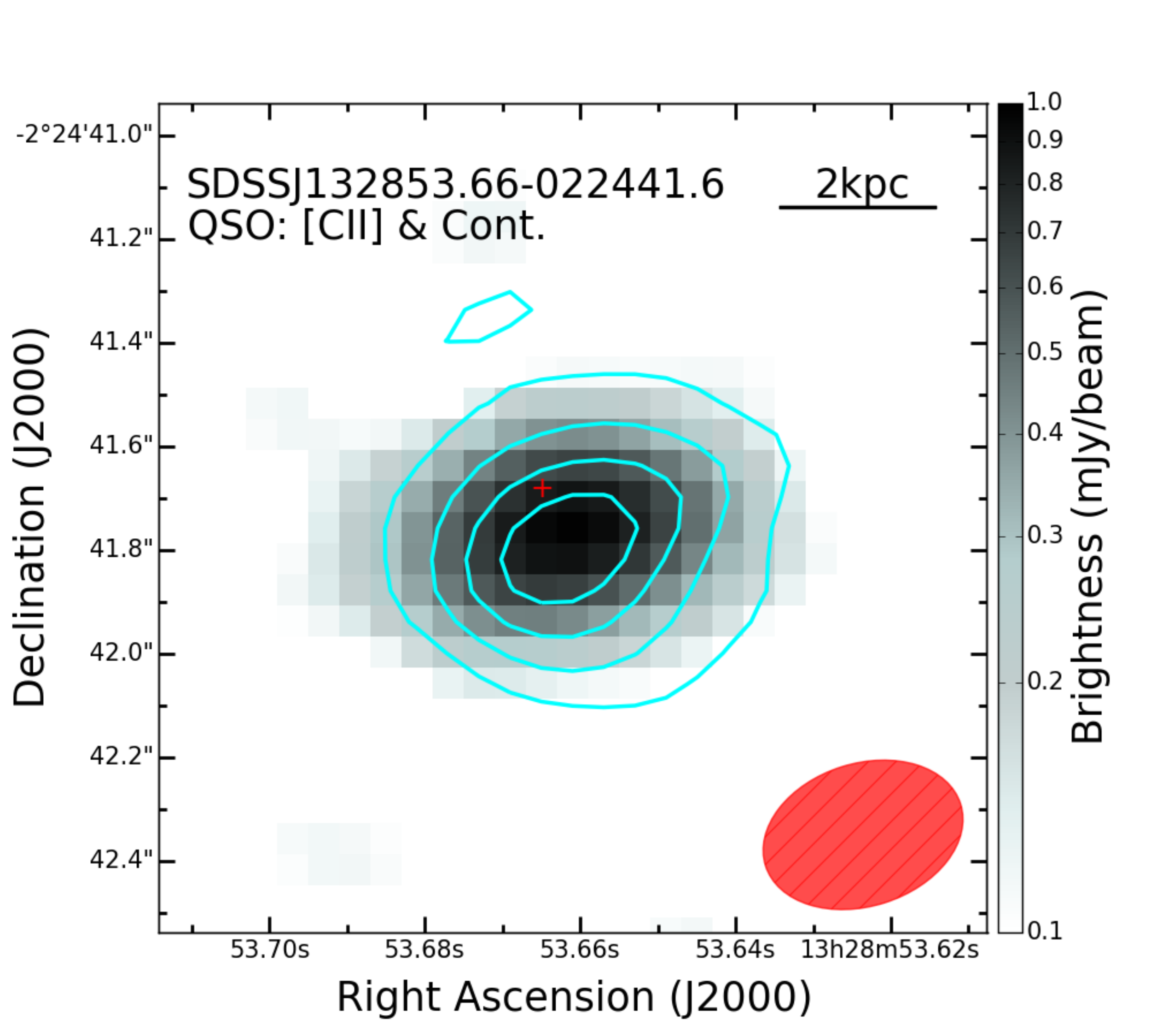}
\includegraphics[width=0.34\textwidth]{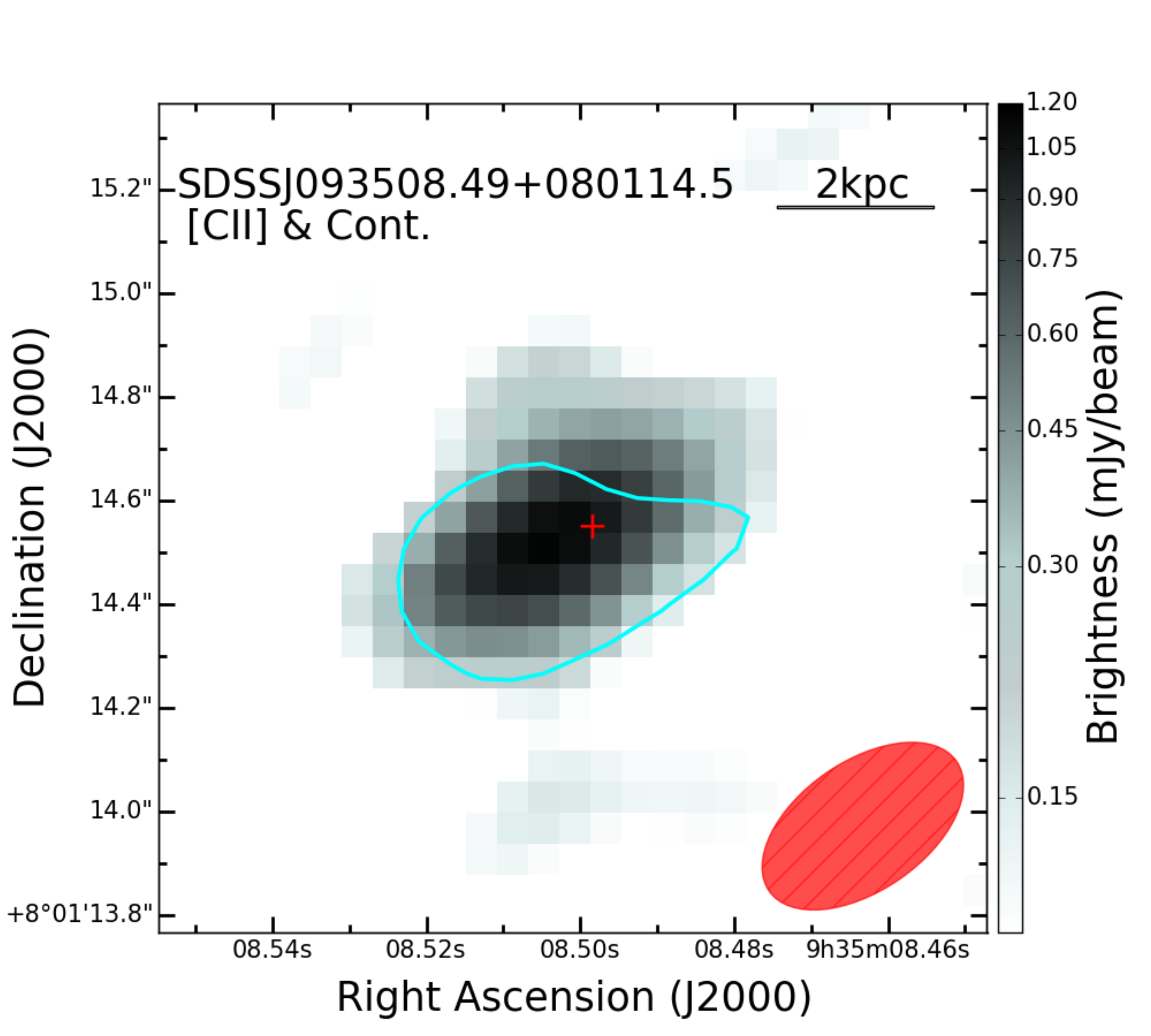}\\
\vspace*{-0.5cm}
\begin{center}
\hspace{1cm}\rule[1ex]{1\textwidth}{0.5pt}\\
\end{center}
\includegraphics[width=0.34\textwidth]{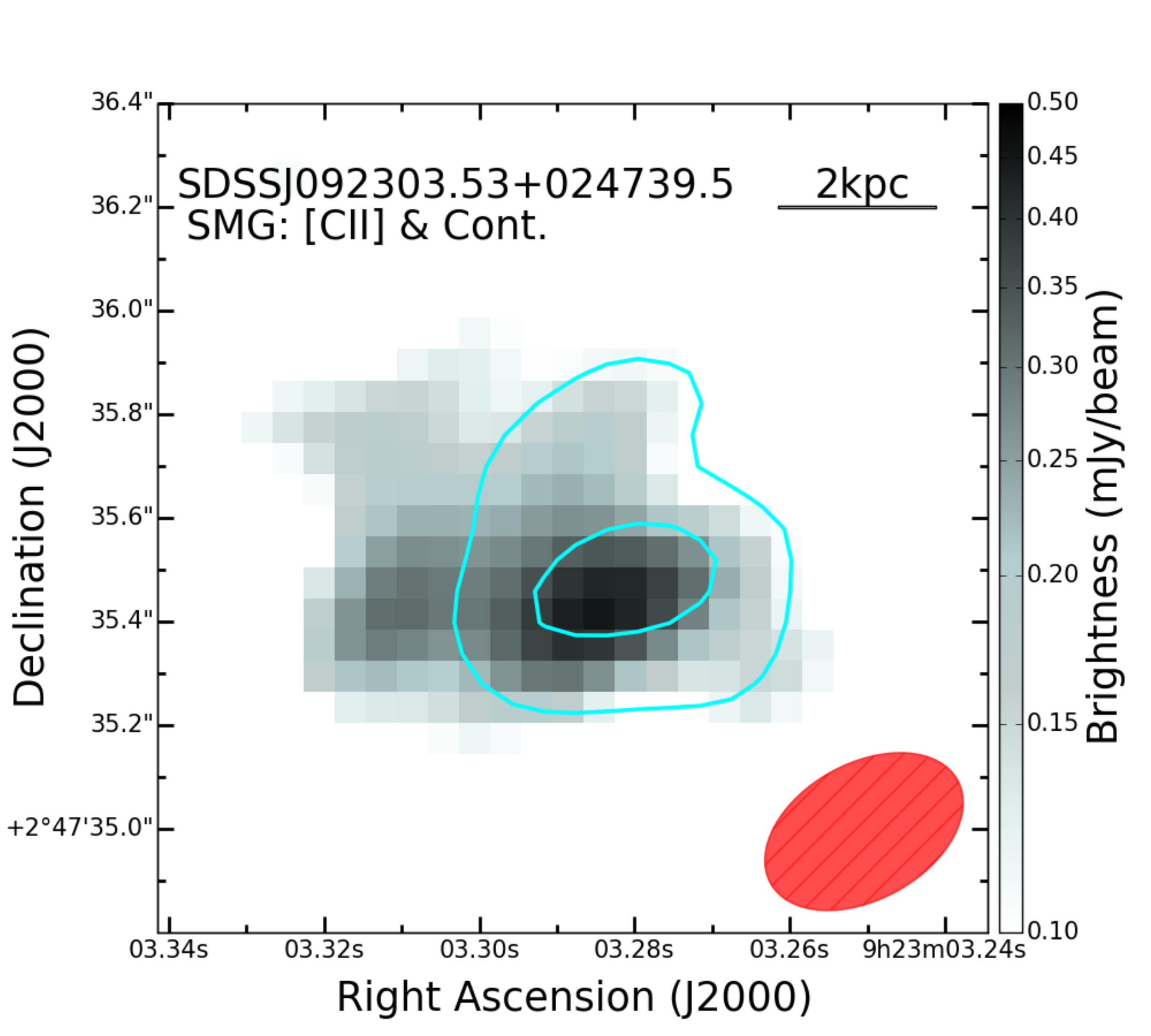}
\includegraphics[width=0.34\textwidth]{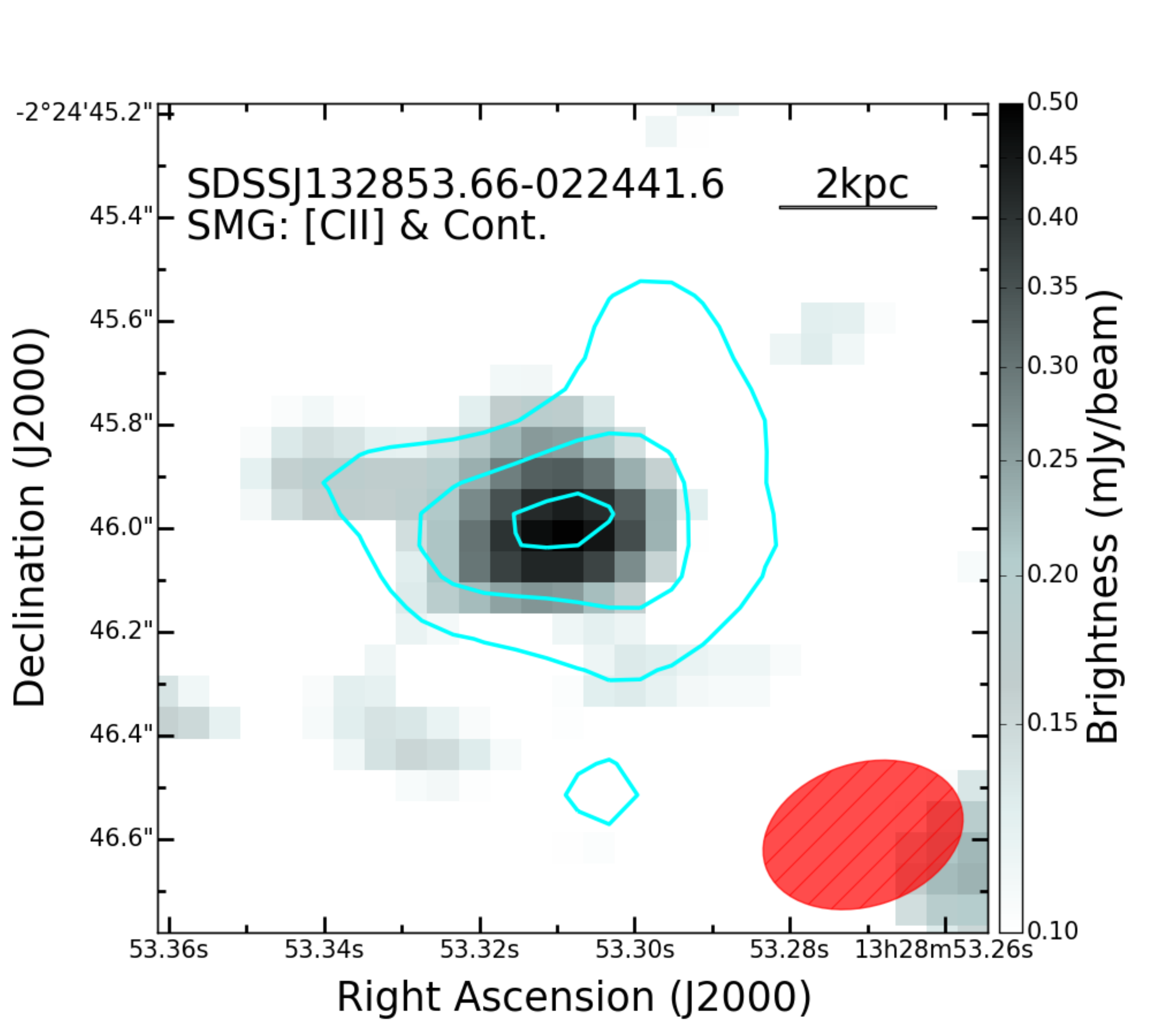} 
\includegraphics[width=0.34\textwidth]{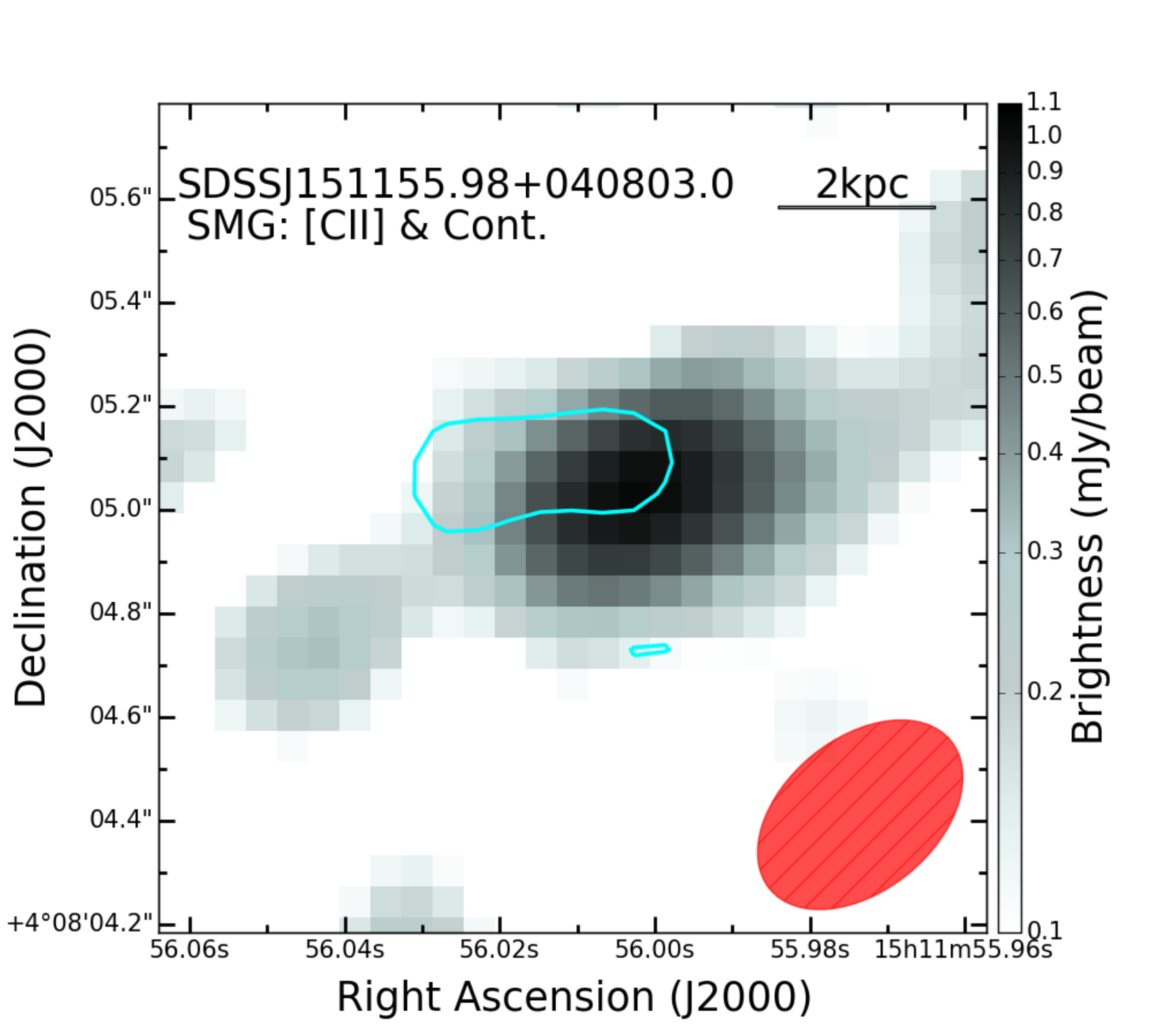} \\
\caption{
Small-scale continuum and \cii\ line emission maps derived from the new ALMA data, for all the sources with clear detection of \cii\ line emission:
the \Nbright\ ``FIR-bright'' sources in our sample (top row), the \Nfaint\ ``FIR-faint'' sources (middle row), and the accompanying SMGs (bottom row).
For each source, the gray-scale map traces the continuum emission, while 
the contours trace the \cii\ line emission (i.e., surface brightness) at significance levels of 3, 6, 9, 12, and 15$-\sigma$.
For each source, the line fluxes used for the contours were extracted from a spectral window spanning $\pm500\,\kms$ around the \cii\ line peak.
Red crosses mark the locations of the quasars' optical emission (taken from the SDSS).
The ALMA beams are shown as red ellipses near the bottom-right of each panel.
}
\label{fig:comb_maps_sm}
\end{figure*}

For one of the FIR-bright sources, J1511, the ALMA data reveal two faint sub-mm-emitting apparent companions, which are detected at the 7-8$\sigma$ level in continuum emission. 
Their centers are spatially separated from the quasar host by about 2\farcs1 and 3\farcs9, which corresponds to 14 and 25 \kpc\ (see \autoref{tab:lum_cii}). 
These two sources are fainter than the quasar host, in the ALMA continuum data, by at least a factor of 5. 
Most importantly, we detect significant \cii\ line emission from the companion closer to the quasar host, as seen in Figures \ref{fig:cii_maps_lg} and \ref{fig:comb_maps_sm}.
The integrated \cii\ line flux suggests that the detection is significant at the $\sim3.5\sigma$ level.
The spectrum we extract for this companion (\autoref{fig:spec_cii}) demonstrates that the line emission, if associated with a redshifted \cii\ transition, is shifted by $\sim$75 \kms\ from the \cii\ line of the corresponding quasar host (see \autoref{tab:lum_cii}).
We do not detect any significant \cii\ line emission from the other, more distant continuum source accompanying J1511.
Moreover, we do not detect any other significant emission in the multi-wavelength data available for the \zfpe\ quasars, for any of these two companions of J1511.
This may suggest that the most distant source seen in the ALMA maps of J1511 is a \emph{projected} companion source, not physically related with the J1511 system.
However, given the low probability of observing such a (second) faint sub-mm source within a single ALMA pointing (see below), it may be indeed tracing a faint, low-mass, and/or low-SFR galaxy that is associated with the quasar host. 
In such a case, the observed properties suggest that it would be a \emph{minor} merger. 
This latter interpretation is further justified by our findings of clearly \emph{interacting} companions, both for J1511 and two of the FIR-faint quasar systems, as described below.
In particular, we note that the interacting companions show clear \cii\ emission, while the distant companion of J1511 does not, despite having comparable FIR continuum fluxes.

We identify two additional emission sources in the maps of two of the FIR-faint sources, J0923 and J1328 -- one around each of these quasars -- which are clearly detected in the ALMA continuum maps.
These sources are detected at significance levels of $\sim3-4\,\sigma$, and have continuum flux densities of 1.2 and 0.7 mJy (for the companions of J0923 and J1328, respectively).
The centers of these accompanying sources are located 5\farcs6 and 6\farcs8 from the quasars hosts, corresponding to about 36.5 and 44.5 kpc.
These two accompanying sources show associated \cii\ line emission, as seen in the bottom panels of \autoref{fig:cii_maps_lg} (and also of \autoref{fig:comb_maps_sm}).
Similarly to the companion of J1511 described above, the spectra we extract for these two companions (\autoref{fig:spec_cii}) demonstrate that the \cii\ line emission is shifted by less than $\sim$450 \kms\ from the \cii\ lines of the corresponding quasar hosts.
We finally note that, similarly to the aforementioned companion of J1511, these two sources are not associated with any other (significant) emission in the multi-wavelength dataset we have available for the \zfpe\ quasars, particularly the \Spitzer\ data (at observed-frame 3.6 and 4.5 \mic; N14).
The (projected) spatial and velocity offsets of all companions are given in \autoref{tab:lum_cii}.

\autoref{fig:comb_maps_sm} presents smaller-scale continuum and line emission maps of all the spectroscopically confirmed systems we identify.
The regions where the continuum and line emission is significantly detected are resolved by a few synthesized beams, in all sources.
As can be clearly seen, the peaks of the \cii\ line emission coincides with the peaks of the dust-dominated continuum emission.
In the quasars hosts, these peaks of continuum and \cii\ emission are also consistent with the locations of the quasars, to within 0\farcs1, as determined from the SDSS optical imaging.

Before moving on with our analysis of the properties of our quasar hosts and companions and our interpretation of these detections in the context of major galaxy mergers (\autoref{subsec:mergers}), we would like to emphasize that the sheer detection of three (and possibly four) faint sub-mm companion sources within our ALMA data are, by itself, highly surprising.
The faintest continuum sources in our data (regardless of \cii\ detection) have flux densities on the order of $\sim$0.5 mJy.
Based on recent deep (ALMA) sub-mm surveys \cite[e.g.,][and references therein]{Karim2013_ALMA_ECDFS,Carniani2015_submm_counts,Aravena2016_HUDF_cont,Fujimoto2016_deep_faint_ALMA}, we would expect on order of $\sim$0.1 such sources per ALMA pointing (i.e., a circular field of view of 18\arcsec).
UV-selected, $z\sim4.8$ SF galaxies are yet more rare. 
The most recent measurements of areal densities in deep fields imply that on the order of 0.01 galaxies with $\sfr\sim100\,\mpyr$ would be observed within a single ALMA \obsband\ pointing \cite[][]{Stark2009,Bouwens2015_hiz_LF}.
Most, but not all such galaxies would have \cii\ detections \cite[e.g.,][]{Capak2015_CII_COSMOS,Aravena2016_HUDF_cii}
Number counts of purely \cii-emitting $z\gtrsim5$ galaxies are highly uncertain, but would probably amount to roughly 0.06 galaxy per each ALMA pointing \cite[e.g.,][]{Aravena2016_HUDF_cii}.
Thus, based on deep (sub-mm) surveys that are designed to detect high-redshift SF galaxies -- our robust detection of continuum-emitting companions in the fields surrounding three of the quasar hosts cannot be attributed to chance coincidence.
Obviously, the robust detection of \cii\ in three of these companions further strengths this conclusion.

We conclude that the host galaxies of all \Ntot\ quasars are clearly detected in the new ALMA data, in both continuum and \cii\ line emission.
We identify three sub-mm galaxies (SMGs hereafter) accompanying three of the quasar systems - one FIR-bright and two FIR-faint (one SMG accompanying each of these quasars).
The companion SMGs are located between roughly 14 and 45 kpc (projected) from the quasars, with relatively small velocity offsets, $\left|\Delta v\right|<450\,\kms$.
The three companion \cii-emitting SMGs -- detected solely through the new ALMA data -- are therefore physically related to, and interacting with the quasar systems.
We finally note that no additional line-emitting sources were found in our ALMA data.

\subsection{\cii\ emission line properties}
\label{subsec:lines}

\begin{figure*}[ht!]
\includegraphics[width=0.34\textwidth]{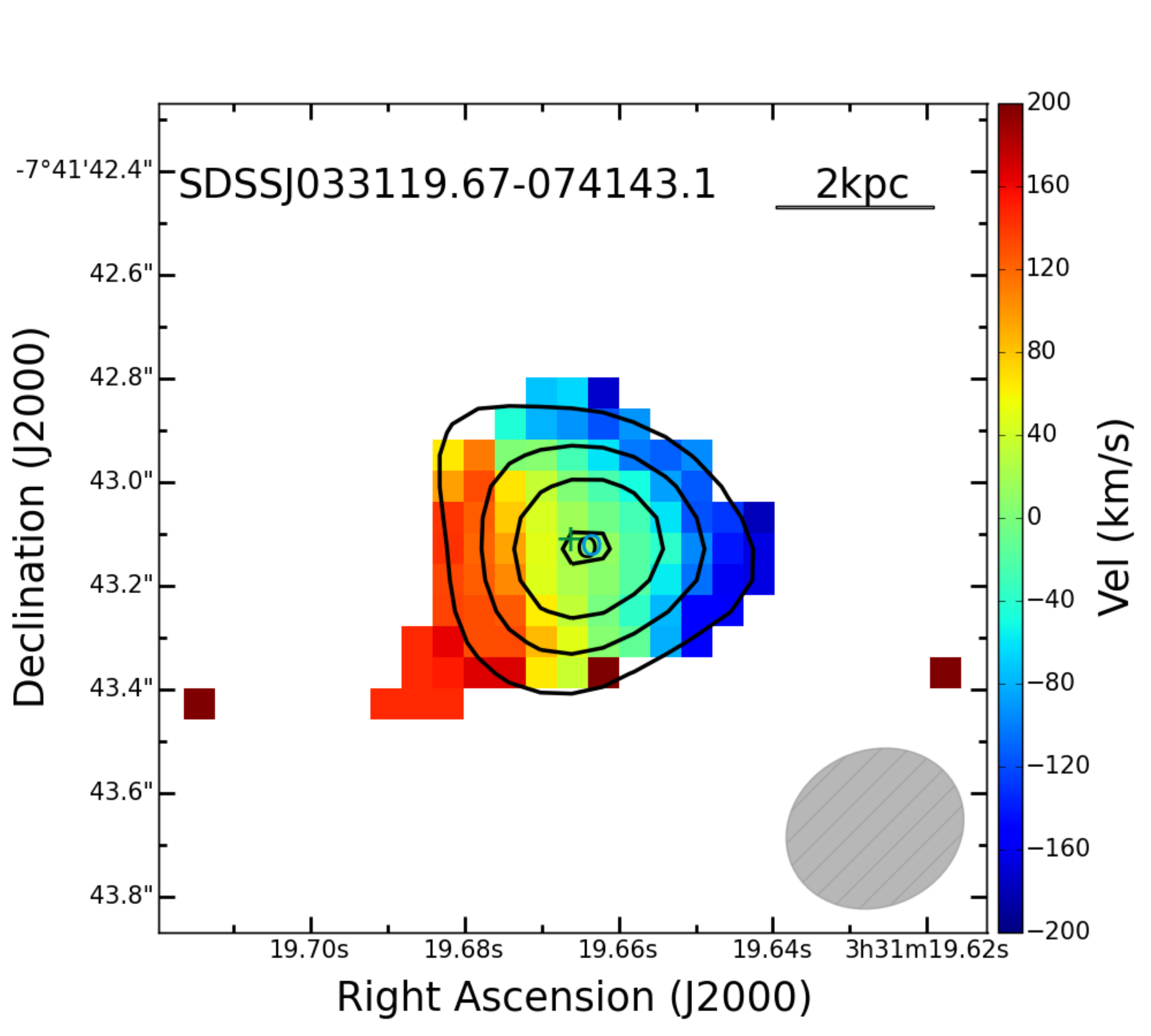} 
\includegraphics[width=0.34\textwidth]{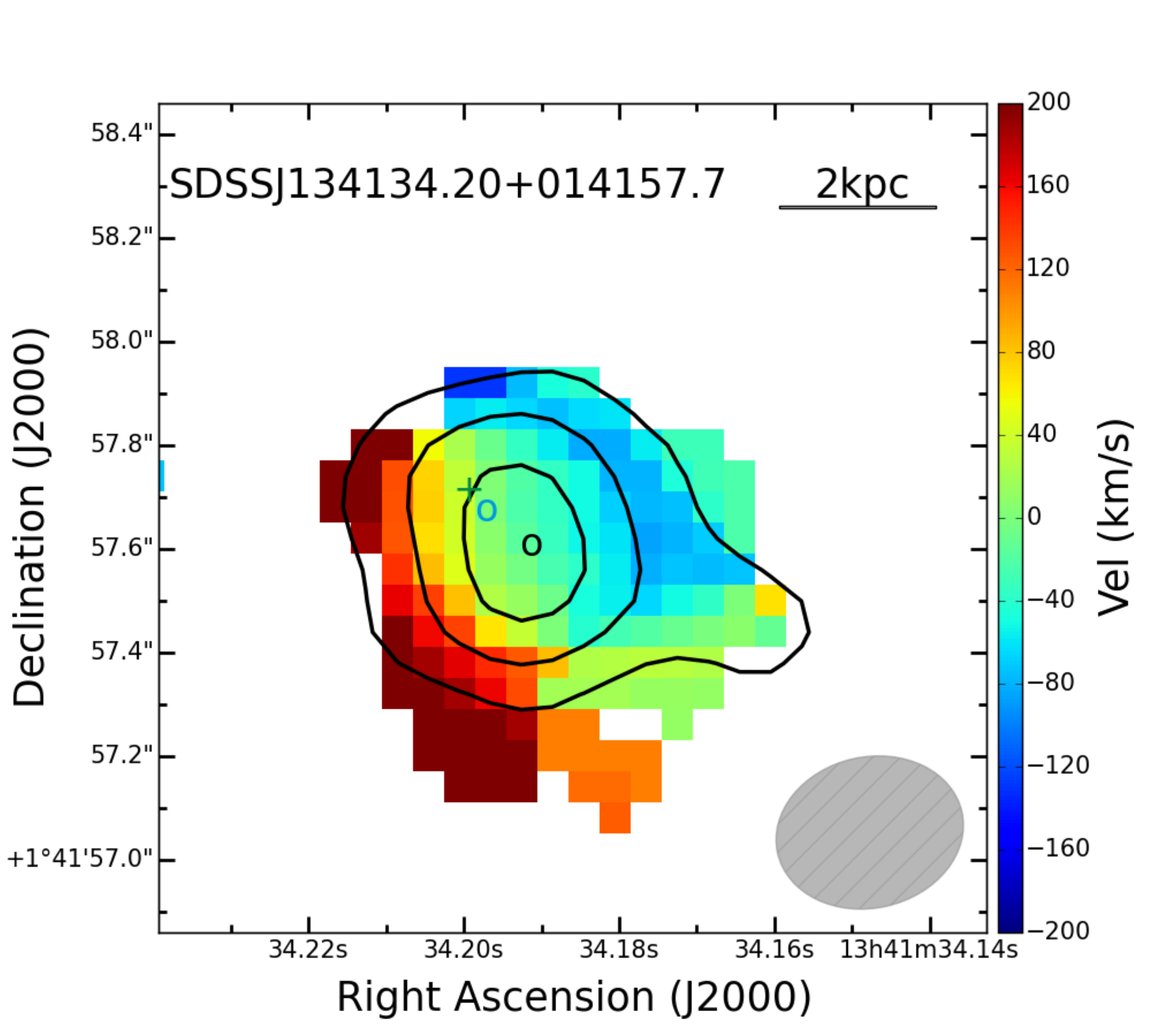}
\includegraphics[width=0.34\textwidth]{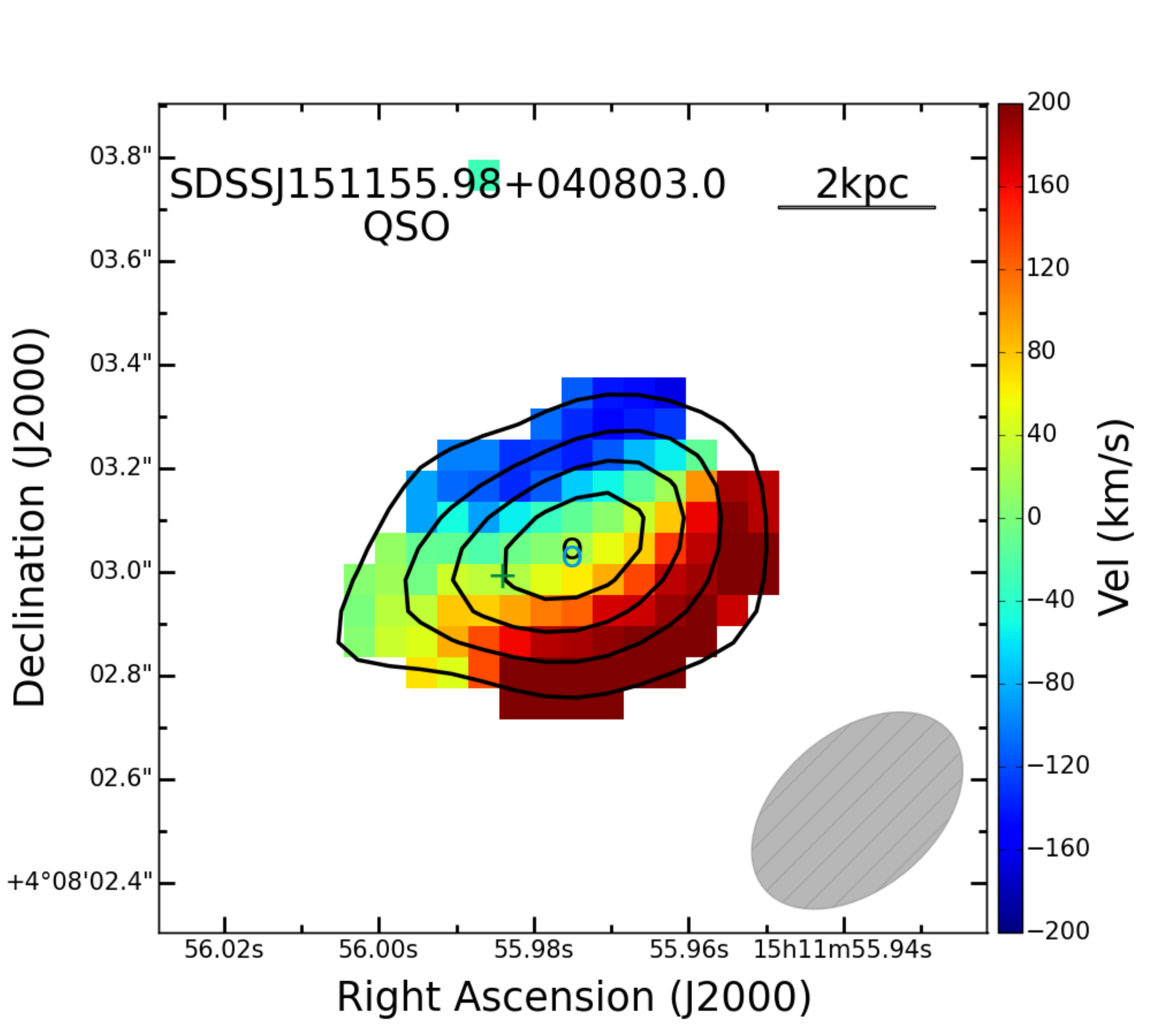} \\
%
%
\includegraphics[width=0.34\textwidth]{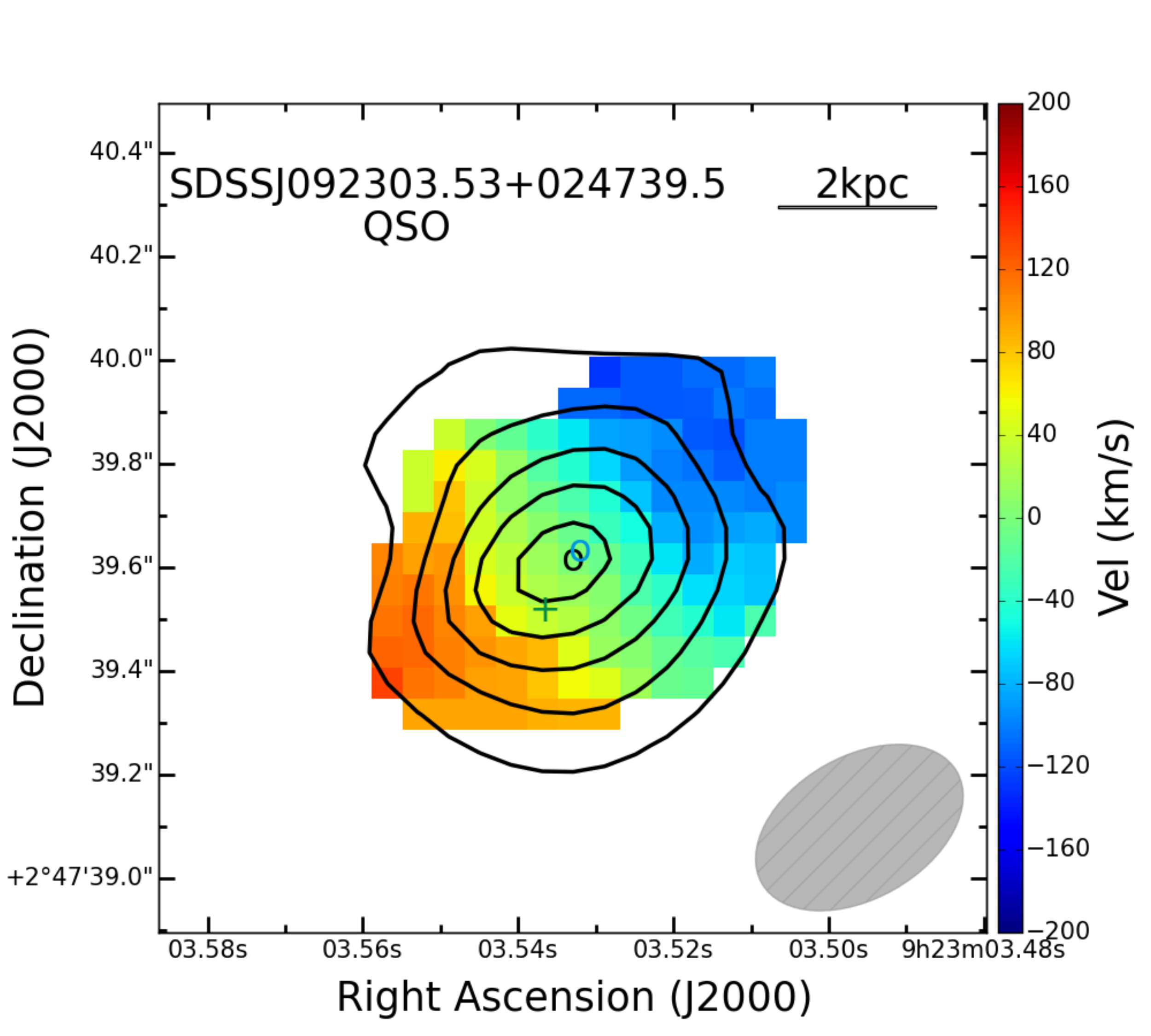}
\includegraphics[width=0.34\textwidth]{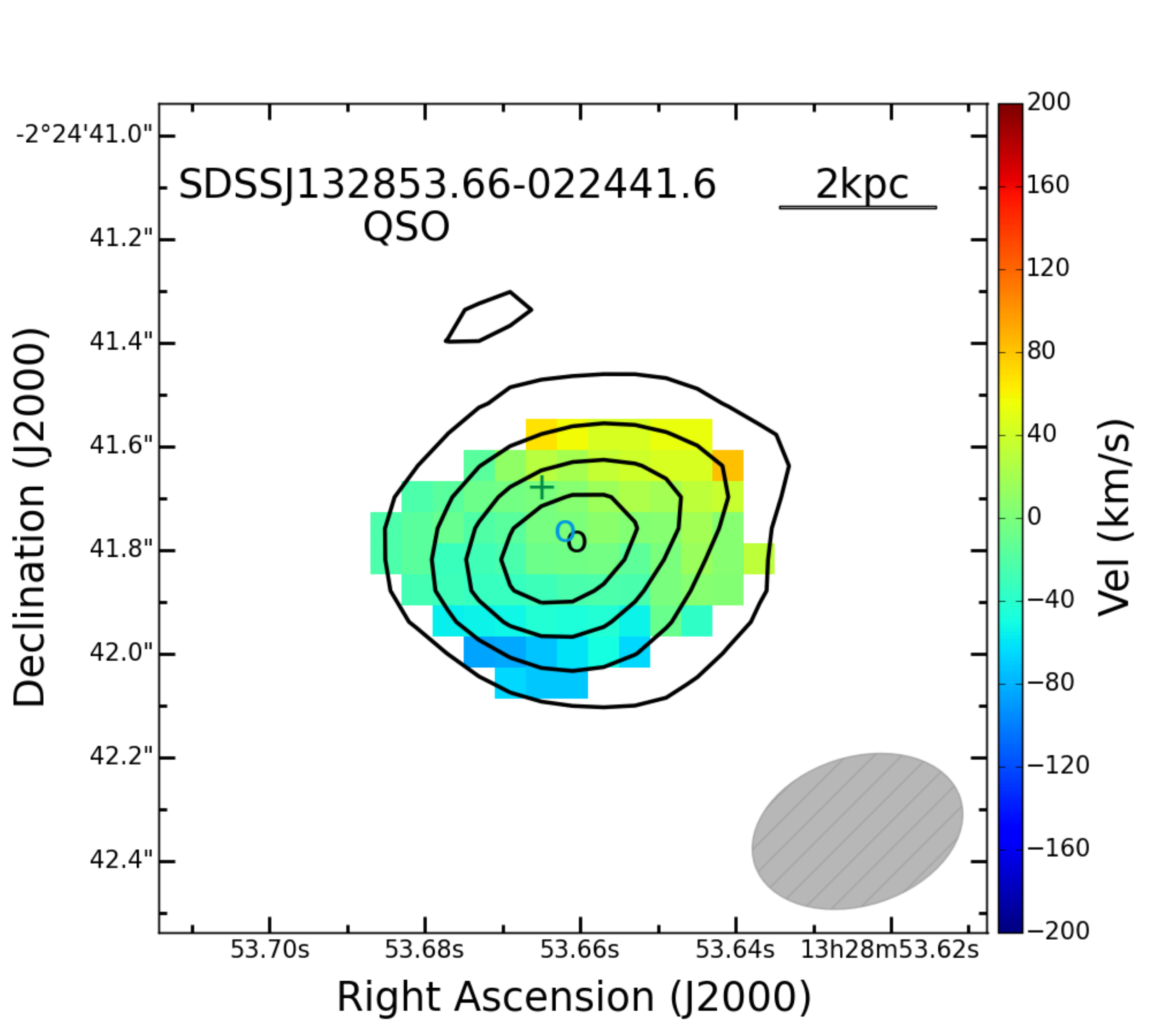}
\includegraphics[width=0.34\textwidth]{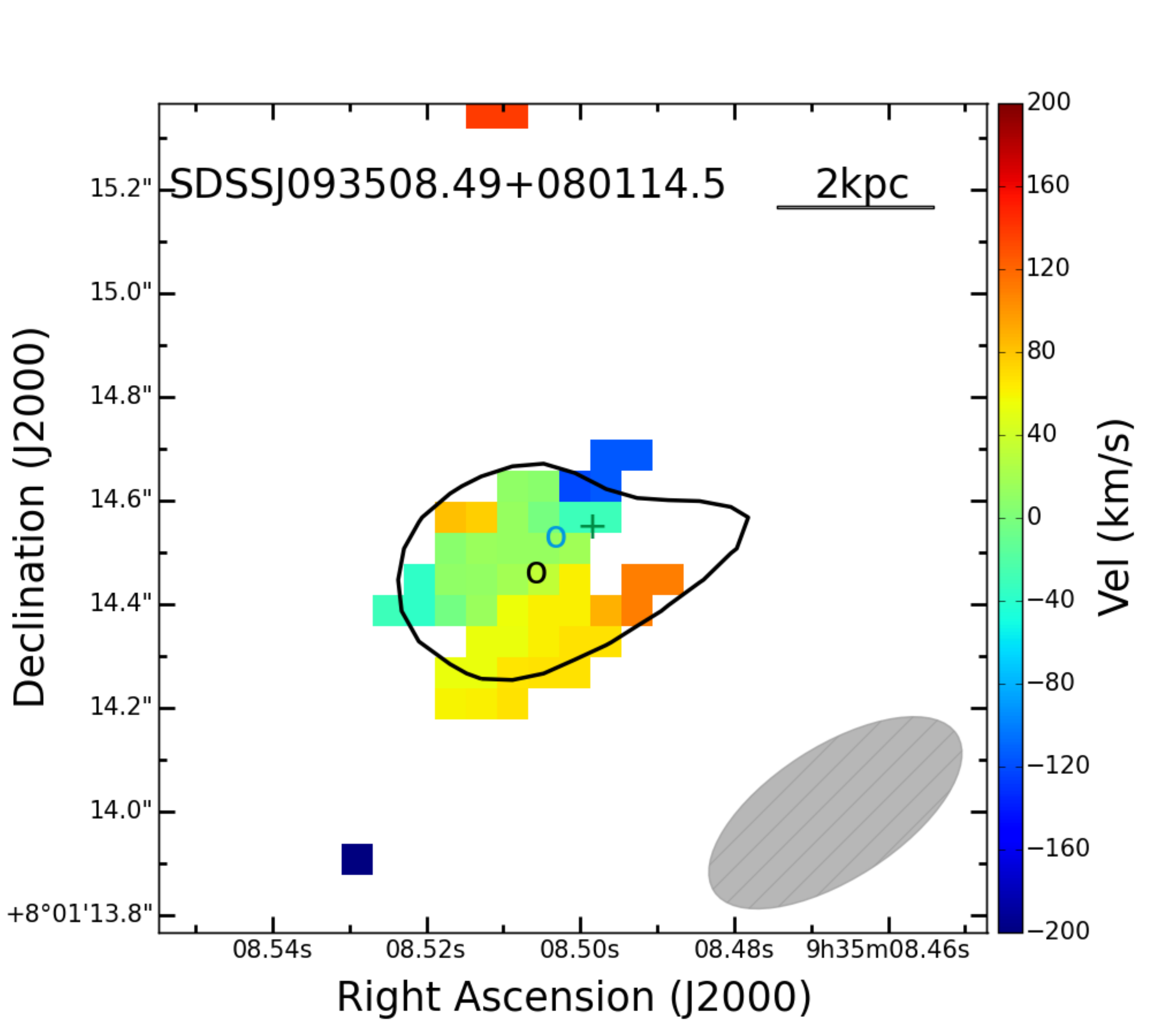}\\
\vspace*{-0.5cm}
\begin{center}
\hspace{1cm}\rule[1ex]{1.\textwidth}{0.5pt}\\
\end{center}
\includegraphics[width=0.34\textwidth]{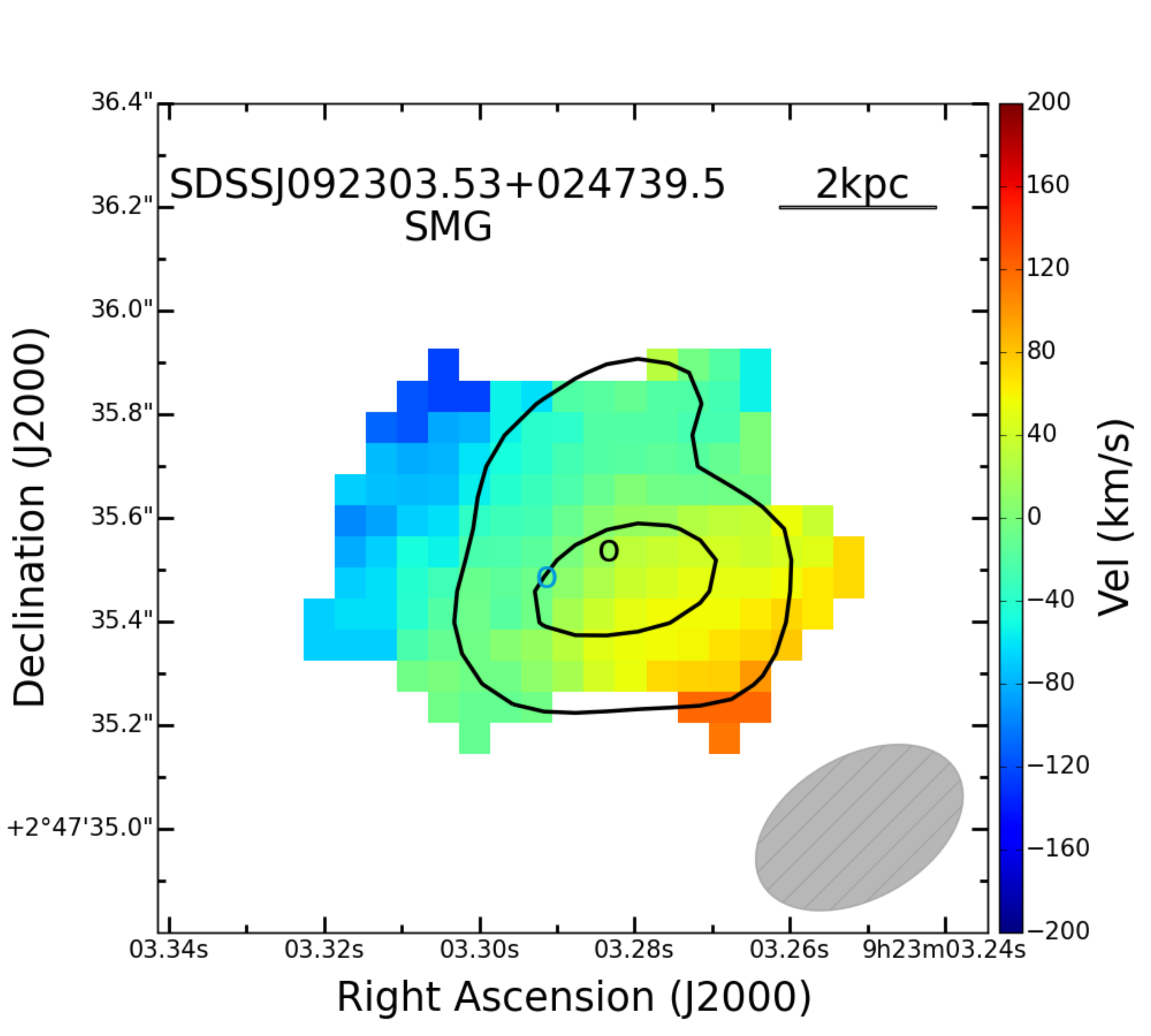}
\includegraphics[width=0.34\textwidth]{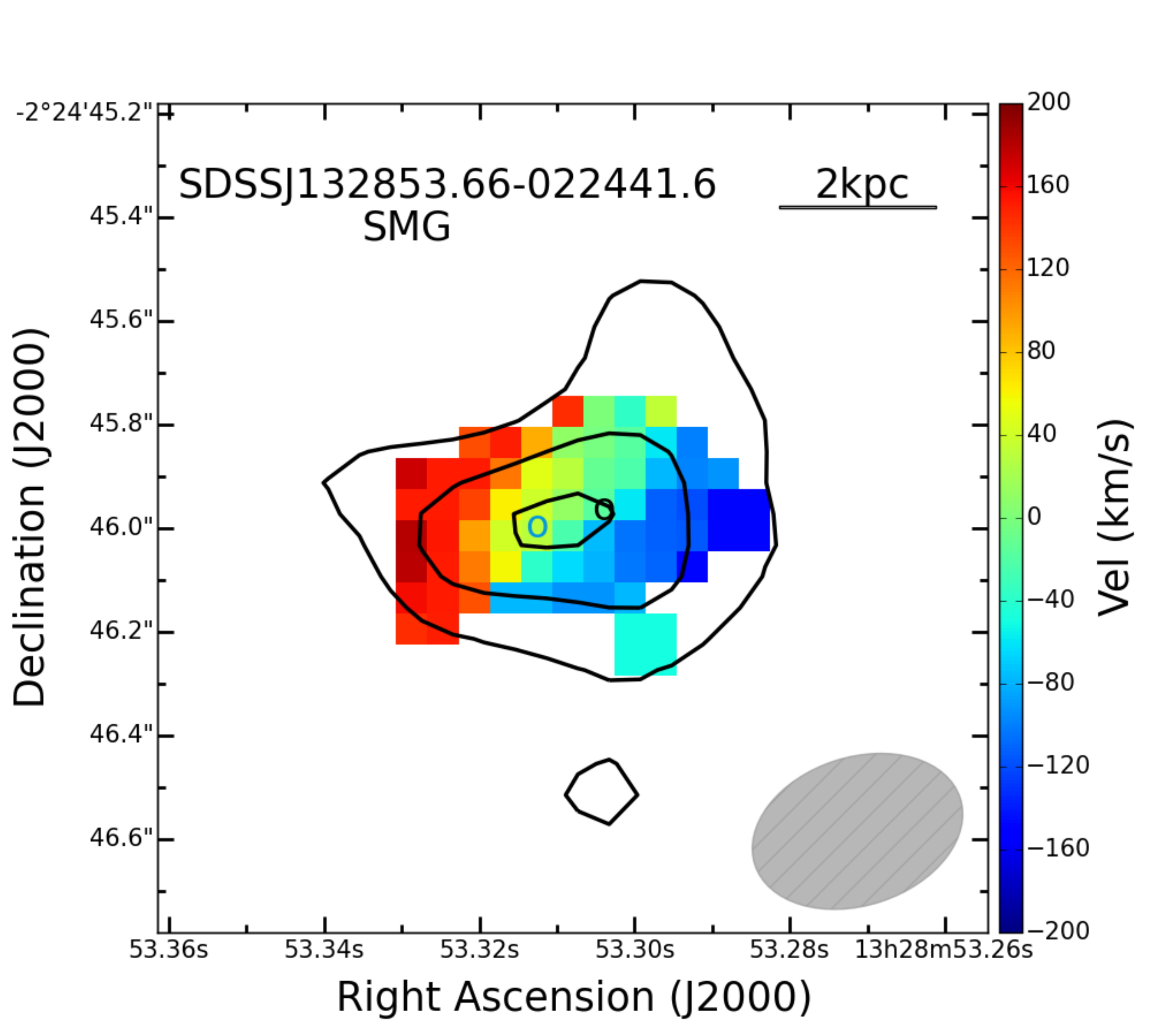}
\includegraphics[width=0.34\textwidth]{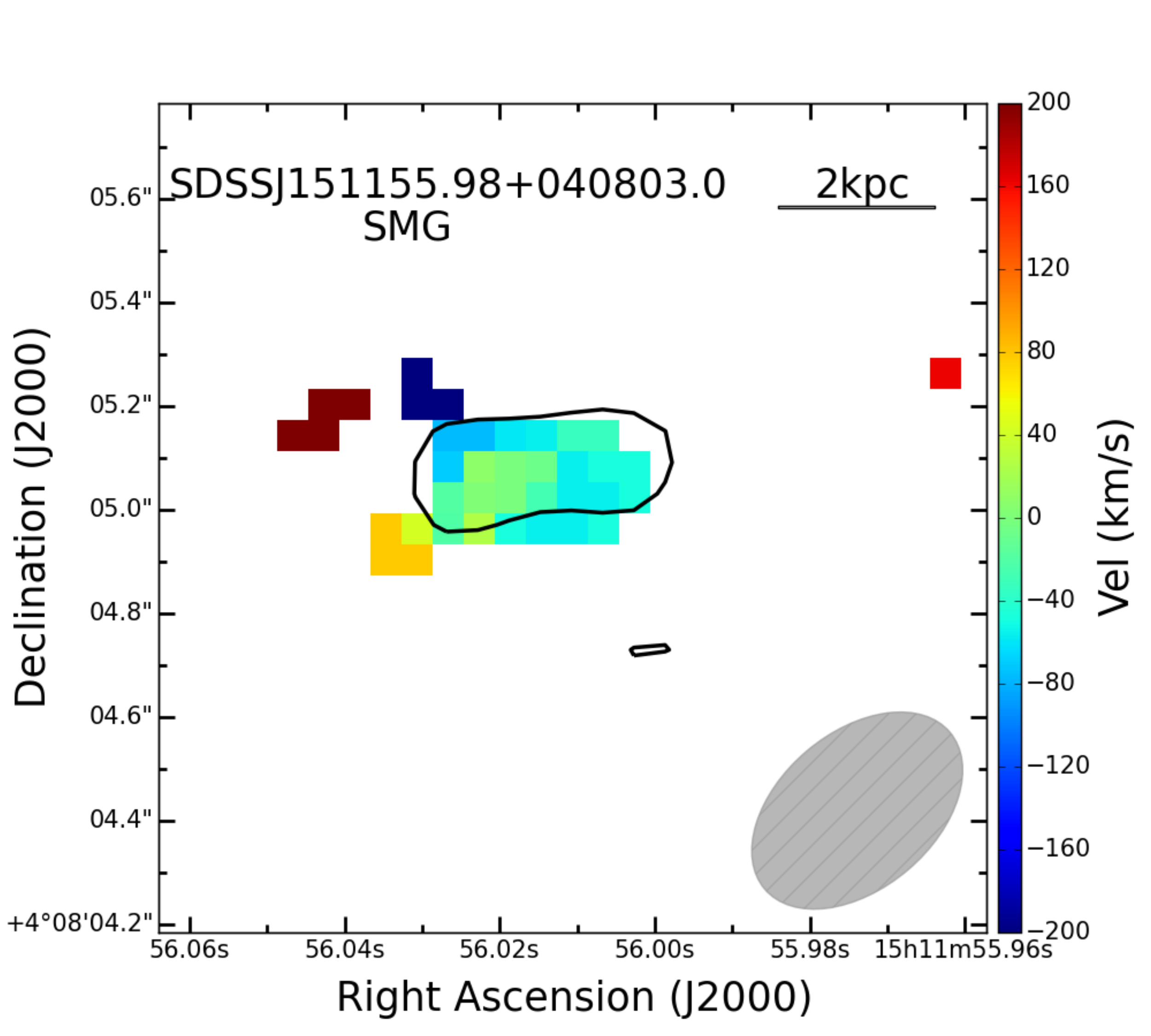}\\
\caption{
\cii\ velocity maps for 
the \Nbright\ ``FIR-bright'' sources in our sample (top row), the \Nfaint\ ``FIR-faint'' sources (middle row), and the companion SMGs (bottom row).
Black contours trace the \cii\ emission line surface brightness at significance levels of 3, 6, 9, 12, and 15$-\sigma$.
Crosses mark the locations of the quasars' optical emission (taken from the SDSS).
Gray circles mark the locations of the peak of the dust (ALMA) continuum emission.
Black circles mark the locations of the peak of the \cii\ emission.
The ALMA beams are shown as hatched gray ellipses near the bottom-right of each panel.
}
\label{fig:velo_maps}
\end{figure*}

As clearly seen in \autoref{fig:spec_cii}, the \cii\ lines we measure in our sample are often offset towards the lower frequency edge of the observed bands - redshifted with respect to the rest-frame UV emission lines of the quasars.
\autoref{tab:line_shifts} compares the redshifts determined from the \cii\ line to those determined from the \mgii\ line and also the enhanced SDSS-based redshift determinations published in \cite{Hewett2010_SDSS_z}. 
The \mgii\ line is believed to be among the best redshift indicators for unobscured AGN, with a scatter of about 200\,\kms\ compared to the systemic redshift \cite[e.g.,][and references therein]{Richards2002_CIV,Shen2016_SDSS_RM_shifts}.
We note that for our \zfpe\ quasars the SDSS-based redshift determinations rely solely on the (rest-frame) UV lines \Lya\ and \CIV, both of which may be problematic for the purpose of redshift determinations.
The \civ\ line is known to present significant blueshifts with respect to systemic redshifts \cite[e.g.,][and references therein]{Shang2007,Richards2011_CIV,Shen2016_SDSS_RM_shifts}, while the profile of the \Lya\ line, and particularly the blue wing, is affected by IGM absorption \cite[e.g.,][]{Becker2013_Lya_forest}.

All quasar hosts present significant shifts of the \cii\ lines, compared to the broad UV emission lines of the quasars.
The \Nbright\ FIR-bright hosts and one of the FIR-faint ones (J0935) show \cii\ emission redshifted by $\sim390-580\,\kms$.
Interestingly, the only sources presenting \cii\ \emph{blue}shifts, of $\sim220$ and $630\,\kms$, are the two FIR-faint sources, J0923 and J1328 respectively, which have interacting SMGs.
Such significant positive velocity shifts of ISM lines, such as \cii, with respect to the quasar's broad emission lines, such as \mgii -- which probe the close vicinity of the SMBHs, are not uncommon in high-redshift quasars.
For example, the recent study of \cite{Venemans2016_z6_cii} shows  \cii\ velocity shifts of $\sim370-1690\,\kms$ in a compilation of seven $z>6$ quasars.
\cite{Willott2015_CFHQS_ALMA} found a shift of $\sim1150\,\kms$ for one $z\simeq6$ quasar, but no significant shift for another.
Negative velocity shifts of several hundred \kms\ were also observed among other high-redshift quasars \cite[e.g.,][]{Wang2013_z6_ALMA,Willott2013_z6_ALMA}.

\autoref{fig:velo_maps} shows the \cii\ velocity maps of the \Ntot\ quasar hosts and the three SMGs accompanying J1511, J0923 and J1328.
The relatively smooth gradients suggest that the \cii\ emission originates from a kiloparsec scale, rotating gas structure, with a rotation axis that coincides with the centers of the galaxies and with the quasars themselves (as shown by the centroid markers). 
The only obvious outlier is the host of J0935, where the \cii\ emission is weaker.
For the FIR-bright quasar hosts, the velocity maps reach maximal velocity values of $\left|\vmax\right|\simeq200\,\kms$, while for the FIR-faint hosts and their accompanying SMGs the corresponding values are significantly lower, $\left|\vmax\right|\simeq100\,\kms$.
This may suggest that the FIR-bright quasar hosts are more massive, or gas rich, than the FIR-faint ones.

\begin{figure*}[ht!]
\includegraphics[width=0.34\textwidth]{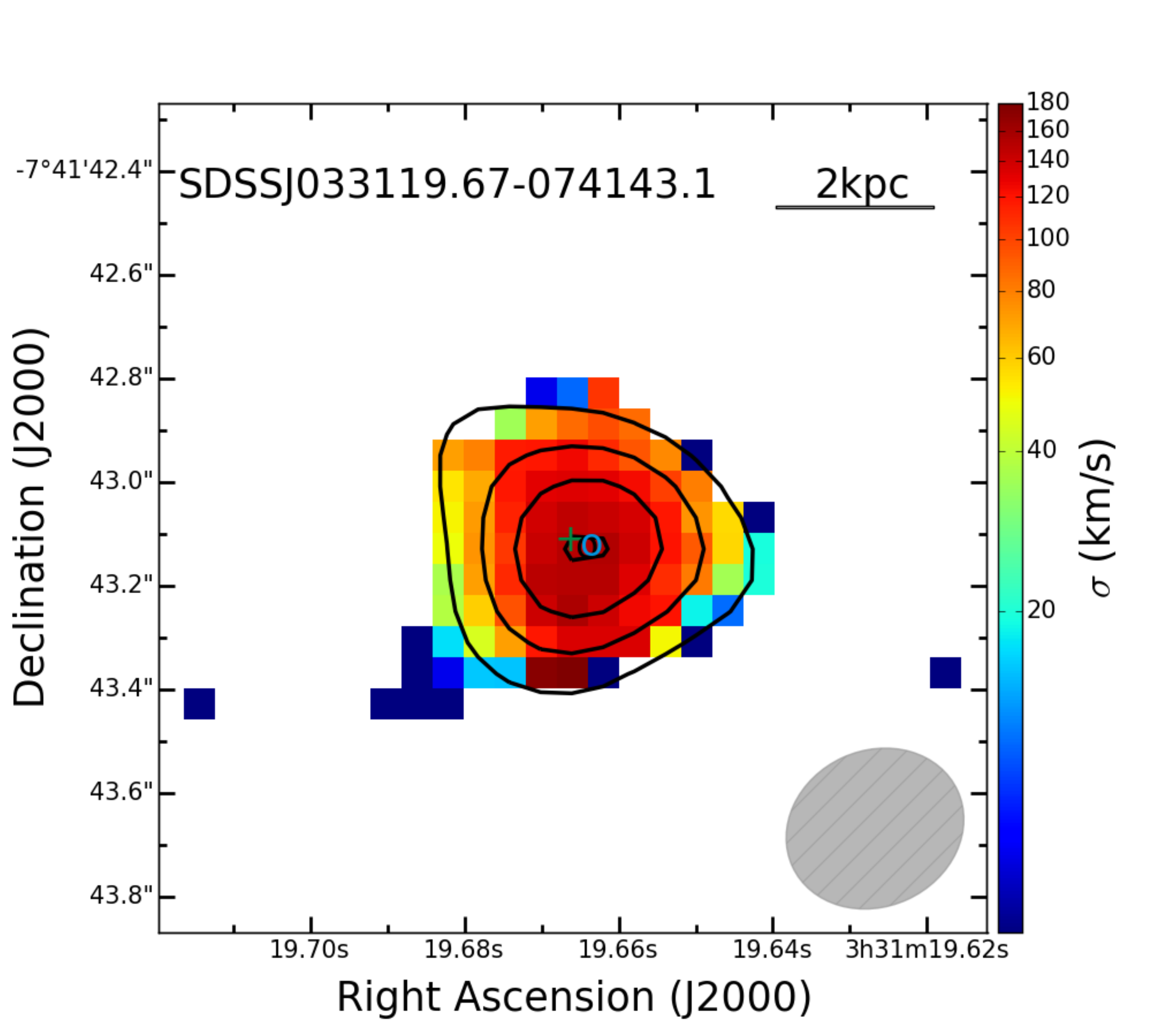} 
\includegraphics[width=0.34\textwidth]{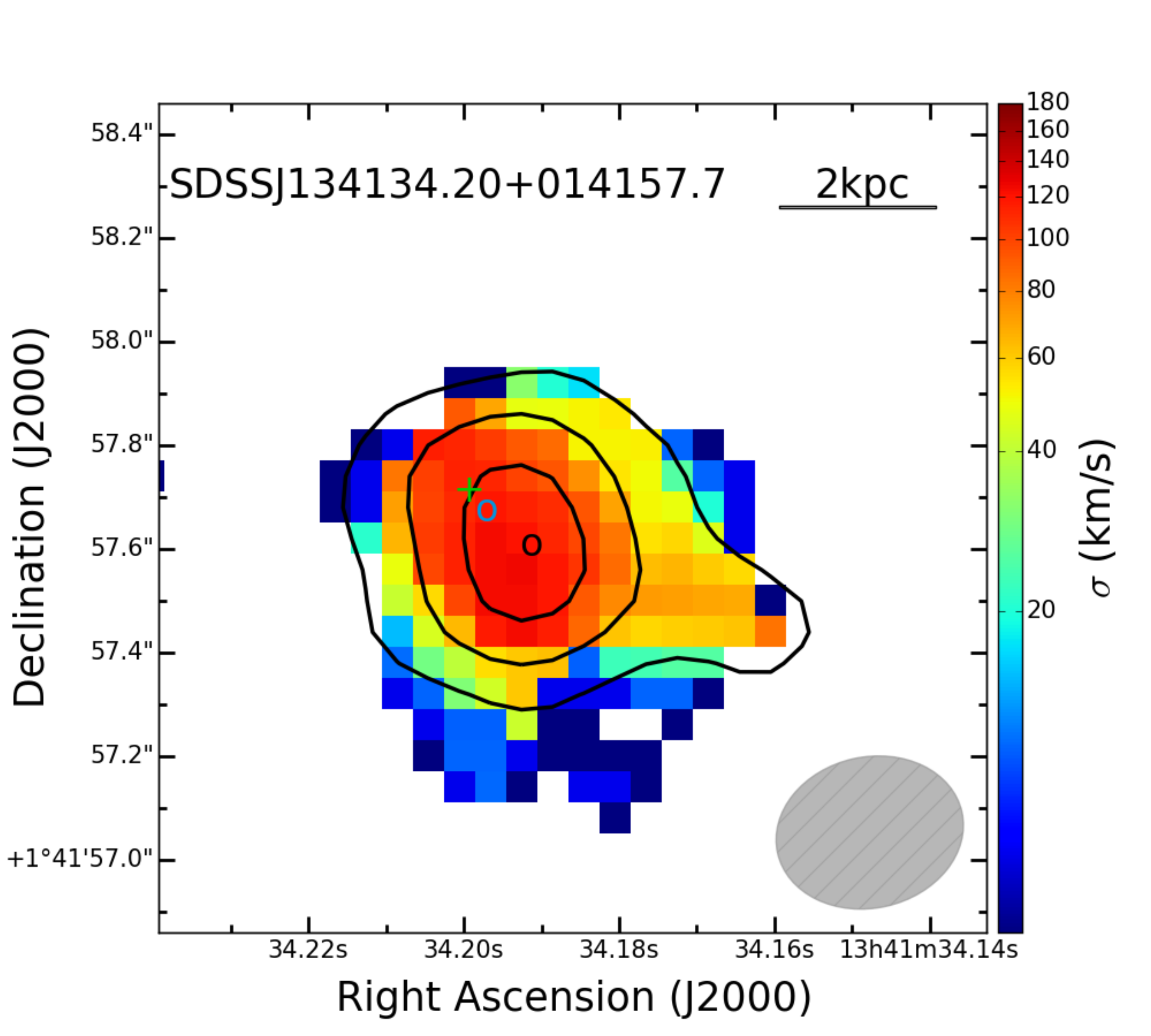}
\includegraphics[width=0.34\textwidth]{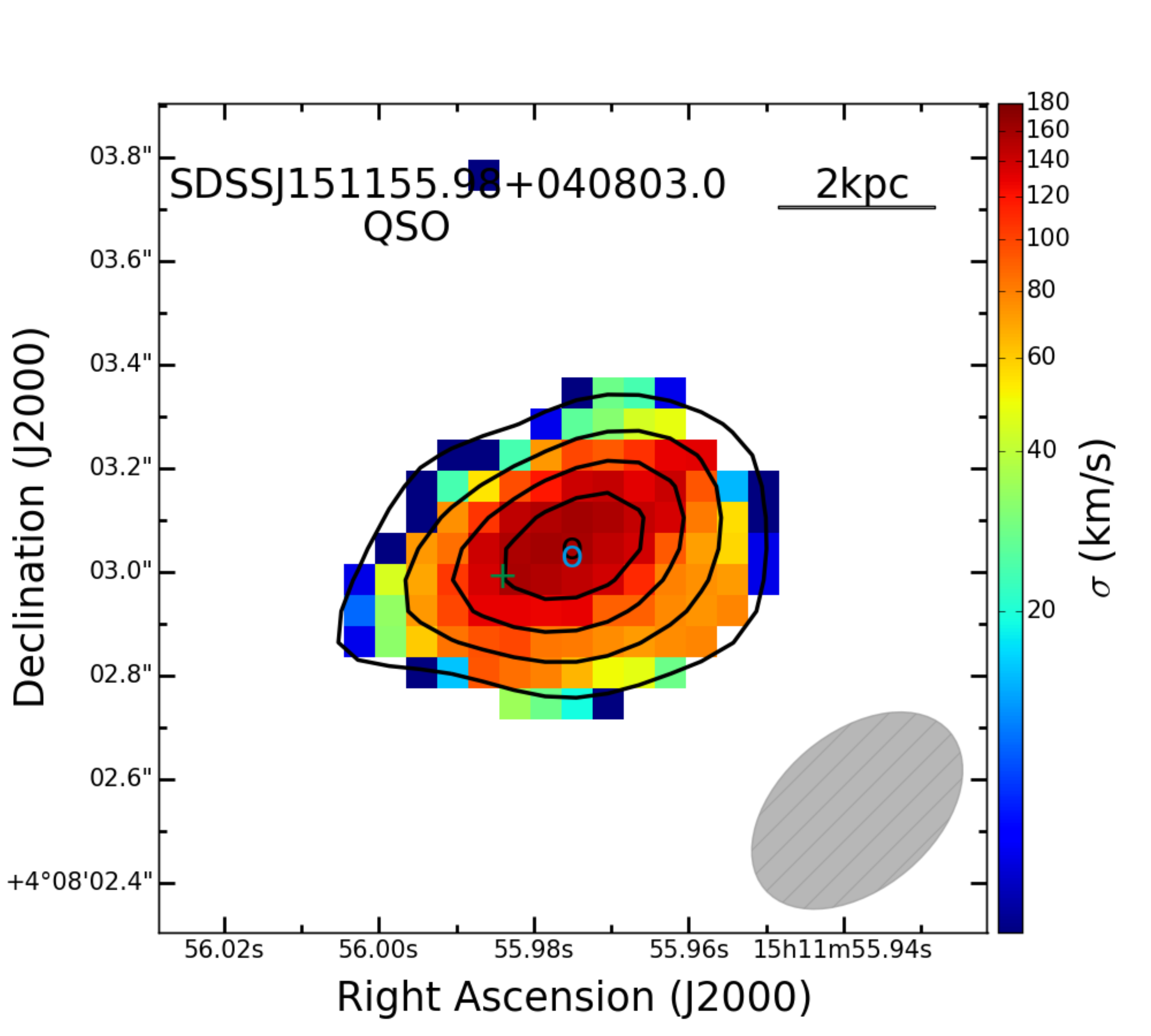} \\
%
%
\includegraphics[width=0.34\textwidth]{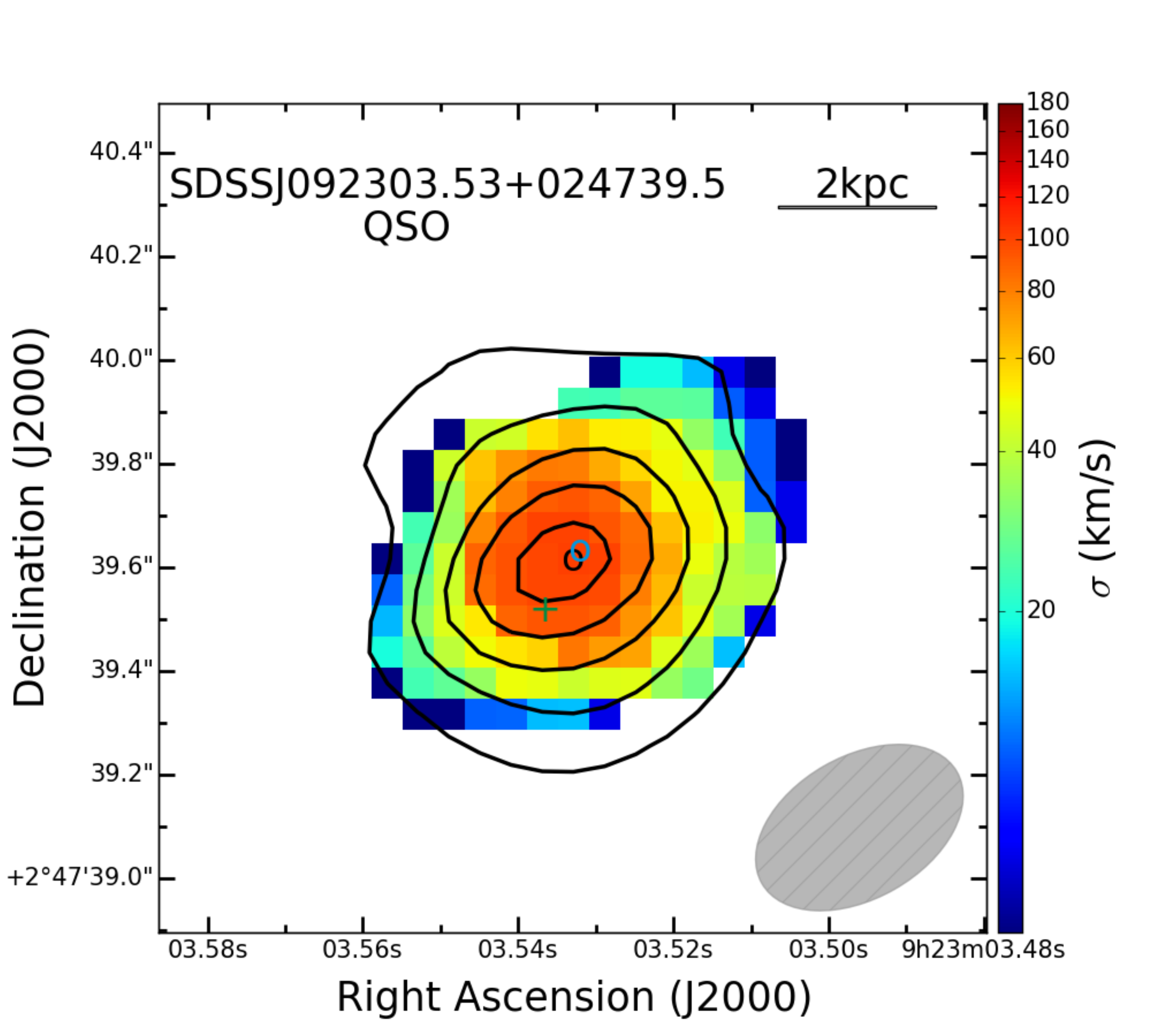}
\includegraphics[width=0.34\textwidth]{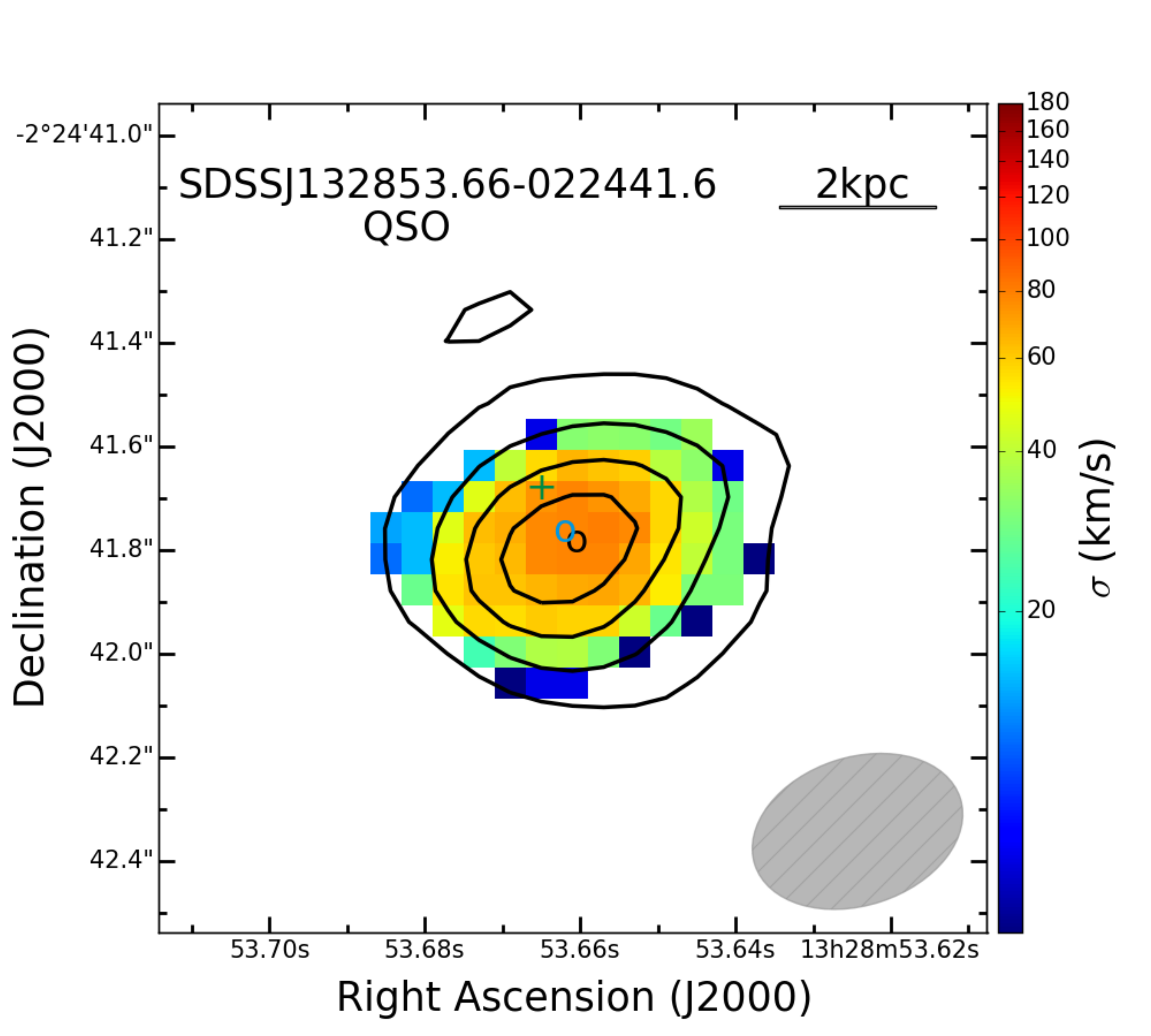}
\includegraphics[width=0.34\textwidth]{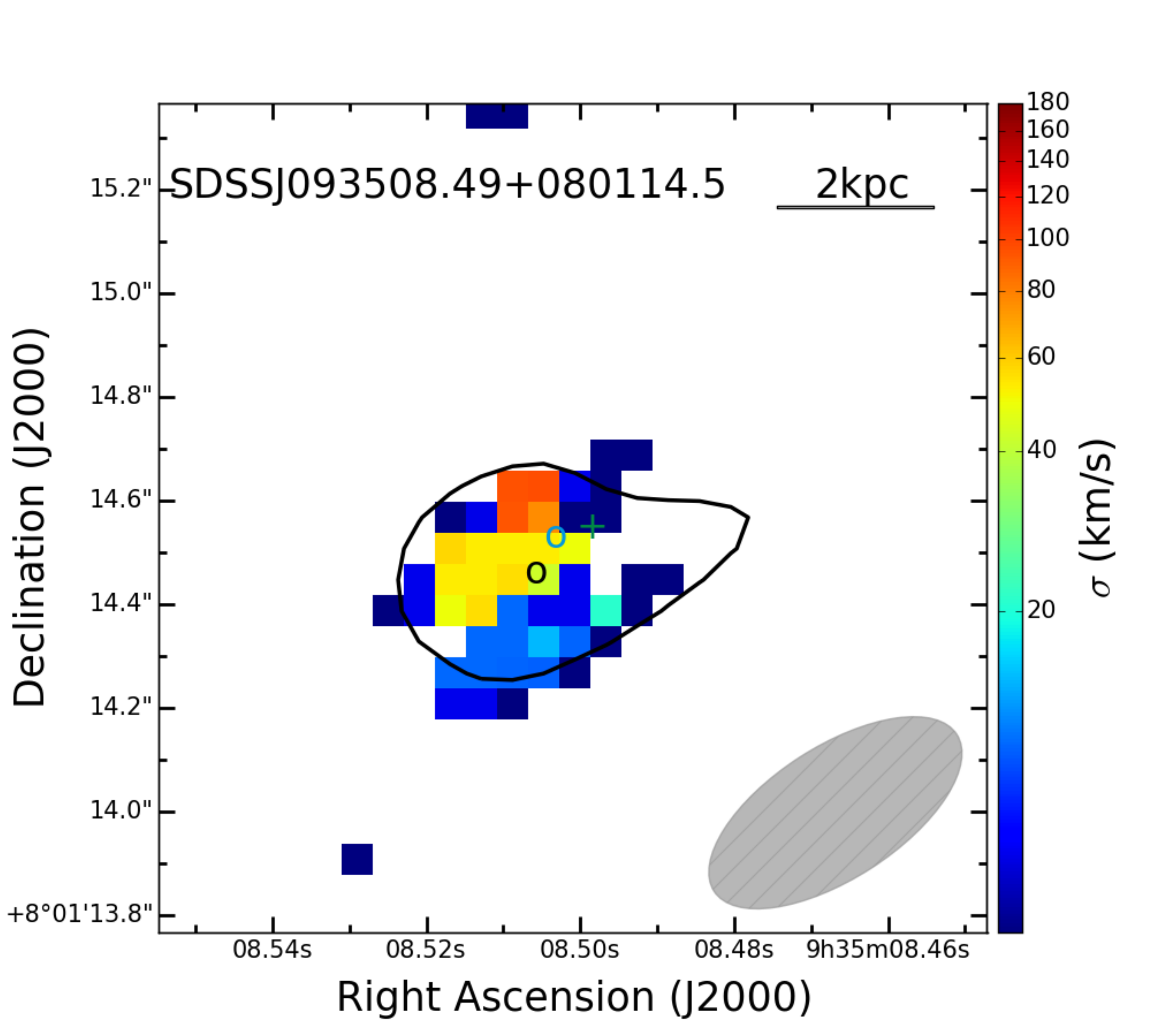}\\
\vspace*{-0.5cm}
\begin{center}
\hspace{1cm}\rule[1ex]{1\textwidth}{0.5pt}\\
\end{center}
\includegraphics[width=0.34\textwidth]{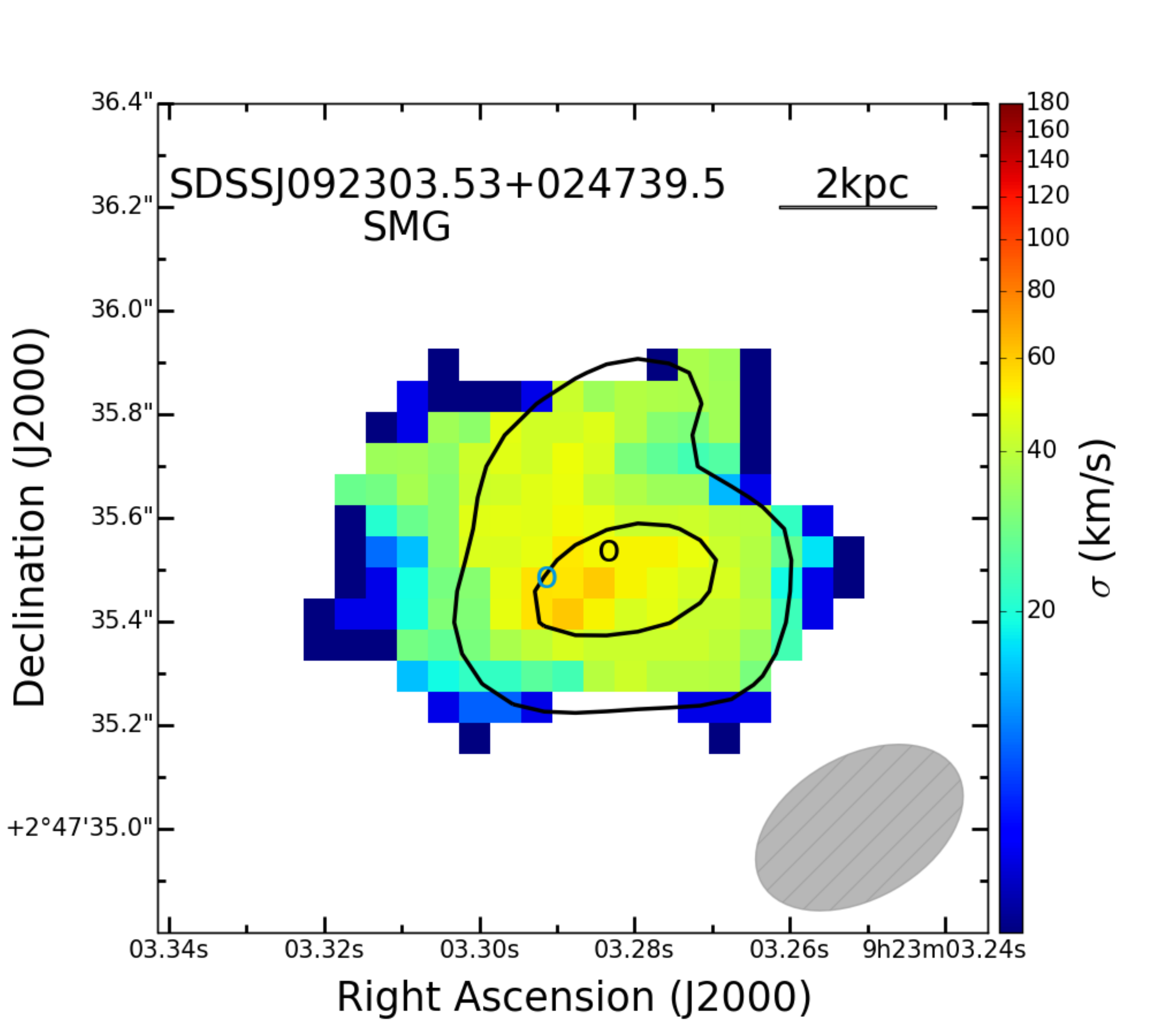}
\includegraphics[width=0.34\textwidth]{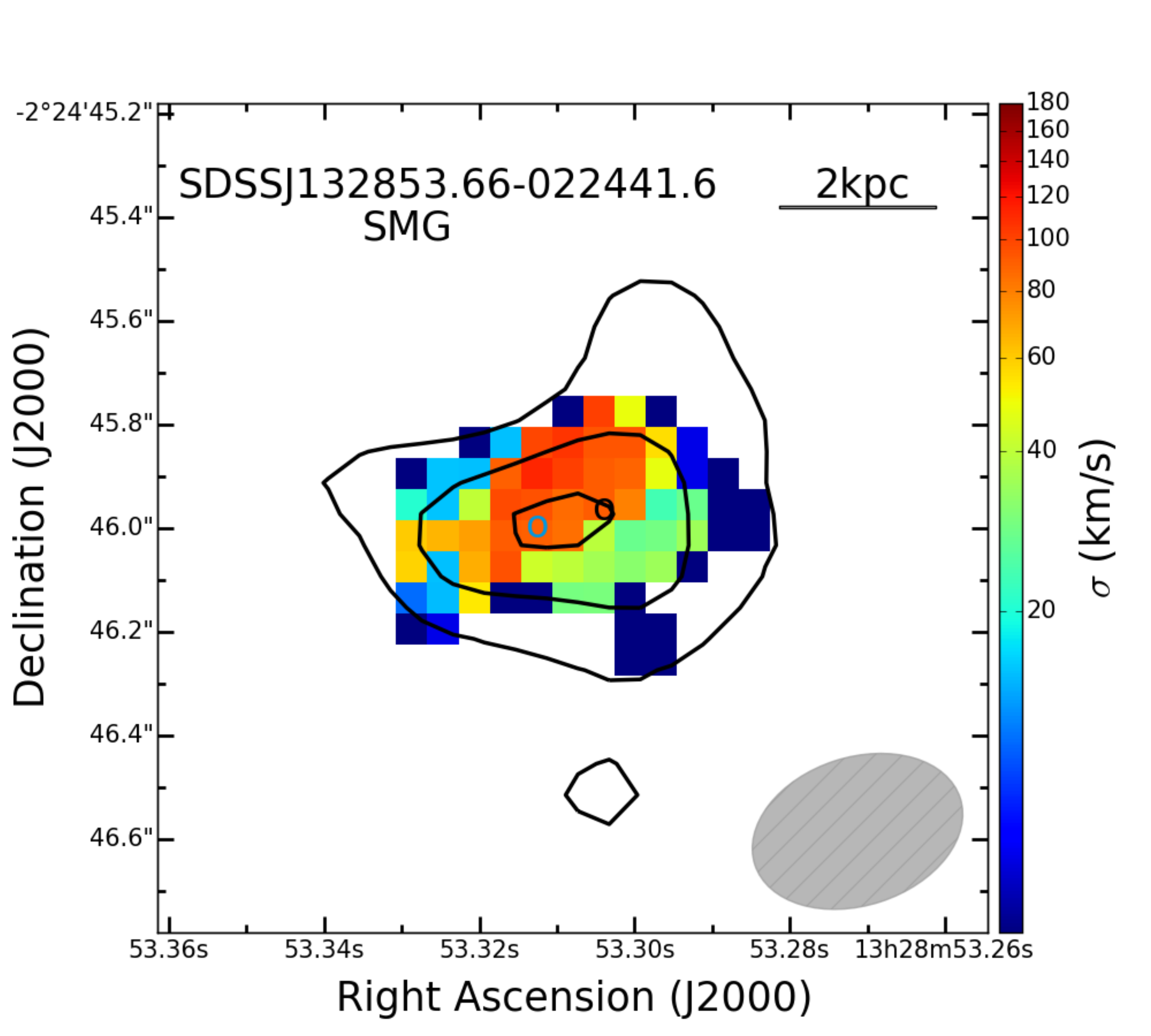}
\includegraphics[width=0.34\textwidth]{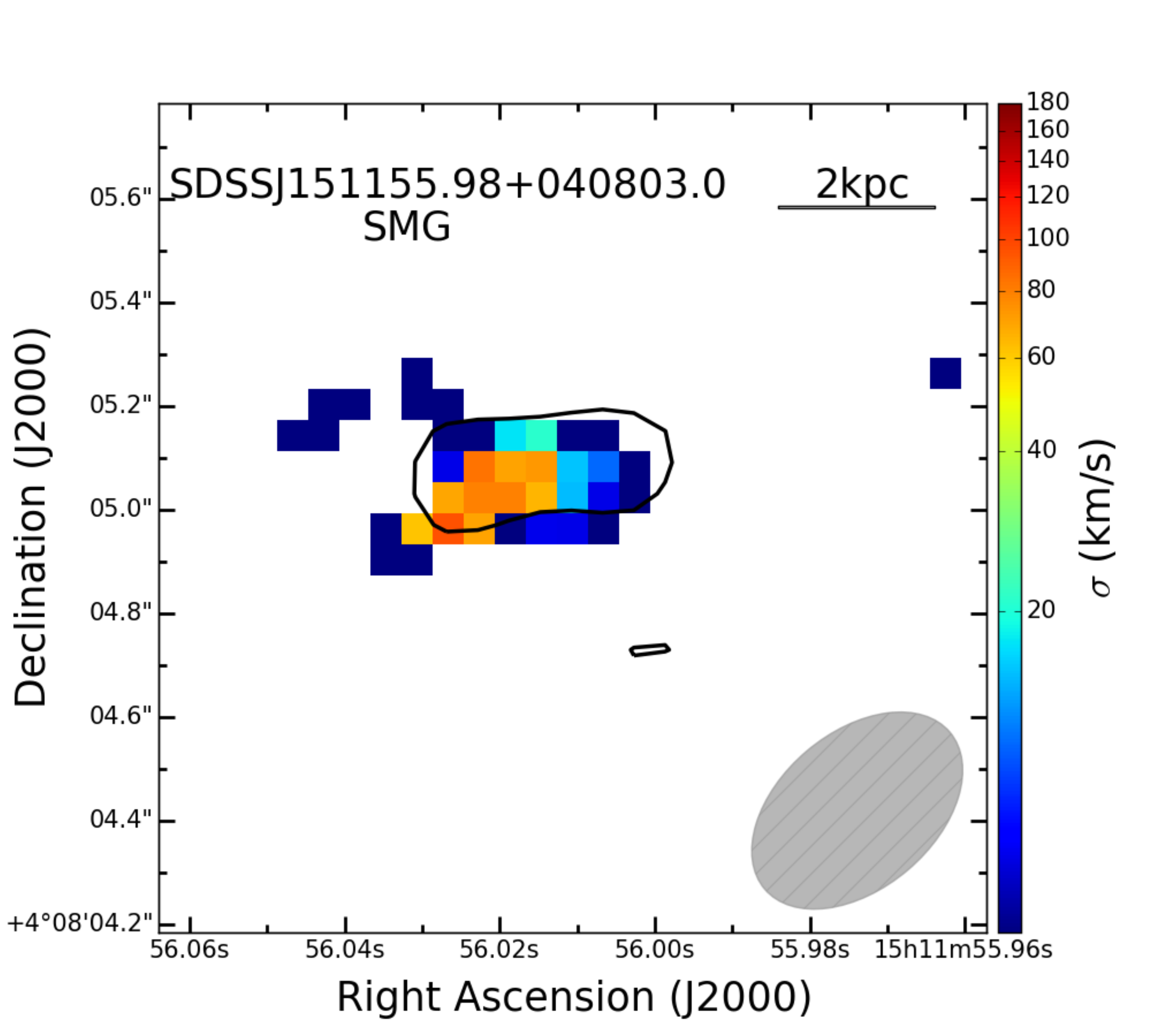} \\
\caption{
\cii\ velocity dispersion maps for the 
the \Nbright\ ``FIR-bright'' sources in our sample (top row); the \Nfaint\ ``FIR-faint'' sources (middle row); and the companion SMGs (bottom row).
Black contours trace the \cii\ emission line surface brightness at significance levels of 3, 6, 9, 12, and 15$-\sigma$.
Crosses mark the locations of the quasars' optical emission.
Gray circles mark the locations of the peak of the dust (ALMA) continuum emission.
Black circles mark the locations of the peak of the \cii\ emission.
The ALMA beams are shown as hatched gray ellipses near the bottom-right of each panel.
}
\label{fig:sigma_maps}
\end{figure*}

The velocity dispersion maps, presented in \autoref{fig:sigma_maps}, show increased velocity dispersions in the centers of all \cii-emitting systems, reaching $\sigma_{\rm v}\sim100\,\kms$.
Given the beam sizes of our ALMA data, this trend is probably mostly driven by the effects of beam smearing, as demonstrated in several detailed studies of the kinematics of high-redshift, sub-mm sources \cite[e.g.,][]{DeBreuck2014_z48_SMG}. 
A clear signature of this effect is a centrally peaked velocity dispersion, elongated along the minor axis of rotation - which is similar to what is seen in some of our sources (e.g., J1341, J1511 and J0923).
Thus, the real underlying central velocity dispersions may be significantly lower than what is seen in \autoref{fig:sigma_maps}, resulting in generally more uniform or flatter velocity dispersion profiles, and thus implying rotation-dominated kinematics with $v/\sigma_{\rm v}\gtrsim1$.
We note that even in such a situation, of $v/\sigma_{\rm v}$ on the order of a few, the kinematics may be significantly affected by a turbulent component, as demonstrated in several recent studies of resolved ISM kinematics in high-redshift galaxies \cite[e.g.,][]{Gnerucci2011_AMAZE_LSD,Williams2014_SINFONI}.
With these limitations in mind, we cautiously conclude that the \emph{outer} parts of the \cii-emitting regions in all sources are dominated by rotation, with $v/\sigma_{\rm v}>1$.
In particular, this is the case in the outer high-velocity regions of the FIR-bright systems, where one could have expected to see evidence for dispersion-dominated gas kinematics in the case these systems were driven by major mergers.

\capstartfalse
\begin{deluxetable}{llccccc}
\tablecolumns{7}
\tablewidth{0pt}
\tablecaption{Redshifts and \cii\ Line Shifts \label{tab:line_shifts}}
\tablehead{
\colhead{sub-sample} &
\colhead{Target} &
\colhead{$z_{\cii}$} &
\colhead{$z_{\rm SDSS}$ \tablenotemark{a}} &
\colhead{$\Delta v _{\rm SDSS}$} &
\colhead{$z_{\mgii}$ \tablenotemark{b}} &
\colhead{$\Delta v _{\mgii}$} \\
 & & & & \kms & & \kms
}
\startdata
Bright & J0331 	&  $ 4.73678 $ &  $ 4.73186 $ &  $ +257 $ &  $ 4.72890 $ &  $ +412 $ \\
~~~~~~ & J1341 	&  $ 4.70030 $ &  $ 4.68172 $ &  $ +981 $ &  $ 4.68944 $ &  $ +573 $ \\
~~~~~~ & J1511 	&  $ 4.67850 $ &  $ 4.67683 $ &  $  +88 $ &  $ 4.66988 $ &  $ +456 $ \\
\hline \\ [-1.75ex]
Faint  & J0923 	&  $ 4.65485 $ &  $ 4.65001 $ &  $ +257 $ &  $ 4.65887 $ &  $ -213 $ \\
~~~~~  & J1328 	&  $ 4.64644 $ &  $ 4.64998 $ &  $ -188 $ &  $ 4.65815 $ &  $ -621 $ \\
~~~~~  & J0935 	&  $ 4.68189 $ &  $ 4.69920 $ &  $ -911 $ &  $ 4.67078 $ &  $ +588 $
\enddata
\tablenotetext{a}{SDSS-based redshifts taken from \cite{Hewett2010_SDSS_z}.}
\tablenotetext{b}{\MgII-based redshifts taken from T11.}
\end{deluxetable}
\capstarttrue

\begin{figure}[ht!]
\centering
\includegraphics[trim={0cm 0cm 7cm 0cm},clip,width=.49\textwidth]{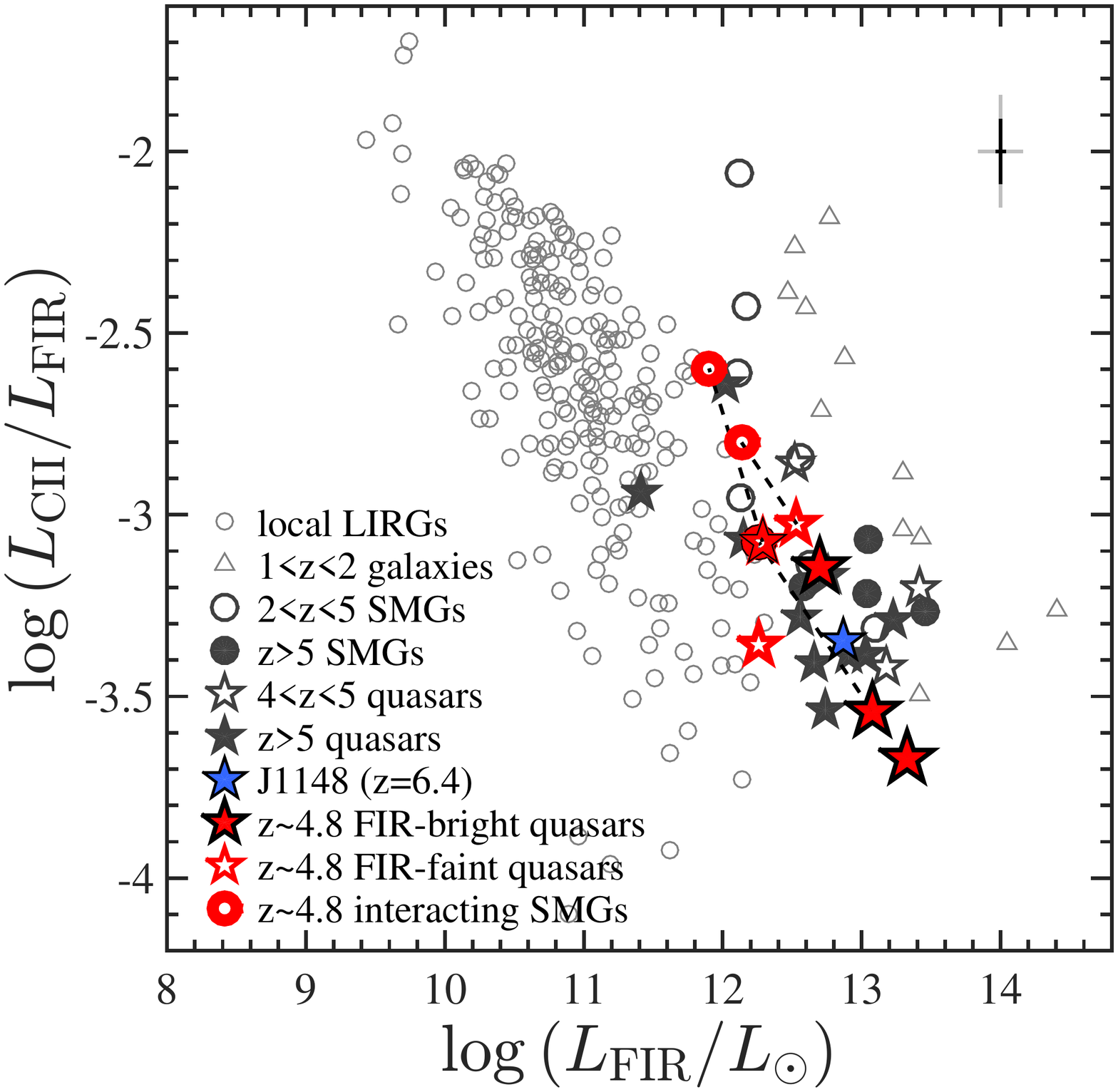} 
\caption{
The \cii\ line luminosity ratio, $L_{\cii} / \LFIR$, vs. \LFIR\ for our sample of \zfpe\ AGN and their SMG companions (red symbols) and several reference samples.
The quasar hosts under study are marked as red stars (further split to FIR-bright and -faint objects), and their interacting companion SMGs are marked as red circles, connected to the respective quasar hosts by dashed lines.
Note that the host galaxy of J1328 and the SMG accompanying J1511 have very similar \Lcii\ and \LFIR, therefore their data points overlap in the parameter space we plot.
Symbols for reference samples follow those in \citet[see their Fig.\ 13]{Cicone2015_J1148}, including local LIRGs \cite[open circles;][]{DiazSantos2013_CII_GOALS}; 
star-forming and/or active galaxies at $1<z<2$ \cite[triangles;][]{Stacey2010}; 
and a compilation of higher-redshift SMGs and quasars, taken from a variety of studies (large circles and stars, respectively; see details in the text).
The crosses at the top-right corner illustrate representative measurement (black) and systematic (gray) uncertainties related to \Lfir\ and \Lcii. 
We plot \Lfir\ systematics of 0.15 dex, reflecting the range covered by the different FIR SEDs we use (i.e., gray-body and template-based; see \autoref{subsubsec:sed_sfrs} for details).
We note that the ratios we measure for the FIR-bright quasar hosts are among the lowest measured to date, particularly at high redshifts.
The interacting quasar hosts have lower \LLfir\ than what is found for the companion SMGs of the same systems.
}
\label{fig:Lcii_Lfir}
\end{figure}

We next turn to the relative \cii\ line strength, traced by the ratio between \cii\ line luminosity and the continuum (rest-frame) FIR luminosity, \LLfir.
\autoref{fig:Lcii_Lfir} shows \LLfir\ vs. \Lfir\ for the \Ntot\ quasar hosts and the \Nsmg\ interacting SMGs in our sample, as well as a large compilation of other galaxies where the \cii\ line was detected. 
The compilation, adapted from \cite{Cicone2015_J1148}, includes inactive star forming galaxies at low and intermediate redshifts \cite[from][]{Stacey2010,DiazSantos2013_CII_GOALS} and SMGs and quasar hosts at $z>4$ \cite[from][]{Maiolino2009_CII_hiz,Ivison2010_z23_SMG,Wagg2010_CII_z44_QSO,Cox2011_z42_SMG,Valtchanov2011_hiz_SMGs,Swinbank2012_z44_SMG,Venemans2012_z71_CII,Venemans2016_z6_cii,Carniani2013_BR1202_CII,Riechers2013_z63_SMG_Nature,Riechers2014_CII_z43_protocluster,Wang2013_z6_ALMA,Willott2013_z6_ALMA,DeBreuck2014_z48_SMG,Neri2014_CII_HDF850.1}. 
For the purpose of this comparative analysis, we calculated the FIR luminosities assuming a gray-body SED with a dust temperature of $T_{\rm d}=47\,{\rm K}$ and a power-law exponent of $\beta=1.6$, scaled to match the continuum emission of each of the ALMA-detected sources. 
These scaled SEDs are then used to calculate the integrated luminosity between $42.5-122.5\,\mic$, $\Lfir\left(42.5-122.5\right)$.
As mentioned in \cite{Cicone2015_J1148}, the measurements of the sources in the compilation were also scaled, to provide consistent estimates of $\Lfir\left(42.5-122.5\right)$.
We present a more detailed analysis of the FIR SEDs of our sources in \autoref{subsubsec:sed_sfrs} below.

For the quasar hosts, the \LLfir\ ratio is in the range of $\left(2.1-9.4\right)\times10^{-4}$, spanning a factor of roughly 4.5.
As \autoref{fig:Lcii_Lfir} shows, the FIR-bright and FIR-faint quasar hosts form a trend of decreasing \LLfir\ with increasing \Lfir, although the range of \LLfir\ for the two sub-samples overlaps. 
Since all quasar hosts have comparable \Lcii, this trend of decreasing \LLfir\ is mostly driven by the increase in \Lfir.
The \Nsmg\ interacting SMGs follow the same trend, extending to lower \Lfir\ and higher \LLfir.
We further verified that the trend of decreasing line-to-continuum ratio with increasing FIR luminosity is also reflected in the \cii\ equivalent widths \cite[EWs; instead of \LLfir; see, e.g.,][]{Sargsyan2014_CII_SFR}.
The EWs of our quasar hosts are in the range ${\rm EW}_{\cii}\simeq 0.2-0.9$ \mic\ ($\sim 370-1700\,\kms$) and follow the same trend with (continuum) FIR luminosity as that found for \LLfir\ -- dropping by about 0.65 dex in EW for a 1 dex increase in (monochromatic) FIR luminosity.
The companion SMGs extend this trend to ${\rm EW}_{\cii}\simeq 2.4$ \mic..
%

\begin{figure*}[ht!]
\includegraphics[trim={0 0.0cm 0 0.95cm},clip,width=\textwidth]{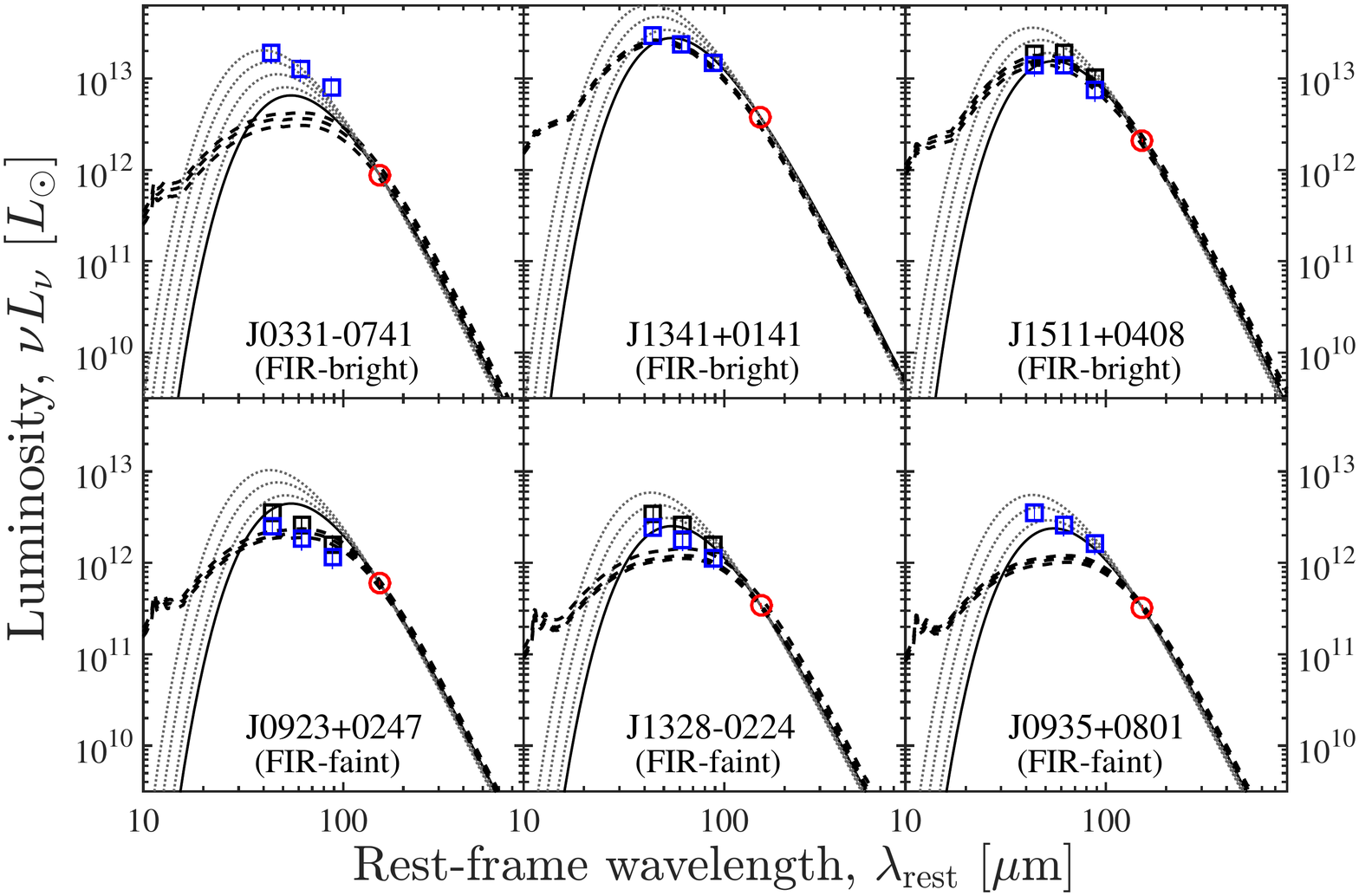} 
\caption{
FIR SEDs for the \Ntot\ quasar host systems in our sample, with the FIR-bright and -faint sources in the top and bottom rows, respectively.
The new ALMA continuum measurements are marked in red.
Blue symbols show the \herschel/SPIRE measurements, scaled to represent the relative contribution of the quasar host, when necessary (i.e., in systems with multiple resolved components). 
In such cases, the black symbols represent the original, un-scaled \herschel\ measurements.
For all \Nfaint\ FIR-faint systems (bottom row), we adopted the measurements of the stacked \herschel\ signal (N14).
For each source, the different lines represent FIR SEDs, normalized by the single ALMA continuum measurements.
Solid thin lines represent a gray-body SED with $T_{\rm d}=47\,{\rm K}$ and $\beta=1.6$.
Thin dotted lines trace alternative SEDs with $T_{\rm d}=50$, 55, and 60 K (and for J0331 also 65 K; all with $\beta=1.6$).
Thick dashed lines illustrate the best-fitting FIR templates from \cite{CharyElbaz2001_FIR}, as well as the two adjacent templates (typically separated by $\sim$0.1 dex in \LTIR).
}
\label{fig:fir_seds}
\end{figure*}

Several recent studies have demonstrated the wide range of possible \LLfir\ in  $z\gtrsim5$ (UV-selected) SF galaxies, covering $\LLfir\sim0.0002-0.02$ - similar to the range observed at lower redshifts \cite[e.g.,][]{Capak2015_CII_COSMOS,Knudsen2016_CII_z6}.
Thus, the so-called ``\cii\ deficit'' observed in high-redshift SMGs and quasar hosts is most probably \emph{not} related to simple observational selection effects in the FIR or sub-mm regime, but rather to the morphology of the SF activity. 
Specifically, high-redshift quasar hosts exhibit more compact starburst-like SF activity, with \LLfir\ ratios as low as observed in lower-redshift ULIRGs.

Interestingly, our measurements show that the \LLfir\ ratio in the three interacting quasar hosts is significantly lower than that found in their companion SMGs.
Specifically, we find that the \LLfir\ ratio in the quasar hosts of J1511, J0923, and J1328 is lower by factors of about 1.7, 3 and 2.9, respectively, compared with the companion SMGs.
As can be clearly seen in \autoref{fig:Lcii_Lfir}, this is consistent with the general trend of decreasing \LLfir\ with increasing \Lfir, observed for our entire sample of quasar hosts and SMGs.
Moreover, this demonstrates that the \cii-deficit in high-redshift quasar hosts is driven by local properties of the ISM and the UV radiation field \emph{within} the host galaxies, and not by larger scale effects.
Indeed, several studies have emphasized that lower \LLfir\ ratios are expected to be found in regions with higher SF densities, similar to starbursts -- consistent with what is observed (e.g., \citealt{DiazSantos2013_CII_GOALS}; see discussion in \citealt{Cicone2015_J1148}).
However, the ALMA data for some of our sources do \emph{not} seem to support this explanation. 
In particular, the hosts of J1328 and J0935 have virtually identical \Lfir\ and \cii-emitting region sizes, but \Lcii\ (and therefore \LLfir) in J1328 is twice as high as in J0935.
Similarly, the hosts of J0331 and J0923 have very similar \Lfir, and \LLfir\ in the latter is higher than in the former by merely 30\%. 
However, the \cii-emitting region in J0923 is larger by a factor of roughly four than the one in J0331.
Thus, our ALMA data do not seem to support a simple scenario where \LLfir\ is mainly controlled by the size of the \cii\ emitting region, however, higher resolution data are needed to critically address this idea.

\subsection{Host galaxy properties}
\label{subsec:hosts}

\subsubsection{FIR SEDs and SFRs}
\label{subsubsec:sed_sfrs}

The (rest-frame) FIR continuum emission observed within the new ALMA data can be used to estimate the total FIR emission and therefore the SFR of the quasar hosts and the accompanying SMGs.
This obviously requires additional assumptions regarding the shape of the FIR SED.
We expect that the AGN-related contribution to the FIR SED is small, on the level of 10\%, at most.
This assumption is based on several detailed studies of the mid-to-far IR SEDs of luminous AGN across a wide luminosity range \cite[e.g.,][]{Schweitzer2006_QUEST,Netzer2007_QUEST,Mor2012_IR,Rosario2012,Netzer2014_z48_SFR,Netzer2016_herschel_hiz,Lutz2016_FIR_size}, although we note that other studies have suggested a higher AGN contribution at FIR wavelengths, particularly at high AGN luminosities \cite[e.g.,][]{Leipski2014,Schneider2015_FIR_z6,Symeonidis2016_FIR_SED_PAHs,Symeonidis2017_FIR_SED_QSOs}.
Given the uncertainties related to the SED shape (see below) and in order to be consistent with other ALMA studies of high-redshift quasar hosts, we choose to neglect the (small) possible AGN contribution to the FIR SEDs of our sources.

We have reconstructed the FIR SEDs of our sources following two different approaches. 
First, we adopted the procedure used in several recent studies of FIR emission in high-redshift quasar hosts \cite[e.g.,][and references therein]{Willott2015_CFHQS_ALMA,Venemans2016_z6_cii} and have assumed a gray-body SED with dust temperature $T_{\rm d}=47\,{\rm K}$ and $\beta=1.6$.
This single temperature dust model is a crude approximation to the more realistic case where dust with a range of temperatures contribute to the observed emission \cite[for a detailed discussion see, e.g.,][and references therein]{Beelen2006_hiz_qsos,Magnelli2012_SMGs_Herschel}.
Second, we have used the grid of FIR SEDs provided by \citet[CE01 hereafter]{CharyElbaz2001_FIR}. 
This grid includes 105 templates spanning a wide range in IR luminosity. 
As the templates have no free parameters\footnote{That is, each template SED is unique in shape and scaling.}, we have simply identified the template that best matches the ALMA continuum measurement (i.e., monochromatic luminosity) for each source. 
This also provides a specific value of total IR (TIR) luminosity, $L_{\rm TIR}\left(8-1000\,\mic\right)$.
In the FIR luminosity regime of interest, the uncertainties on the ALMA continuum measurements are typically consistent with, or smaller than, the differences between adjacent FIR templates.
The main limitation of this approach is that the template library was constructed to account for low-redshift SF galaxies. 
However, the small number of available data points does not warrant the use of other sets of templates.

In \autoref{fig:fir_seds} we present the ALMA continuum measurements and the \herschel\ data from N14, along with the two previously mentioned types of FIR SEDs.
The low spatial resolution of the \herschel\ data means that these flux measurements also include the emission from any accompanying continuum sources for the systems where these are resolved by ALMA.
We therefore also show in \autoref{fig:fir_seds} a scaled-down version of the \herschel\ data points, assuming the relative flux densities of the different neighboring sources follow those of the ALMA measurements.
As \autoref{fig:fir_seds} shows, the gray-body SEDs are generally in good agreement with the previous \herschel\ data.
In particular, in the observed-frame 350 \mic\ band (rest-frame wavelength of roughly 60 \mic), the ALMA-based SEDs for two of the FIR-bright systems (J1341 and J1511) differ from the \herschel\ data by less than 0.05 dex (for either the gray-body or CE01 SEDs). 
For the third FIR-bright object, J0331, the luminosities expected from the ALMA-based SEDs are significantly lower, by factors of about 2 and 3, than the fluxes observed with \herschel\ for the gray-body and CE01 SEDs, respectively.\footnote{Similarly, the CE01 template that fits the \herschel\ data alone (as reported in N14) over-predicts the ALMA flux by a factor of 2.4.}
Since this source has no detectable companions in the ALMA data, we suggest that this discrepancy can only be attributed to the uncertainty on the exact shape of the SED, and perhaps to the limited quality of the \herschel\ data. 
Below we investigate how a gray-body SED with a different (higher) temperature may account for this discrepancy.

For the FIR-faint sources, the comparison between the ALMA-based SEDs and the \herschel\ data is obviously less straightforward, as the \herschel\ data represent the stacked signal coming from a much larger sample of sources for which the existence of companions that may contribute to the FIR fluxes is unclear.
Nonetheless, the agreement between the new ALMA data and the \herschel\ stacking measurements is respectable.
For two of the sources (J1328 and J0935), the ALMA-based gray-body SEDs are consistent within 0.13 dex of the \herschel\ data (again at observed-frame 350 \mic).
For the third source (J0923), the gray-body SED over-predicts the luminosity at 350 \mic\ by 0.37 dex, while the CE01 SED agrees with the \herschel\ data to within less than 0.1 dex.
As a sanity check, if one ignores the scaling factors related to the companions (in J0923 and J1328), that is - assume that the stacked \herschel\ signal was not significantly affected by companions, then the differences between the ALMA-based gray-body SEDs and the \herschel\ data become somewhat smaller.

We have also experimented with both colder and warmer gray-body SEDs, always ``anchored'' to the new ALMA data.
The colder SEDs, with $T_{\rm d}=40\,{\rm K}$ and $\beta=1.5$, identical to those used in N14, systematically \emph{under}-predict the \herschel\ data, by 0.1-0.3 dex (at 350 \mic; the discrepancies obviously decrease at observed-frame 500 \mic).
Among the warmer SEDs, shown as dotted lines in \autoref{fig:fir_seds}, we note that an extremely warm gray-body with $T_{\rm d}=65\,{\rm K}$ provides a better agreement between the ALMA and the \herschel\ data for J0331. 
However, this high temperature is beyond what is typically observed, even among the most luminous FIR sources \cite[e.g.,][and references therein]{Lee2013_FIR_SEDs,Magnelli2014_PEP_HerMES}. 
This may suggest that, in this source, the radiation emerging from the vicinity of the SMBH does contribute to heating of galaxy-scale dust, which is not associated with SF regions in the host \cite[e.g.,][]{Schneider2015_FIR_z6}.
We also note that choosing a higher-luminosity template from the CE01 grid would not resolve the discrepancy between the ALMA and \herschel\ data for this source, as the highest-luminosity CE01 template that closely matches the \herschel\ data would significantly over-predict the ALMA continuum measurement. 
We conclude that both the CE01 and fiducial gray-body FIR SEDs (i.e., with $T_{\rm d}=47\,{\rm K}$ and $\beta=1.6$) provide generally good agreement between the ALMA and \herschel\ data. We therefore chose to use these sets of SEDs in what follows.

We next use the two types of FIR SEDs to calculate the total IR luminosities between $8-1000\,\mic$, $L_{\rm TIR}\left(8-1000\,\mic\right)$ for all sources -- both quasar hosts and accompanying SMGs.\footnote{We also use these SEDs to calculate $\Lfir\left(42.5-122.5\right)$, as mentioned in \autoref{subsec:lines}.}
From these, we estimate SFRs following ${\rm SFR}/\mpyr=L_{\rm TIR}\left(8-1000\,\mic\right)/10^{10}\,\Lsol$ (following the assumed Chabrier IMF).
Importantly, we note that the agreement between the integrated TIR luminosities obtained with the two types of SEDs is remarkably good.
For the six quasar hosts, the \LTIR\ estimates based on the CE01 template SEDs are lower than those based on gray-body SEDs by merely 0.1 dex (median value).
For the \Nsmg\ interacting SMGs, the difference is only slightly larger, $\sim$0.13 dex.
The SFRs we obtain for the quasar hosts following this procedure span a wide range of $\sim260-3040\,\mpyr$ for the gray-body SEDs or $\sim190-3530\,\mpyr$ for the CE01 SEDs (see \autoref{tab:gal_props}).
The SFRs of the FIR-faint sources, $\sim260-490\,\mpyr$ (or $190-360\,\mpyr$ using CE01), are in excellent agreement with the value found from stacking analysis performed in N14, using \emph{all} the \herschel-undetected sources in the parent \zfpe\ sample, of roughly $440\,\mpyr$ (see also \citealt{Netzer2016_herschel_hiz} for a slightly lower value).
As noted above, the low spatial resolution \herschel\ stack included the FIR emission from both the quasar hosts and the accompanying SMGs. 
The \emph{total} SFRs of the different sub-components in the FIR-faint systems ($\sim680$, $390$, and $260\,\mpyr$) are, again, consistent with the stacking result.
For the FIR-bright sources, the SFRs we derive based on the ALMA data are high, ranging from $\sim710-3040\,\mpyr$ for the fiducial gray-body FIR SEDs or $\sim630-3530\,\mpyr$ for the CE01 ones. 
We note that the lowest SFRs among this sub-sample are those of J0331, where an extremely warm gray-body SED is required to match the \herschel\ data.
Using the $T_{\rm d}=65\,{\rm K}$ gray-body SED we obtain $\log\left(\LTIR/\Lsol\right)=13.35$ and $\sfr=2225\,\mpyr$.
These SFRs are in excellent agreement with those derived in N14, when comparing similar SEDs.
The differences between the CE01-based IR luminosities (and therefore SFRs) among the FIR-bright quasar hosts are of 0.1 dex, at most.
If we instead consider the gray-body SEDs used in N14, which assumed $T_{\rm d}=40\,{\rm K}$ and $\beta=1.5$, and employ these SED parameters to our ALMA data, then the resulting SFRs are, again, in excellent agreement with the N14 ones for two of the systems (J1341 and J1511).
As mentioned above, such a cold SED is inconsistent with the data available for J0331, and the ALMA-based \Lfir\ for this source is lower than the one obtained in N14 by 0.4 dex.

We conclude that the new ALMA continuum measurements are broadly consistent with the \herschel-based ones (presented in N14) and indicate that all quasar hosts and accompanying SMGs harbor significant SF activity. 
The exact values of SFR obviously depend on the assumed shape of the FIR SED, but are in good agreement with the ones derived from the \herschel\ data.
The only outlier is the FIR-bright system J0331, where the ALMA-based FIR SED and SFR estimates are found to be considerably lower than the \herschel-based ones.
Despite these systematic uncertainties, the new ALMA data strongly support the picture that the FIR-bright sources among the T11 sample of \zfpe\ quasars have extreme SFRs, exceeding $\sim1000\,\mpyr$, while the FIR-faint sources have lower, though still intense SFRs, on the order of $\sim200-400\,\mpyr$.

Most star-forming galaxies are found to populate the so-called ``main sequence of star formation'' on the SFR-\mstar\ plane (SF-MS hereafter).
Although this relation is not yet well established at $z>4$, the most recent results from deep surveys suggest a relation of roughly 
\begin{equation}
 \sfr \simeq 300 \left(\frac{\mstar}{10^{11}\,\Msol}\right)^{0.8}\,\,\mpyr\, ,
\label{eq:sf_ms}
\end{equation}
with an intrinsic scatter of $\sim0.2-0.3$ dex \cite[e.g.,][]{Lee2012_MS,Speagle2014,Steinhardt2014_SPLASH_MS}.
The SFRs we find for the FIR-faint quasar hosts, and for the companion SMGs, are therefore consistent with those of typical massive, high-redshift SF galaxies, with $\mstar\simeq10^{11}\,\Msol$.
Several recent studies have highlighted the fact that such SFRs can be sustained without invoking major mergers and instead be driven by the accretion of cold gas onto these galaxies \cite[e.g.,][]{Bouche2010_bathtub,Lilly2013}.
On the other hand, it is challenging to account for the extremely high SFRs found for the FIR-bright sources by assuming MS hosts. 
Such an assumption would require stellar masses in excess of $\sim5\times10^{11}\,\Msol$.   
This would imply that the quasar hosts are among the most massive and rarest galaxies ever observed, at any redshift \cite[e.g.,][]{Baldry2012_GAMA_MF,Ilbert2013_UltraVISTA}, with number densities on the order of $\Phi\ltsim10^{-6}-10^{-5}\,\NDunit$ at $z\sim5$ \cite[e.g.,][]{Duncan2014_hiz_SMF,Stefanon2015_hiz_SMF}. 
In this context, we note that the quasars we study here are among the most luminous in the universe (by selection) and therefore also represent a population of rare objects, with number densities on the order of $\Phi\ltsim10^{-8}\,\NDunit$ \cite[e.g.,][]{Richards2006_QLF,McGreer2013_QLF_z5}.
Alternatively, it is possible that the FIR-bright sources are hosted in galaxies with masses that are comparable to those of the FIR-faint systems, but located well above the SF-MS. 
Such systems are typically associated with short periods of intense starburst activity. 
In particular, SFRs on the order of $>1000\,\mpyr$ are often observed in ``classical'' (i.e., luminous) SMGs, where they are attributed to late stages of major mergers of massive gas-rich galaxies \cite[see, e.g., the recent review by][and references therein]{Casey2014_SMGs_rev}. 
In the next section we use the available \cii\ data to constrain the (dynamical) masses of the quasar hosts, and of their companions, and return to the question of their location in the SFR-mass plane. 
We discuss the relevance of the major merger interpretation for our sample in \autoref{subsec:mergers}.

\subsubsection{Host and companion galaxy masses}
\label{subsubsec:mdyn}

\capstartfalse
\begin{deluxetable*}{lllccccccccc}
\tablecolumns{12}
\tablewidth{0pt}
\tablecaption{Galaxy Properties \label{tab:gal_props}}
\tablehead{
\colhead{sub-sample} &
\multicolumn{2}{c}{Target} &
\colhead{$\log\Ltir$ \tablenotemark{a}} &
\colhead{$\log\Ltir$ \tablenotemark{b}} &
\colhead{SFR \tablenotemark{a}}  &
\colhead{SFR \tablenotemark{b}}  &
\colhead{$\log\mdyn\,\sin^2 i$} &
\colhead{$\log\mdyn$ \tablenotemark{c}} &
\colhead{$\log\mbh$ \tablenotemark{d}} &
\colhead{$\mdyn/\mbh$} &  
\colhead{$\dot{M}_*/\Mdotbh$ \tablenotemark{e}} \\
        & ID     & comp.       & (\Lsol)        & (\Lsol)   & (\mpyr)       & (\mpyr)  &  (\Msol)  & (\Msol)   & (\Msol)  &           & 
}
\startdata
Bright 	& J0331  &  QSO        & $~12.85^{\rm f}$ & $ 12.80 $ & $~715^\dag $  & $  625 $ & $ 10.58 $ & $ 10.78 $ & $ 8.83 $ & $  88 $ & $  32 $ \\
~~~~~~ 	& J1341  &  QSO        & $ 13.48 $        & $ 13.55 $ & $ 3035 $      & $ 3529 $ & $ 10.67 $ & $ 10.86 $ & $ 9.82 $ & $  11 $ & $ 122 $ \\
~~~~~~ 	& J1511  &  QSO        & $ 13.23 $        & $ 13.34 $ & $ 1696 $      & $ 2180 $ & $ 10.80 $ & $ 10.85 $ & $ 8.42 $ & $ 264 $ & $ 191 $ \\
~~~~~~ 	& ~~~~~  &  SMG        & $ 12.42 $        & $ 12.28 $ & $  261 $      & $  191 $ & $ 10.77 $ & $ 10.78 $ & $  ... $ & $ ... $ & $ ... $ \\
~~~~~~ 	& ~~~~~  &  B$^{\rm g}$& $ 12.51 $        & $ 12.39 $ & $  326 $      & $  246 $ & $  ...  $ & $  ...  $ & $  ... $ & $ ... $ & $ ... $ \\
\hline \\ [-1.75ex]
Faint  	& J0923  &  QSO        & $ 12.69 $      & $ 12.56 $ & $  488 $      & $  361 $ & $ 10.51 $ & $ 10.87 $ & $ 8.68 $ & $ 158 $ & $  48 $ \\
~~~~~~ 	& ~~~~~  &  SMG        & $ 12.29 $      & $ 12.16 $ & $  195 $      & $  144 $ & $ 10.20 $ & $ 10.33 $ & $  ... $ & $ ... $ & $ ... $ \\
~~~~~~ 	& J1328  &  QSO        & $ 12.44 $      & $ 12.31 $ & $  277 $      & $  206 $ & $ 10.12 $ & $ 10.78 $ & $ 9.08 $ & $  50 $ & $  20 $ \\
~~~~~~ 	& ~~~~~  &  SMG        & $ 12.06 $      & $ 11.86 $ & $  114 $      & $   73 $ & $ 10.82 $ & $ 11.03 $ & $  ... $ & $ ... $ & $ ... $ \\
~~~~~~ 	& J0935  &  QSO        & $ 12.42 $      & $ 12.28 $ & $  261 $      & $  191 $ & $ 10.40 $ & $ 10.57 $ & $ 8.82 $ & $  56 $ & $  16 $
\enddata
\tablenotetext{a}{Calculated assuming a gray-body SED with $T_{\rm d}=47\,{\rm K}$ and $\beta=1.6$.}
\tablenotetext{b}{Calculated from the best-fit template SED of \cite{CharyElbaz2001_FIR}.}
\tablenotetext{c}{Calculated using the inclination-angle corrections derived from the sizes of the \cii-emitting regions.}
\tablenotetext{d}{Black hole masses taken from T11.}
\tablenotetext{e}{Calculated assuming the CE01-based SFRs, $\Mdotbh= \left(1-\eta\right) \Lbol/\eta c^2$, and $\eta=0.1$.}
\tablenotetext{f}{For J0331 the \herschel\ data suggest a significantly hotter gray-body SED with $T_{\rm d}=65\,{\rm K}$. This would result in $\log\left(\Ltir/\Lsol\right)=13.35$ and $\sfr=2225\,\mpyr$.}
\tablenotetext{g}{The FIR luminosities for the faint companion of J1511 that lacks \cii\ detection assume the same \cii-derived redshift as the quasar host.}

\end{deluxetable*}
\capstarttrue

The \cii\ measurements may be used to estimate the \emph{dynamical} masses (\mdyn) of the quasar host galaxies and the companion SMGs.
For this purpose, we employ the same prescription as used in several recent studies of \cii\ (and CO) emission in high-redshift sources, which assumes the ISM is arranged in an inclined, rotating disk  \cite[e.g.,][and references therein]{Wang2013_z6_ALMA,Willott2015_CFHQS_ALMA,Venemans2016_z6_cii}:
\begin{equation}
\mdyn = 9.8 \times 10^{8} \left(\frac{D_{\cii}}{\kpc}\right) \left[\frac{\fwcii}{100\,\,\kms}\right]^2\,\sin^2\left(i\right) \,\,\Msol \,\, .
\label{eq:mdyn_fwcii}
\end{equation}
In this prescription, $D_{\cii}$ is the size (deconvolved major axis) of the \cii-emitting region (as tabulated in \autoref{tab:lum_cii}).
The $\sin\left(i\right)$ term reflects the inclination angle between the line of sight and the polar axis of the hosts' gas disks, in which the circular velocity is given by $v_{\rm circ}=0.75\times\fwhm/\sin\left(i\right)$.
Practically, under the assumption of an inclined disk, the inclination angle is often derived from the (resolved) morphology of the line-emitting region, following $\cos\left(i\right)=\left(a_{\rm min}/a_{\rm maj}\right)$, where $a_{\rm min}$ and $a_{\rm maj}$ are the semi-minor and semi-major axes of the \cii\ emitting regions, respectively.

Such dynamical mass estimates carry significant uncertainties due to the different assumptions required to derive them and given the kind of data available for our systems.
In particular, our ALMA data may not be able to detect the more extended lower surface brightness \cii-emitting regions, thus underestimating $D_{\cii}$, and consequently \mdyn. 
In this sense, the \mdyn\ estimates derived from our ALMA data would only trace the very central \cii-emitting regions, on scales of a few kiloparsecs.
Faint extended \cii\ emission may also affect the inclination corrections, though this would probably be a subtle effect. 
More importantly, the inclination corrections for marginally resolved extended sources are somewhat sensitive to non-circular beam shapes, as is the case with some of our data.
On the other hand, deeper and higher resolution data may also reveal non-rotating ISM components, thus significantly altering the \mdyn\ estimates.
As noted in \autoref{subsec:lines}, the high central velocity dispersions we observe in our sources are most probably driven by the limited spatial resolution of our ALMA data and not by such non-rotating components.
Even if the ISM is indeed mostly found in a rotating disk, then determining the underlying (central) velocity dispersion would result in lower $v_{\rm circ}$, and therefore lower \mdyn.
Some of these effects are clearly demonstrated whenever increasingly deeper observations were obtained for some $z\gtrsim5$ sub-mm sources (e.g., \citealt{Maiolino2012_J1148_feedback,Cicone2015_J1148} for J1148 at $z=6.4$;  \citealt{Gallerani2012} for BRI~0952-0115 at $z=4.4$; and \citealt{Carniani2013_BR1202_CII} for BR~1202-0725 at $z=4.7$).
A more detailed discussion of these, and other uncertainties related to \mdyn\ estimates, is also given in \cite{Valiante2014_z6}.

Notwithstanding these uncertainties and caveats, we proceed to estimate the dynamical masses of the sources in our sample, using the prescription given in \autoref{eq:mdyn_fwcii}.
For the quasar hosts, the dynamical masses are in the range of $\mdyn\,\sin^2\left(i\right)\simeq\left(1-4.4\right)\times10^{10}\,\Msol$ (see \autoref{tab:gal_props}).
The three companion SMGs have $\mdyn\,\sin^2\left(i\right)=5.8\times10^{10}$, $1.57\times10^{10}$, and $6.6\times10^{10}\,\Msol$ for the SMGs accompanying J1511, J0923, and J1328, respectively.
We further derive rough estimates of the inclination angles based on the observed morphology of the \cii\ emission in all \cii-emitting systems (given in \autoref{tab:lum_cii}).
For the quasar host of J1511 and the accompanying SMG, where the \cii-emitting regions are not formally resolved, we use the upper limits on $a_{\rm min}$ and $a_{\rm max}$.
The inclination angles we deduce for our sources are in the range $i\sim28-80\deg$.
Taking these inclination corrections into account, we obtain dynamical masses of $\mdyn\simeq\left(3.7-7.5\right)\times10^{10}\,\Msol$ for the quasar hosts, 
while for the companion SMGs we have $\mdyn=6\times10^{10} $, $2.1\times10^{10}$, and $10.7\times10^{10}\,\Msol$ (for the SMGs accompanying J1511, J0923, and J1328, respectively).
We find no significant difference between the dynamical masses of FIR-bright and FIR-faint systems.
In what follows, we use these inclination-corrected estimates of \mdyn.

We first note that the dynamical masses of the quasar hosts cover a very narrow range, $\sim\left(3.7-7.5\right)\times10^{10}\,\Msol$ (i.e., spanning roughly a factor of 2), and five of the six systems have $\mdyn\sim\left(6-7.5\right)\times10^{10}\,\Msol$ (i.e., spanning less than 0.1 dex). 
Interestingly, the latter mass is in excellent agreement with the observed ``knee'' of the \emph{stellar} mass function in SF galaxies ($M^*$), which is known to show very limited evolution up to at least $z\sim3.5$ \cite[e.g.,][]{Ilbert2013_UltraVISTA,Muzzin2013}.
We also note that the dynamical masses of the interacting SMGs differ from those of the corresponding quasar hosts by factors of about 0.85, 0.3, and 1.8 (for the J1511, J0923, and J1328 systems, respectively). 
Given the uncertainties on our \mdyn\ estimates mentioned above, these mass ratios are consistent with our interpretation of these interacting systems being \emph{major} galaxy mergers (see \autoref{subsec:mergers} below).

To complement our estimates of dynamical masses, we also derive rough estimates of the \emph{dust} and \emph{gas} masses in our sources.
Dust masses are estimated assuming that the FIR continuum fluxes measured from our ALMA data are emitted by optically thin dust, following an SED with $T_{\rm d}=47\,{\rm K}$ and $\beta=1.6$, and further assuming an opacity coefficient of $\kappa_\lambda=0.77\left(850\,\mic/\lambda_{\rm rest}\right)^\beta$ (following \citealt{Dunne2000_SCUBA_Md}, for consistency with \citealt{Venemans2016_z6_cii}; see also, e.g., \citealt{Beelen2006_hiz_qsos}).
The dust masses we derive are in the range $M_{\rm dust}\sim\left(0.4-4.8\right)\times10^{8}\,\Msol$ for the quasar hosts and $\left(0.2-0.4\right)\times10^{8}\,\Msol$ for the companion SMGs.
Importantly, the dust masses of the quasar hosts comprise $<1\%$ of the dynamical masses. 
This qualitative result is virtually independent of the significant uncertainties involved in the dust mass estimates (due to the assumptions on the SEDs and on $\kappa_\lambda$).
Rough estimates of gas masses can then be inferred by assuming a (uniform) gas-to-dust ratio of 100.
These are rather conservative estimates, as several recent studies have shown that the gas-to-dust ratio in high-redshift hosts may be significantly lower \cite[e.g., as low as $\sim20-60$;][and references therein]{Ivison2010_z23_SMG,Banerji2017_redQSOs_ALMA}.
For most of the systems, and particularly the quasar hosts, the gas masses comprise $<20\%$ of \mdyn\ and reach $\sim60\%$ in only one quasar host (J1341). 
Adopting the aforementioned lower gas-to-dust ratios would obviously result in yet lower gas-to-dynamical mass ratios.
We conclude that the \mdyn\ estimates of our sources are dominated, to a large degree, by the stellar components within the galaxies.

Using our estimates of \mdyn\ as proxies for \mstar, we again find that all the FIR-bright systems are found well above the SF-MS, offset from the relation in \autoref{eq:sf_ms} by at least 0.5 dex (J0331) and by up to 1.2 dex (J1341). 
On the other hand, all the FIR-faint quasar hosts, as well as two of the three accompanying SMGs (those of J1511 and J0923), are consistent with the SF-MS, being within about 0.2 dex of the aforementioned relation, which is consistent with the intrinsic scatter associated with it.
%

\subsubsection{SMBH-host galaxy relations}
\label{subsubsec:mm}

\begin{figure}[t!]
\centering
\includegraphics[trim={1.4cm 0cm 7.5cm 0cm},clip,width=0.475\textwidth]{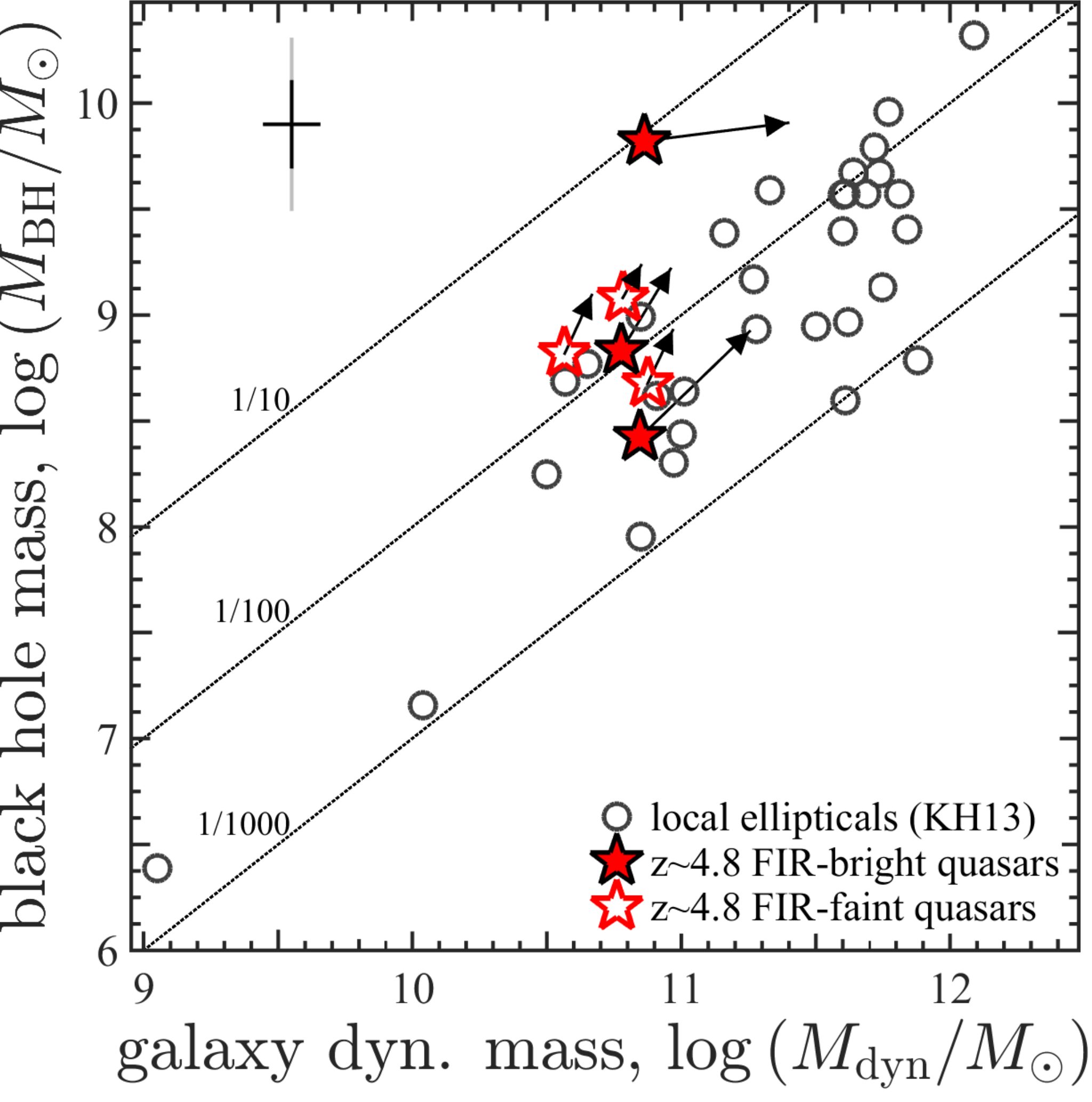} 
\caption{
Black hole masses, \mbh, vs. host galaxy dynamical masses, \mdyn, for our sample of \zfpe\ quasars (red stars), compared with a sample of $z\simeq0$ elliptical galaxies \cite[taken from][black circles]{KormendyHo2013_MM_Rev}. 
The dotted diagonal lines trace different constant BH-to-host mass ratios.
The crosses at the top-left corner illustrate representative measurement uncertainties (black) on both properties, and systematic uncertainties (gray) uncertainties on \mbh\ (of 0.4 dex). 
The significant systematic uncertainties on \mdyn\ (not shown) are more complicated, involving the poorly-constrained \cii\ emission sizes and inclinations (see \autoref{subsubsec:mdyn} for details).
Arrows indicate the possible evolution in both the BH and stellar components, assuming constant mass growth rates over a period of 50 Myr.
We note that all BH masses would increase by 0.24 dex if the calibration of \cite{TrakhtNetzer2012_Mg2} is adopted (see text for details).
Our quasars cover a wide range, with most systems being consistent with the ratio observed in the local universe and some exceeding $\mbh/\mdyn\sim1/100$.
The extreme object J1341, which has $\mbh/\mdyn\simeq1/10$ at \zfpe, is expected to evolve towards the locally observed ratio.
}
\label{fig:Mbh_Mdyn}
\end{figure}

We now turn to compare the mass growth rates and the masses of the SMBHs powering our quasars relative to those of the stellar populations in their host galaxies. 

For the quasar host galaxies, we assume that the mass grows only due to the formation of new stars at a rate determined by the CE01-based SFRs (see \autoref{subsubsec:sed_sfrs} above).
For the SMBHs, the growth rates are calculated assuming $\Mdotbh= \left(1-\eta\right) \Lbol/\eta c^2$, where \Lbol\ is the bolometric luminosity, estimated from the rest-frame UV continuum emission (see T11), and the radiative efficiency is assumed to be $\eta=0.1$.
We find that all systems have $\Mdotbh/\dot{M}_* > 1/200$ (see \autoref{tab:gal_props}), with the highest SFR systems J1341 and J1511 having $\Mdotbh/\dot{M}_* \simeq 1/190$ and $1/120$, respectively. 
The lower-SFR systems have growth-rate ratios as high as $\Mdotbh/\dot{M}_* \simeq 1/30$ (median value). 
These growth-rate ratios are consistent with those derived in N14 \cite[and in][]{Netzer2016_herschel_hiz}, which is expected given the consistency between the new ALMA data and the previous \herschel\ measurements.

As for the mass comparison, we rely on the dynamical mass estimates derived above and the \mgii-based BH masses available from T11.
These \mbh\ estimates used the calibration by \cite{McLure_Dunlop2004}. 
The more recent calibration by \cite{TrakhtNetzer2012_Mg2} would have increased \mbh\ by a factor of 1.75 ($\sim$0.24 dex), but we chose not to use it for the sake of consistency with our own previous work and with other samples of $z\gtsim5$ quasars \cite[see also][]{Shen_dr7_cat_2011,Mejia2016_XS_MBH}.
\autoref{tab:gal_props} lists \mbh\ and the BH-to-host mass ratios, which are in the range $\mbh/\mdyn\sim1/260-1/10$.
\autoref{fig:Mbh_Mdyn} shows our estimates of \mdyn\ and \mbh, along with some similar estimates in local galaxies.
For this, we use the subset of elliptical galaxies tabulated in \cite{KormendyHo2013_MM_Rev}, for which we take $\mhost=\mbul$.\footnote{Excluding NGC~4486B.} 
We note that \mbul\ itself, which is indeed the focus of many studies of various types of local galaxies \cite[see an extensive discussion in][]{KormendyHo2013_MM_Rev}, is not accessible with our data.

The mass and growth-rate ratios we derive can be compared to the typical ratios between the \emph{masses} of galaxies and their central SMBHs, as observed in the local universe. 
In the \mbh\ regime of our quasars, these are in the range $\mbh/\mstar\simeq1/300-1/200$ \cite[see \autoref{fig:Mbh_Mdyn}; e.g.,][]{MarconiHunt2003,HaringRix2004,Sani2011_MM,KormendyHo2013_MM_Rev}.
As \autoref{fig:Mbh_Mdyn} shows, this is broadly consistent only with the lower end of the $\mbh/\mdyn$ range we find in our quasars.
Moreover, for four of the \Ntot\ quasar hosts we find $\mbh/\mdyn\gtsim1/90$, which is significantly higher than the locally observed value. 
We stress that the BH-to-\emph{stellar} mass ratios of our sources would be even higher, thus increasing the discrepancy with the local value, recalling the alternative \mbh\ calibration mentioned above and recalling that $\mstar<\mdyn$ (\autoref{subsubsec:mdyn}).
At the same time, we have shown that \mdyn\ itself may be overestimated.
Similarly high BH-to-host mass ratios were derived for other luminous $z\gtrsim5$ quasars using similar data and methods (see, e.g., \citealt{Walter2009_CII_J1148,Venemans2012_z71_CII,Willott2015_CFHQS_ALMA,Venemans2016_z6_cii}, and references therein, but also \citealt{Lyu2016_FIR_SED}). 
This adds to the evidence for a general trend of increasing $\mbh/\mhost$ with increasing redshift, out to $z\sim2-3$, that is supported by several studies with direct estimates of \mmsmall\ \cite[e.g.,][and references therein]{Decarli2010_MM_evo,Merloni2010,Bennert2011,Bongiorno2014_MM,Trakhtenbrot2015_CID947}, as well as indirect arguments \cite[][]{Netzer2003_MM,Trakht_Netzer_MM_2010,Caplar2015_coeval}.
It should be noted, however, that the high AGN luminosities of our quasars (i.e., above the break in the quasar luminosity function) may mean that the high $\mbh/\mhost$ values we find do not represent the general (active) SMBH population at $z\sim5$ \cite[see, e.g.,][]{Willott2005_z6_comp,Lauer2007_MM_bias,Caplar2015_coeval}.

Turning back to the \mmdotsmall\ ratios, the high-SFR systems in our sample are broadly consistent with what is expected if one assumes that SMBHs and their hosts grow ``in tandem'',  obeying the BH-to-stellar mass ratio observed in high-mass systems in the local universe at all times.
On the other hand, the high \mmdotsmall\ ratios found for the rest of the sample (four of six quasars) suggests that in these systems the SMBHs are bound to \emph{over}grow the stellar populations.
To illustrate the possible evolutionary scenarios for our SMBHs and host galaxies, we plot in \autoref{fig:Mbh_Mdyn} the expected mass growth in both the BH and stellar components, assuming the observed (linear) growth rates are sustained for a short period of 50 Myr \cite[as in][]{Venemans2016_z6_cii}.
Within this short timespan, most systems will remain broadly consistent with the local range of \mmsmall\ ratios.
If the SMBHs would stop growing within a comparably short time-scale, at about $\mbh\sim10^{9}\,\Msol$, then their host galaxies would have to experience only a slightly longer period of  SF activity to reach the corresponding local mass ratios ($\mmsmall\sim1/300-1/200$).
However, if the BH growth rates are maintained over longer periods, and/or if the SMBHs are instead assumed to be growing exponentially (i.e., at constant \lledd, instead of constant \Lbol), then the SMBHs would significantly overgrow their hosts while approaching the highest BH masses known, $\mbh\sim10^{10}\,\Msol$, and moving away from the local BH-host relation. 
In this mass regime, the corresponding local mass ratios are on the order of $\mmsmall\sim1/100$, which would still require significant SF activity in our objects.

Our conclusions regarding the relative growth rates are in line with the recent results of \cite{Netzer2016_herschel_hiz}. 
This study used \herschel\ data for a large sample of some of the most luminous quasars at $z\gtrsim2$ (including all of our sources) and found that the vast majority of systems has $\mmdotsmall \gtsim 1/100$, while a small fraction had $\mmdotsmall \sim 1/200-1/100$.
Assuming that the most luminous quasars at $z\gtrsim2$ form a continuously evolving population, the \cite{Netzer2016_herschel_hiz} study also suggested that the epoch of fast SMBH growth traced by our sources would extend to $z\sim2-3$, perhaps at low duty cycles (see T11), while the intense SF activity seen in some of the quasar hosts may decrease shortly after $z\sim5$.
In the context of the \zfpe\ quasars we study here, this may indeed mean that our SMBHs would reach the high end of the known \mbh\ range, with $\mbh\sim10^{10}\,\Msol$. 
The (decreasing) SF activity is still required to suffice to reach $\mmsmall\sim1/100$.

We finally highlight the exceptional properties of the J1341 system, which has a high of $\mbh=6.6\times10^{9}\,\Msol$, an extremely high BH-to-host mass ratio of $\mmsmall\sim1/10$ and a low mass-growth ratio of $\mmdotsmall\sim1/190$.
All this suggests that the SMBH is approaching its final mass, while the host galaxy is forming stars at an intense rate and the system would likely expected to approach the high mass end of the $z\sim0$ mass ratio (see \autoref{fig:Mbh_Mdyn}). 
This is similar to the over-massive BH CID-947, recently identified at $z\simeq3.3$, and speculated to have experienced an earlier episode of fast Eddington-limited growth, to reach the observed $\mbh\simeq7\times10^{9}\,\Msol$ and $\mmsmall\simeq1/10$ \cite[][]{Trakhtenbrot2015_CID947}.\footnote{It is, however, worth bearing in mind that the two systems (J1341 at \zfpe\ and CID-947 at $z\simeq3.3$) are drawn from parent samples of markedly different number densities.}
J1341 may therefore be illustrative of the scenario of early, fast SMBH growth to the highest known mass well before $z\sim4.5$, with somewhat longer timescale stellar growth, to eventually reach $\mmsmall\sim1/100$.
%

\subsection{Major mergers among hosts of fast-growing SMBHs}
\label{subsec:mergers}

Given the results of \herschel\ analysis available prior to the new ALMA observations, the naive expectation for the sample under study was that the FIR-bright sources are powered by major mergers between gas-rich galaxies, while the FIR-faint sources are evolving secularly or, perhaps, are related to a \emph{later} evolutionary phase, where the accreting SMBHs may have already affected the SF in their hosts (see N14 for a detailed discussion).
Moreover, there is strong evidence that the occurrence rate of mergers among AGN increases with increasing AGN luminosity and redshift \cite[][]{Treister2012_mergers}.
Based on these trends, a high occurrence rate of mergers, in excess of $50\%$ and perhaps as high as $80\%$, is expected for our \zfpe, high-\Lagn\ (and high-\lledd) quasars.
The new high-resolution ALMA data allow us to critically revisit these ideas. 
The small size of our sample naturally limits the scope of our interpretation, however we note that most of the previous studies addressing these questions at comparably high redshifts included yet fewer objects and/or relied on lower-quality data.

At face value, our new ALMA data clearly show that a significant fraction of \zfpe\ luminous quasars -- 50\% in our small sample -- are interacting with companion galaxies of comparable mass, thus supporting the idea that major mergers are a dominant driver of the intense BH and SF activity.
Moreover, we note that this may constitute a lower limit on the real fraction of interacting quasar hosts, when considering the possibility that additional companions are locate just outside of the ALMA fields (i.e., separated by $\gtrsim50$ kpc); that additional close companions are too faint to be detected in our ALMA data (i.e., have $\sfr\ll100\,\mpyr$) and/or that some of the isolated quasars are actually in the final stages of a merger with interaction signatures that can only be detected with higher resolution data.

However, the properties of the SMBHs in our sample, their hosts and companion galaxies under study, highlight several shortcomings of the simplistic merger-driven growth scenario.
First, two of the three quasars with robust detections of physically interacting companions (i.e., SMGs) are actually among the FIR-\emph{faint} sources, while only one of the FIR-bright sources has an interacting companion (J1511). 
Second, the velocity maps of all the FIR-bright systems, and particularly those that lack companions, show evidence for ordered rotation, around an axis that coincides with the centroids of the host galaxies (and quasars' locations; see \autoref{subsec:lines} and \autoref{fig:velo_maps}). 
The coincidence of the region of zero velocity with the centroids of the SF activity in the hosts suggests that the redshifted and blueshifted \cii\ emitting regions are \emph{not} tracing (smaller) coalescing galaxies.
Moreover, as noted in \autoref{subsec:lines}, the outer parts of the ISM in the FIR-bright systems appear to be rotation dominated.
This is also seen among the FIR-faint systems, although to a lesser extent.

We also note that the speculation made in N14 that the FIR-faint systems are found in a \emph{later} evolutionary stage is disfavored by the new ALMA data, as (two of) the FIR-faint systems are seen to be in a rather early stage of a major merger. 
Their lower SFRs will therefore likely increase as the interacting galaxies coalesce.
In principle, a possible interpretation for our new ALMA data may have been that the SMBH activity in the FIR-faint sources is driven by the ``first passage'' of the interacting quasar hosts with their SMG companions. 
Indeed, the high SFRs of the interacting galaxies in the J1511 system may be indicative of what the FIR-faint systems would undergo in later stages of the interaction. 
However, most simulations of major mergers suggest that the SMBH would produce a luminous quasar, with high \lledd, at the final coalescence phase, and \emph{not} in the first passage phase.
These same simulations also suggest that the first passage enhances SFR in the interacting hosts, to levels comparable to those found in the later, final coalescence phase (see, e.g., \citealt{Blecha2011,DeBuhr2011,Sijacki2011,VanWassenhove2012_mergers}, but also \citealt{Volonteri2015_SF_BH_mergers} and \citealt{Gabor2016_GASOLINE_RAMSES}).
Our sample, on the other hand, shows similarly intense SMBH growth, with $\Lagn\simeq10^{47}\,\ergs$ and $\lledd\sim0.5-1$, for both high- and low-SFR quasar hosts and among systems with and without an interacting companion.

Any direct comparison with merger simulations is further complicated by the possibility that the companion SMGs we detect at larger separations of $\sim40\,\kpc$ may have already experienced a much closer passage to the quasar hosts and are observed close to their apocenter. 
The intense SMBH growth may have been triggered during this pericenter passage, when tidal forces were maximal.
In this context, we note that any observations of interacting galaxies would be biased toward large separations, due to the longer periods spent at increasingly larger separations.
Indeed, an inspection of several of the aforementioned simulations suggests that the interacting galaxies are separated by $\gtrsim20\,\kpc$ for over 80\% of the simulated merger. First and second passages are extremely short, taking up $\sim5\%$ of the time.\footnote{It is worth noting that many simulations are set up in a way that will lead to a merger within a relatively short time, due to computational limitations. The fraction of time spent at large separations, on the order of $\sim40\,\kpc$, may therefore be even higher.}

We therefore conclude that our ALMA data provide compelling evidence for significant galaxy--galaxy interactions (major mergers) in some, but not all, quasar hosts. 
Moreover, the links between these interactions and the intense accretion onto the SMBHs remains unclear.
For the FIR-bright sources, we caution that even with the new ALMA data, we cannot completely disprove the possibility that \emph{all} these systems are indeed observed in the advanced stages of a major merger.
In particular, the evidence for rotation in the ISM of some quasar hosts does not by itself disprove a merger scenario, 
as several studies of low-redshift mergers (i.e., ULIRGs) have identified nuclear structures of rotating molecular gas, on scales of a few kiloparsecs (see, e.g., \citealt{Downes1998,Tacconi1999_NGC6240,Ueda2014_molecular_mergers} and the discussion in \citealt{Hodge2012_SMG_z4}).
Simulations of (low-redshift) major mergers have demonstrated that such gas disks may indeed form several hundreds of Myr after the ``peak'' of the merger, once the coalesced nucleus has relaxed \cite[e.g.,][]{Barnes2002_merger_disks,Springel2005_merger_spiral}. 
On the other hand, the interacting companions we identify among the FIR-faint systems cannot be related to the SF and SMBH activity in a straightforward way, given their large separations.
The only way in which \emph{all} the quasars in our sample can be explained as triggered by major merger is if the FIR-bright systems that lack an interacting companion (namely J0331 and J1341) are observed in the final coalescence stage, while the two interacting FIR-faint quasar hosts have already experienced a close first passage that triggered the SMBH growth and are observed close to their apocenter. 
Even this scenario fails to account for one of the FIR-faint systems (J0935), which has no companions out to $\sim50\,\kpc$.

Several sub-mm studies published in recent years have probed the existence of SF galaxies that are accompanying, or indeed interacting with, the hosts of $z\gtrsim4.5$ quasars.
A prominent example is the interacting system BR1202-0725 at $z=4.7$, which was shown to consist of a pair of interacting SMGs, separated by 25 kpc, one of which hosting a luminous quasar, with an additional (marginal) detection of a faint AGN in the other one \cite[][]{Iono2006_z47,Wagg2012_z47_QSO,Carniani2013_BR1202_CII}.
However, such close interactions are found to be quite rare among high-redshift quasars.
Several deep optical imaging campaigns in the fields around the known $z\simeq6$ quasars provide no or little robust evidence for physically associated companions \cite[e.g.,][]{Willott2005_z6_comp}.
Moreover, the sub-mm (ALMA and PdBI) studies published for most of the known $z\gtrsim6$ quasars \cite[by][]{Wang2013_z6_ALMA,Willott2013_z6_ALMA,Willott2015_CFHQS_ALMA,Venemans2016_z6_cii} -- which could have detected companions with SFRs comparable to those we find at separations of up to $\sim$100 kpc -- report \emph{no} such companions.

Taken at face value, this may suggest a fast increase in the occurrence of mergers among luminous quasar hosts between \zsix\ and \zfpe\ -- that is, within $\sim$350 Myr. 
We note, however, that this discrepancy may be driven by the limited sensitivity and/or limited spatial resolution of the available sub-mm (ALMA) data for some \zsix\ quasars (e.g., $\sim$4 kpc in \citealt{Wang2013_z6_ALMA}). 
Additional deep observations of these \zsix\ sources at sub-kiloparsec resolution may indeed resolve this discrepancy.

To conclude the discussion of major mergers, we recall that our SMBHs -- as well as the other $z\gtrsim5$ quasars mentioned above -- had to grow continuously and at high rates since very early epochs to account for their high masses (see T11, and references therein).
Given the number of interacting companions we find and their separations from the quasars' hosts, it seems unlikely that this kind of mergers can be the only driver of such a prolonged period of fast growth.  
All of the above suggests the epoch of fastest growth of the most massive BHs is driven, at least in part, by mechanisms that are not related to major mergers - such as direct accretion of IGM gas \cite[e.g.,][]{Dekel2009_cold_streams,DiMatteo2012,Dubois2012_hiz_inflows}, minor mergers, and/or galaxy-scale instabilities of the gas or stellar components  \cite[e.g.,][]{Springel2005_sims,Bournaud2011_VDI_sims,Bournaud2012_VDI_obs}.

Finally, we note that even if the companion SMGs we find are not directly related to the fueling of the fast-growing \zfpe\ SMBHs, their presence seems to support the idea that rapid early BH growth preferentially takes place in dense large-scale environments \cite[e.g.,][]{DiMatteo2008,Sijacki2009,Costa2014_z6_env_sims,Dubois2012_hiz_inflows}.
To date, such evidence has been highly elusive, with several observational campaigns looking for over-densities of (rest-frame) UV-bright galaxies around luminous high-redshift quasars yielding highly ambiguous results (see, e.g., \citealt{Overzier2006_z52,Kim2009_idrops_z6,Utsumi2010_env_J2329,Husband2013} for several examples of such over-dense environments, but also \citealt{Willott2005_z6_comp,Banados2013_z57_env,Simpson2014_ULASJ1120_env} and some of the systems in \citealt{Kim2009_idrops_z6} for the contrary).
Our analysis demonstrates that such studies may be significantly biased against dust-obscured high-redshift SF galaxies, thus underestimating the real (over-)density of galaxies around high-redshift quasars.
Indeed, the three interacting companions we identify were \emph{not} identified in the \Spitzer\ images of the respective quasars (N14).
A complete census of the cosmic environments of high-redshift, fast-growing SMBHs would therefore require a multi-wavelength approach, covering scales of up to a few arc-minutes, and in spectral regimes that are not affected by dust obscuration. 
This can be done, for example, using compact sub-mm arrays.

\section{Summary and Conclusions}
\label{sec:summary}

We have presented ALMA \obsband\ observations of \Ntot\ luminous quasars at \zfpe, drawn from a sample of 40, UV-selected SDSS quasars with a wealth of supporting multi-wavelength data.
The data probe the rest-frame far-IR continuum emission that arises from cold dust, heated by SF in the host galaxies of the quasars, as well as the \CII\ emission line that originates from the cold phase of the hosts' ISM.
The ALMA observations resolve the continuum- and line-emitting regions on scales of $\sim2$ \kpc.
Our main findings are as follows.

\begin{enumerate}
 
 \item 
 All quasar hosts are clearly detected and resolved, in both continuum and \cii\ line emission. 
 The continuum emission suggests intense SF, with the FIR-bright sources reaching $\sim1000-3000\,\mpyr$, consistent with \herschel\ observations of these systems. 
 The quasar hosts exhibit evidence for massive, rotation-dominated gas structures.
 
 \item 
 Three quasar hosts -- one FIR-bright and two FIR-faint systems -- are accompanied by spectroscopically confirmed, interacting companions, with separations in the range $\sim14-45$ \kpc\ and within $\left|\Delta v\right|\ltsim450\,\kms$.
 The companions themselves are forming stars at rates of a few hundred \mpyr, slightly lower than the quasar hosts with which they interact.
 
 \item 
 The remaining quasar hosts -- two FIR-bright and one FIR-faint -- lack significant companions. 
 This, combined with the evidence for rotation, may suggest that processes other than major mergers are driving the significant SF activity and fast SMBH growth in these systems.
 
 \item
 The dynamical masses of the quasar hosts, estimated from the \cii\ lines, are within a factor of $\sim$3 of the masses of the interacting companions, supporting our interpretation of these interactions as major mergers.

 \item 
 The \cii-based dynamical masses also show that the FIR-faint systems are consistent with the ``main sequence'' of star-forming galaxies, while the FIR-bright systems are located above it. 
 
 \item
 Compared with the BH masses, the \cii-based dynamical host masses are generally lower than what is expected from the locally observed BH-to-host mass ratio. 
 In some of the systems, this discrepancy may grow further, given the high accretion rates of the SMBHs.
 
 \item 
 The \LLfir\ ratios in the quasar hosts are consistent with those found in other $z\gtrsim5$ quasar hosts and SMGs and follow the observed trend of declining \LLfir\ with increasing \Lfir.
 Although our data suggest that the \cii\ deficit is most probably driven by mechanisms or properties that are intrinsic to the quasar hosts, we do not find evidence for the compactness of the SF regions being the driver of the $\LLfir-\Lfir$ trend.
 
\end{enumerate}

Our analysis clearly demonstrates the wide variety of host galaxy properties, particularly in terms of SFRs and of possible SMBH fueling mechanisms, among a relatively uniform population of the fastest-growing SMBHs in the early gas-rich universe.
It appears that vigorous SMBH growth is not necessarily accompanied by extreme SF activity (i.e., above what is found in inactive SF galaxies) and that galaxy--galaxy interactions are not a necessary condition for either of the two processes.
This broadly supports a scenario where intense SMBH and stellar growth in the early universe is driven by secular processes, such as large-scale flows of cold gas, that penetrate into the centers of massive dark matter halos and/or gas or stellar instabilities on smaller scales.

Our results motivate several paths for follow-up studies to address and test the predictions of the different fueling mechanisms.
To robustly determine the role of mergers in the systems that lack companions quasars would require the detection of tidal features (e.g., using \jwst\ imaging), or mapping the ISM kinematics at higher resolution and/or to larger scales (i.e., with deeper ALMA observations).
Obviously, a critical test of the relevance of mergers to the general population of high-redshift quasars necessitates a significantly larger sample, with observations that cover a large field of view while maintaining a high spatial resolution.
This can be achieved by extending our analysis to additional $z\sim5$ and $z\sim6$ quasars.
We have recently guaranteed cycle-4 ALMA \obsband\ time to observe 12 additional \zfpe\ quasars from the T11 sample, which would allow us to study the host galaxies and close environments of a total of 18 fast-growing, \zfpe\ SMBHs.

\vskip 1cm 
\section*{}

\acknowledgements

We thank the anonymous referee for the very constructive comments that helped us to improve our manuscript.
We thank K.\ Schawinski, L.\ Mayer, R. Teyssier, P. Capelo, M. Dotti and D.\ Fiacconi for useful discussions.
This paper makes use of the following ALMA data: ADS/JAO.ALMA\#2013.1.01153.S. 
ALMA is a partnership of ESO (representing its member states), NSF (USA) and NINS (Japan), together with NRC (Canada), NSC and ASIAA (Taiwan), and KASI (Republic of Korea), in cooperation with the Republic of Chile. The Joint ALMA Observatory is operated by ESO, AUI/NRAO, and NAOJ.
This work made use of the MATLAB package for astronomy and astrophysics \cite[][]{Ofek2014_matlab}.
H.N.\ acknowledges support by the Israel Science Foundation grant 284/13.
C.C.\ gratefully acknowledges support from the Swiss National Science Foundation Professorship grant PP00P2\_138979/1. 
C.C.\ also acknowledges funding from the European Union's Horizon 2020 research and innovation programme under the Marie Sklodowska-Curie grant agreement No 664931.
R.M.\ acknowledges support by the Science and Technology Facilities Council (STFC) and the ERC Advanced Grant 695671 ``QUENCH''.

\noindent
{\it Facility:} \facility{ALMA (\obsband)}\\
{\it Software:} CASA





\end{document}